\documentclass[natbib]{svjour3}                     % onecolumn (standard format)
\journalname{Space Science Reviews ~~~~~~~~~~~~~~~~~~~ DOI 10.1007/s11214-015-0141-3 ~~~~~~~~~~~~~~~~~~~}
\usepackage{times, graphicx}
\usepackage{cool} % For partial derivatives macros
\usepackage{breqn} % For the dmath environment (multiline equations)
\usepackage{bm} % bold math
\usepackage{fullpage}
\newcommand{\kms}{km\,s$^{-1}$}
\newcommand{\mms}{Mm\,s$^{-1}$}
\newcommand{\kgm}{kg\,m$^{-3}$}

\newcommand{\solphys}{\textit{Sol. Phys.}}
\newcommand{\apj}{\textit{ApJ}}
\newcommand{\apjs}{\textit{ApJ Supp.}}
\newcommand{\apjl}{\textit{ApJL}}
\newcommand{\aap}{\textit{A\&A}}
\newcommand{\araa}{\textit{ARA\&A}}
\newcommand{\pasj}{\textit{PASJ}}
\newcommand{\apss}{\textit{Ap\&SS}}
\newcommand{\jcomputphys}{\textit{J. Comput. Phys.}}
\newcommand{\sovast}{\textit{Sov. Astron.}}
\newcommand{\na}{\textit{New A.}}
\newcommand{\nat}{\textit{Nature}}
\newcommand{\ssr}{\textit{Space Sci. Rev.}}
\newcommand{\procspie}{\textit{Proc. SPIE}}
\newcommand{\mnras}{\textit{MNRAS}}
\newcommand{\zap}{\textit{ZAp}}
\newcommand{\ao}{\textit{Appl. Opt.}}
\newcommand{\bullamer}{\textit{Bull. Amer. Meteor.}}
\newcommand{\skytel}{\textit{S\&T}}
\newcommand{\nad}{{Na~{\sc{i}}~D$_{1}$}}
\newcommand{\cak}{{Ca~{\sc{ii}}~K}}
\newcommand{\cah}{{Ca~{\sc{ii}}~H}}
\newcommand{\mgb}{{Mg~{\sc{i}}~b$_{2}$}}

\newcommand{\vAi}{\ensuremath{v_{A_i}}}

\newcommand{\vAe}{\ensuremath{v_{A_e}}}

\newcommand{\vSi}{\ensuremath{v_{s_i}}}

\newcommand{\vSe}{\ensuremath{v_{s_e}}}

\newcommand{\va}{\ensuremath{v_{A}}}
\newcommand{\vs}{\ensuremath{v_{S}}}
\newcommand{\vav}{\ensuremath{\bm{v}_{A}}}

\newcommand{\vsm}{\ensuremath{v_{slow}}}
\newcommand{\vfm}{\ensuremath{v_{fast}}}
\newcommand{\vsmv}{\ensuremath{\bm{v}_{slow}}}
\newcommand{\vfmv}{\ensuremath{\bm{v}_{fast}}}
\newcommand{\Bp}{\ensuremath{\delta\!\bm{B}}}
\newcommand{\pp}{\ensuremath{\delta\! p}}
\newcommand{\pTp}{\ensuremath{\delta\! p_T}}
\newcommand{\rhop}{\ensuremath{\delta\! \rho}}
\newcommand{\vp}{\ensuremath{\delta\! \bm{v}}}

\newcommand{\planeExp}{\ensuremath{e^{i(\bm{k} \cdot \bm{x} - \omega t)}}}
\newcommand{\Beq}{\ensuremath{\bm{B}_0}}
\newcommand{\peq}{\ensuremath{p_0}}
\newcommand{\rhoeq}{\ensuremath{\rho_0}}
\newcommand{\vph}{\ensuremath{v_{ph}}}

\newcommand{\Mi}{\ensuremath{m_i}}
\newcommand{\Me}{\ensuremath{m_e}}

\newcommand{\vTi}{\ensuremath{v_{T_i}}}

\newcommand{\vTe}{\ensuremath{v_{T_e}}}
\newcommand{\twopartdef}[4]
{
	\left\{
		\begin{array}{ll}
			#1 & \mbox{for } #2 \\
			#3 & \mbox{for } #4
		\end{array}
	\right.
}

\begin{document}
\title{Multiwavelength studies of MHD waves in the solar chromosphere}
\subtitle{An overview of recent results}

\titlerunning{MHD waves in the solar chromosphere} % if too long for running head

\author{D. B. Jess, R. J. Morton, G. Verth, V. Fedun, \\
S. D. T. Grant \& I. Giagkiozis}

\authorrunning{D. B. Jess et al.} % if too long for running head

\institute{D. B. Jess \& S. D. T. Grant \at
Astrophysics Research Centre\\
School of Mathematics and Physics\\
Queen's University Belfast\\
Belfast BT7 1NN, Northern Ireland, UK \\
%Tel.: +44 28 9097 3045\\
%Fax: +44 28 9097 3997\\
\email{d.jess@qub.ac.uk, sgrant19@qub.ac.uk}           %  \\
\and
R. J. Morton  \at
Department of Mathematics \& Information Sciences \\
Northumbria University \\
Newcastle Upon Tyne, NE1 8ST, UK \\
\email{richard.morton@northumbria.ac.uk}
\and
G. Verth \& I. Giagkiozis  \at
Solar Physics and Space Plasma Research Centre (SP$^{2}$RC) \\
The University of Sheffield \\
Hicks Building, Hounsfield Road \\
Sheffield, S3 7RH, UK \\
\email{g.verth@sheffield.ac.uk, i.giagkiozis@sheffield.ac.uk}
\and
V. Fedun  \at
Space Systems Laboratory \\
Department of Automatic Control and Systems Engineering \\
University of Sheffield \\ 
Sheffield, S1 3JD, UK \\
\email{v.fedun@sheffield.ac.uk}
\and
I. Giagkiozis \at
Complex Optimization and Decision Making Laboratory \\
Automatic Control and Systems Engineering Department \\ 
University of Sheffield \\ 
Sheffield, S1 3JD, UK \\
\email{i.giagkiozis@sheffield.ac.uk}
}

\date{Received: 7 November 2014 / Accepted: 15 February 2015}

\maketitle

\begin{abstract}
The chromosphere is a thin layer of the solar
atmosphere that bridges the relatively cool
photosphere and the 
intensely 
heated transition region and corona. 
Compressible and incompressible waves propagating 
through the chromosphere can supply significant 
amounts of energy to the interface region and corona.
%Compressible and
%incompressible waves propagating through the
%chromosphere have long been
%considered viable transportation mechanisms to
%provide energy to the outer regions
%of the solar atmosphere. 
In recent years an abundance
of high-resolution observations
from state-of-the-art facilities have provided new and
exciting ways of disentangling
the characteristics of oscillatory phenomena propagating
through the dynamic chromosphere.
Coupled with rapid advancements in
magnetohydrodynamic wave theory, we are now in
an ideal position to thoroughly investigate the role
waves play in supplying energy to
sustain chromospheric and coronal heating. Here,
we review the recent progress made in
characterising, categorising and interpreting oscillations
manifesting in the solar
chromosphere, with an impetus placed on their
intrinsic energetics.

\end{abstract}

\section{Introduction}
\label{intro}
Ever since the Sun's corona was found to be dominated by emission lines
characteristic of multi-million degree temperatures, it was obvious that the
heating of the plasma was not dominated by purely thermodynamic processes.
As a result, research quickly built momentum in an attempt to understand which
non-thermal processes, especially those of magnetic origin, were responsible
for the continual supply of energy. This has since become known as
the ``coronal heating problem''.
%Originally, during the total solar eclipse of 1869~August~7
%which passed over Russia and North America, \citet{You71} detected a faint emission line
%in his solar spectrum at $5303$~{\AA}. This emission feature was untraceable to currently known
%elements, and was thus designated as a potentially new chemical element, {\it{coronium}}, which was
%subsequently renamed {\it{newtonium}} at the turn of the 20$^{th}$ century \citep{Men03}. It was
%not until the 1930s that researchers finally uncovered the origin of the mysterious $5303$~{\AA}
%emission line. Concurrent work by \citet{Gro31, Gro39} and \citet{Edl43, Edl45} identified the
%emission feature as belonging to highly-ionised iron (Fe$^{13+}$), thus revolutionising our
%appreciation of the Sun's temperature structure. However, at the same time this newfound
%knowledge instilled a paradoxical dilemma that has been plaguing astronomers and physicists
%ever since: {\it{How is the Sun's corona heated to (and maintained at) multi-million degree
%temperatures?}}
Over the years, efforts to provide a conclusive heating mechanism for the outer solar atmosphere
have produced two (seemingly) distinct classes of theory: magnetic reconnection and waves.
In the former, it is suggested that regular reconfigurations of the embedded
magnetic field lines will produce extreme localised heating through the conversion of magnetic
energy into heat \citep{Pri86, Pri99}. Large-scale flare events are one of the most dramatic
eruptive phenomena on our Sun that can be triggered by magnetic reconnection,
often releasing in excess of $10^{31}$~ergs
of energy during a single event. However, the relative rarity of these large-scale flares
means that they cannot provide the necessary basal heating that the outer solar atmosphere
requires to maintain its multi-million degree temperatures. Instead, it has been suggested that
rapidly occurring, small-scale flare events, or ``nanoflares'' with individual energies
$\sim10^{24}$~ergs, may occur with such regularity in the solar atmosphere that
they can provide the continual source of heat required to maintain the elevated
temperatures \citep{Parker88}. Unfortunately, however, the small spatial sizes and radiative
signatures of such events places them within or below the noise threshold of current
observations \citep{Ter11}, and therefore only tentative evidence exists to support their presence
in the outer solar atmosphere \citep{Kli01, Bra12, Tes13, Jes14}.

\begin{figure}
\centerline{\includegraphics[width=0.97\textwidth]{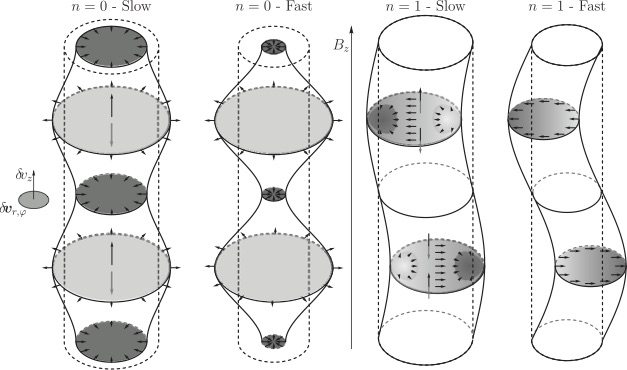}}
\caption{Schematic diagrams of MHD waves in synthetic cylindrical waveguides.
The velocity perturbations in the $z$ direction, denoted by $\delta\! v_z$, are
depicted by vertical arrows within the tube while the velocity perturbations in
the $(r,\varphi)$-plane, $\delta\! \bm{v}_{r,\varphi}$, are illustrated using horizontal
arrows. The horizontal plane cuts on the flux tubes illustrate the density perturbations,
with darker and brighter shades signifying higher and lower densities, respectively, 
with respect to equilibrium. The two schematics to the left represent
slow and fast sausage modes (with azimuthal wavenumber $n=0$), while the two figures to the right represent
slow and fast kink modes ($n=1$). Notice that for the slow modes the main component
of the velocity perturbation is in the $z$ direction (plasma-$\beta<1$), which is associated with,
in contrast to the fast modes, stronger density perturbations.
\label{fig:chromosphere:cartoon}}
\end{figure}

\vspace{3mm}
On the other hand, wave heating theories
can be substantiated by a vast number of publications detecting oscillatory phenomena throughout
the solar atmosphere since the early 1960s \citep{Lei60, Lei62, Noy63}.
Purely wave-based heating requires that waves, generated near the solar surface through
the continual convective churning of plasma, propagate upwards, dissipate a considerable portion
of their energy in the chromosphere, and {\it{still}} have sufficient energy remaining to
heat the corona. However, the solar atmosphere is highly magnetic in nature. 
Localised magnetic
field strengths often exceed $1000$~G, and can even exceed $6000$~G in 
extreme cases \citep{Liv06}, resulting in the oscillatory modes becoming
highly modified, producing anisotropic waves that can be accurately modelled using
magnetohydrodynamic (MHD) approximations
\citep[][to name but a few]{Rob81a, Rob81b, Edw83, Cal86, Has87, Goo92, Nak95, Erd06a, Erd06b, Erd07b, Erd10, Ver08}.
In the MHD approximation there exist three types of waves,
Alfv{\'{e}}n (see Figure~{\ref{fig:movie:alfven}}), \textit{fast} and \textit{slow} magnetoacoustic waves
\citep[Figure~{\ref{fig:chromosphere:cartoon}};][]{Erd97, Zhu99, Nak05}. A number of
wave modes have been observed at discrete layers of the solar atmosphere, ranging from the
deepest depths of the photosphere through to the outermost extremities of the corona
\citep[][to name but a few of the hundreds of examples to date]
{Ulr70, Pen93, Asc99, Asc02, Asc04, Nak99, Bal11, Mor11, Scu11, Jes12b, Sri13, Lun14}.
However, the goal is now to utilise multiwavelength observations to be able to
track the waves as a function of height, ultimately allowing researchers to diagnose
changes in wave energy and look for the corresponding signatures of localised
atmospheric heating. In the past flare and wave heating mechanisms have often been
considered as opposing and deeply conflicting viewpoints. However, in more recent years
with the advent of higher sensitivity instrumentation, it has become apparent that not only
can eruptive flare events trigger oscillatory phenomena
\citep[e.g.,][]{Ver04, Wan04, DeM07, VanD07, VanD09, Jes08b, Lun08, Sri13a, Yuan13},
but that waves interacting with magnetic field lines can also induce the instabilities
necessary to incite reconnective phenomena \citep[e.g.,][]{Iso06, Iso07, Jes10c, Li12, Jac13, She14}. 
Furthermore, \citet{Che10, Che11} have also demonstrated how 
MHD waves can be initiated within large-scale coronal streamers 
following the impact of a rapidly propagating coronal 
mass ejection, suggesting how 
oscillatory motion can be triggered over an incredibly wide 
range of spatial scales.
Therefore, it does not seem inconceivable that the once opposing viewpoints may actually
work in harmony to sustain the basal heating required to balance atmospheric radiative losses.

\vspace{3mm}
While the majority of research over the last 70 years has been dedicated
to the understanding of multi-million degree coronal signatures, it is the solar chromosphere
that provides more tantalising prospects for rapid advancements in astrophysical understanding.
Even though the chromosphere is only a thin layer spanning approximately $1000$~km,
it may play a pivotal role in our understanding by acting as the interface between the
relatively cool photospheric plasma and the super-heated corona. Furthermore, whilst
the chromosphere is only heated to a few thousand
degrees above the corresponding photospheric layer, the relatively high
densities found within the chromosphere, compared to those in the corona,
means that it requires at least double the
energy input to balance its radiative losses \citep[Table~{\ref{Withbroe}},][]{Wit77, And89}.
Typical chromospheric radiative losses are on the order of
$10^{6}$--$10^{7}$~erg{\,}cm$^{-2}${\,}s$^{-1}$ (or $1000$--$10{\,}000$~W{\,}m$^{-2}$), compared
with values of $10^{4}$--$10^{6}$~erg{\,}cm$^{-2}${\,}s$^{-1}$ (or $10$--$1000$~W{\,}m$^{-2}$)
for the solar corona \citep{Wit77}.
As a consequence, the solar chromosphere is universally recognised as an important layer
when attempting to constrain any potential energy transfer mechanisms between the photosphere
and the corona.

\vspace{3mm}
Seismological approaches have long been used to characterise solar 
atmospheric structuring through the analysis of propagating and 
standing wave motion. Dating back to the mid 1970s, the first detection 
of a truly global solar pressure oscillation inspired researchers to use 
such data to investigate the properties of the solar interior, hence 
initiating the field of helioseismology \citep{Hill75, Brown78}. Then, 
following the launch in the 1990s of (at the time) high resolution satellite imagers 
capable of observing the Sun's corona, numerous examples of 
wave and oscillatory behaviour were detected through EUV diagnostics 
\citep[e.g.,][to name but a few of the early examples]{Ofman1997, Def98, Asc99, Nak99}. 
This led researchers to probe the detected oscillatory 
phenomena in order to better understand coronal parameters that 
were unresolvable using traditional imaging and/or spectroscopic 
approaches, thus creating the field of coronal seismology 
\citep[see, e.g., the review paper by][]{Nak05}. Coronal seismology 
has proven to be a powerful tool, with vast numbers of high-impact 
publications produced to date, including those related to the uncovering 
of magnetic fields \citep{Nak01}, 
energy transport coefficients \citep{Asc03} and sub-resolution fine-scale 
structuring \citep{King03}. Ultimately, the goal is to employ such 
seismological techniques in order to better understand the energy 
dissipation rates within the corona, and therefore determine the 
specific role MHD wave and oscillatory phenomena play in providing 
heat input to the outer solar atmosphere. Of course, a natural extension 
is to apply such innovative approaches to the solar chromosphere, a 
region that is rife with ubiquitous wave activity. This form of analysis 
has only recently risen to the forefront of chromospheric research, 
aided by the recent advancements made in telescope facilities, 
instrumentation and theoretical knowledge.

\vspace{3mm}
From a purely theoretical and modelling point of view, 
the chromosphere presents a substantially different 
plasma environment for MHD wave modes compared 
to the corona. The coronal plasma regime modelled 
for such waves often assumes a one-fluid, low 
plasma-beta and fully-ionized plasma. In contrast, 
realistic MHD modelling of chromosphere 
should be multi-fluid, finite plasma-beta and include 
the additional effects of partial ionisation and
 radiative transfer under non-local thermodynamic equilibrium 
(non-LTE) conditions 
\citep{Han07}. However, even in this more complex 
plasma environment, on observable MHD time/length 
scales, the particular defining properties of different 
wave modes in fine-scale magnetic flux tubes 
remain unchanged.  However, such modes, including 
torsional Alfv{\'{e}}n, sausage and kink, could be 
subject to frequency-dependent effects not 
encountered in the corona 
\citep[e.g., ion-neutral damping;][]{Soler13, Soler15}.
This has important implications for understanding the 
true nature of wave-based heating in the chromosphere. 
Furthermore, such frequency dependent effects must 
also be taken into account when performing remote 
plasma diagnostics from MHD wave mode 
observations, i.e., {\it{chromospheric seismology}}.

\vspace{3mm}
Over the last decade there has been a significant number of 
reviews published that document the abundance of 
MHD wave phenomena in the outer solar atmosphere. Such 
detailed overviews include quasi-periodic \citep{Nak05SSR}, 
standing \citep{Wan11}, magnetoacoustic \citep{VanD09b, DeM09} 
and Alfv{\'{e}}n \citep{Mat13} waves. However, the majority of these 
reviews are solely focused on coronal oscillations, and as a result, 
choose to ignore the presence of MHD waves occurring in the lower 
regions of the solar atmosphere. Older review articles have touched 
on the manifestation of waves and oscillations in the 
solar chromosphere, including those that discussed observations of 
spicules \citep{ZAQERD2009}, filaments \citep{Lin11} and 
more-general chromospheric plasma 
\citep[e.g.,][]{Fri72, Bon81, Nar90, Nar96, Tar09}. 
However, since the confirmation of omnipresent waveforms in 
the chromosphere is a relatively recent achievement, until now 
there has been a distinct lack of a dedicated and wide-ranging 
review article that details both the observational and theoretical 
advancements made in chromospheric wave studies. As a result, 
we now take the opportunity to gather recent observational 
and theoretical publications and provide the solar physics 
community with a thorough overview of ubiquitous MHD 
wave phenomena intrinsic to the solar chromosphere.

\begin{table}[!t]
\begin{center}
\footnotesize
\caption{Energy losses experienced in quiet Sun, coronal hole and active
region locations at both coronal and chromospheric heights. Regardless of the solar
location it is the chromosphere that displays the greatest energy losses. Table adapted from \citet{Wit77}.
\label{Withbroe}}
\begin{tabular}{lccc}
~&~&~ \\
\hline
\vspace{-3mm} \\
Parameter			& Quiet		& Coronal	& Active  	 	\\
					& Sun		& hole 		& region		\\
\vspace{-3mm} \\
\hline
\hline
\vspace{-2mm} \\
Transition layer pressure (dyn{\,}cm$^{-2}$) 	& $2\times10^{-1}$	& $7\times10^{-2}$	& $2$	\\
Coronal temperature (K at $r\approx1.1{\,}R_{\odot}$)	& $1.1-1.6\times10^{6}$ 	& $10^{6}$		& $2.5\times10^{6}$	\\
Coronal energy losses (erg{\,}cm$^{-2}${\,}s$^{-1}$)   &   &                   &                                \\
\hspace{5mm}Conductive flux $F_{c}$    & $2\times10^{5}$   & $6\times10^{4}$   & $10^{5}-10^{7}$                \\
\hspace{5mm}Radiative flux $F_{r}$     & $10^{5}$          & $10^{4}$          & $5\times10^{6}$                \\
\hspace{5mm}Solar wind flux $F_{w}$    & $<5\times10^{4}$  & $7\times10^{5}$   & $<10^{5}$                      \\
\hspace{5mm}Total corona loss $F_{c}+F_{r}+F_{w}$   & $3\times10^{5}$          & $8\times10^{5}$    & $10^{7}$  \\
Chromospheric radiative losses (erg{\,}cm$^{-2}${\,}s$^{-1}$)	&					&				&           \\
\hspace{5mm}Low chromosphere           & $2\times10^{6}$   & $2\times10^{6}$   & $>10^{7}$		\\
\hspace{5mm}Middle chromosphere        & $2\times10^{6}$   & $2\times10^{6}$   & $10^{7}$			\\
\hspace{5mm}Upper chromosphere         & $3\times10^{5}$   & $3\times10^{5}$   & $2\times10^{6}$		\\
\hspace{5mm}Total chromospheric loss   & $4\times10^{6}$   & $4\times10^{6}$   & $2\times10^{7}$		\\
\hline
~&~&~ \\
\end{tabular}
\normalsize
\end{center}
\end{table}

%%%%%%%%%%%%%%%%%%%%
%%%%%%%%%%%%%%%%%%%%
%%%%%%%%%%%%%%%%%%%%
%%%%%%%%%%%%%%%%%%%%
%%%%%%%%%%%%%%%%%%%%
%%%%%%%%%%%%%%%%%%%%
%%%%%%%%%%%%%%%%%%%%
%%%%%%%%%%%%%%%%%%%%
%%%%%%%%%%%%%%%%%%%%
%%%%%%%%%%%%%%%%%%%%

\clearpage
\newpage
\section{Observational \& Theoretical Difficulties}
\label{difficulties_section}

Even though the Sun's chromosphere has been identified as a key area of interest by the
solar physics community, it is unfortunately an incredibly difficult portion of the
atmosphere to observe and interpret efficiently. Firstly, the chromosphere is predominantly
observed through a collection of deep absorption lines in the optical portion of the
electromagnetic spectrum. 
These features include the Fraunhofer absorption lines
of Ca~{\sc{ii}}~H~\&~K ($3933-3968$~{\AA}),
Mg~{\sc{i}}~b$_{1,2,4}$ ($5167-5184$~{\AA}),
H$\beta$ ($4861$~{\AA}),
Na~{\sc{i}}~D$_{1,2}$ ($5889-5895$~{\AA}) and
H$\alpha$ ($6563$~{\AA}), in addition to 
%the Ba~{\sc{ii}}~4554~{\AA} resonance line, 
some near-UV and UV spectral signatures
such as the Mg~{\sc{ii}}~{\it{h}}~\&~{\it{k}} lines ($2795-2803$~{\AA}),
the C~{\sc{iv}} resonant doublet ($1548-1550$~{\AA}) and H~{\sc{i}}~L$\alpha$ ($1216$~{\AA}). 
Observing such deep, dark optical absorption cores results in minimal
light levels reaching the telescope detectors once atmospheric (if using a
ground-based facility), telescope, lens,
filter and camera transmission factors have been taken into consideration.
\citet{Jes10b} derived photon count-rate statistics for a number of chromospheric
spectral profiles and indicated that $\ll1${\%} of the incident flux on Earth's
atmosphere is converted into counts at the imaging detector. As a result,
longer exposure times need to be employed to maintain adequate signal-to-noise ratios.
This can have the adverse effect of blurring any rapidly evolving underlying
chromospheric features such as spicules, mottles, fibrils and jets. Furthermore,
it is impossible (engineering wise) to fabricate an infinitesimally narrow bandpass filter
that would only capture the deepest core of the chromospheric spectral line. Typical
Lyot--type filter widths are on the order of $200$~m{\AA}~FWHM, with some more specialised
spectral imagers including the Interferometric BIdimensional Spectrometer
\citep[IBIS;][]{Cav06} and the CRisp Imaging SpectroPolarimeter
\citep[CRISP;][]{Sch08} achieving pass-bands as narrow as $50$~m{\AA}~FWHM.
Nevertheless, line core intensities often contain significant photospheric flux
leaking into the filter pass-bands, creating a complex puzzle as to which
features and measurements correspond to photospheric and/or chromospheric structures
\citep{Hal08}. To complicate matters yet further, upwardly or downwardly propagating
material will induce intrinsic Doppler shifts into the spectroscopic line profiles, thus
causing the static wavelength filters to sample features far out into the spectral wings
(which contain significant photospheric continua), rather than the true chromospheric
absorption core. 
Indeed, employing a narrowband ($80$~m{\AA}) Lyot filter capable 
of imaging the wings of both the H$\beta$ Fraunhofer and 
Ba~{\sc{ii}}~4554~{\AA} resonance lines, \citet{Sut01} revealed how such 
Doppler shifts permeate all high-resolution 
lower atmospheric observations, thus complicating the source of 
fine-scale intensity fluctuations. 
\begin{figure}[!t]
\begin{center}
\includegraphics[width=0.9\textwidth,clip=]{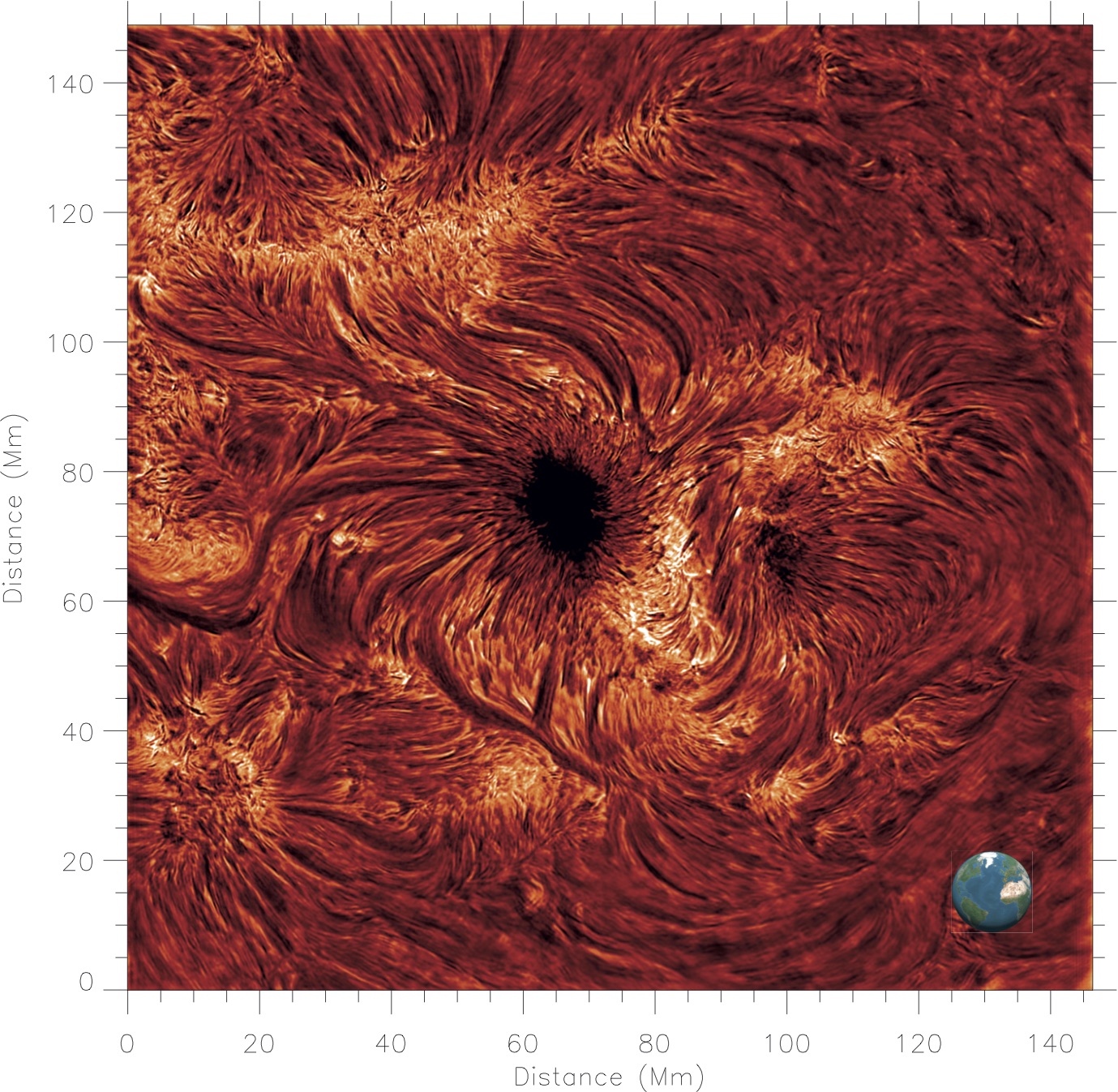}
\caption{An image of a solar active region acquired through a narrowband
$0.25${\,}{\AA} H$\alpha$-core filter. Employing a new generation of
large format, low noise CMOS sensors, it is now possible to obtain chromospheric
fields-of-view in excess of $200'' \times 200''$ (at the diffraction limit) with
frame rates exceeding $60$~s$^{-1}$. A scale representation of the
Earth is depicted in the lower-right section of the image.
This snapshot, courtesy of D.~B.~Jess, was acquired using an Andor Technology 
4.2~MP Zyla CMOS detector (15~ms exposure time at a frame rate of 64~s$^{-1}$) 
at the Dunn Solar Telescope, NM, USA.
\label{Jess_sCMOS}}
\end{center}
\end{figure}

\vspace{3mm}
Chromospheric densities experience a significant decrease from their corresponding
photospheric counterparts, and as a result radiative transition rates generally
dominate over collisional rates \citep{Uit95, Uit97, Uit01, Uit02}. This makes the
chromosphere a non-local thermodynamic equilibrium (non-LTE) environment,
resulting in the need for full radiative transfer modelling of all simulated processes.
Moving away from $1$D hydrostatic models, \citet{Kle76, Kle78} and
\citet{Car92, Car95, Car97}, to name but a few, have demonstrated the strenuous
computational requirements necessary for efficient $1$D modelling in full non-LTE.
However, as time progressed, it became clear that even $2$D non-LTE models
\citep[e.g.,][]{Van02, Car12} of the solar atmosphere were not entirely representative of
the observed chromospheric structures \citep{dela12, Lee12}. In the modern era,
full $3$D non-LTE modelling has only been made possible by the continual
computational improvements in both speed and storage delivered to end users.
Nevertheless, even $3$D non-LTE simulations of chromospheric processes
have significant caveats attached, manifesting as uncertainties in the
multi-level atomic transitions, atmospheric mixing-lengths, non-gray radiative
transfer components and sensitivities to
asymmetric spectral line profiles \citep{Cun07, Caf11, Bee12, Lee12, Pra13}.
Therefore, while the use of $3$D non-LTE simulations to assist with the
interpretation of chromospheric phenomena and wave energy transportation is
beneficial, the complex nature of the chromosphere itself introduces
considerable difficulties when attempting to efficiently and accurately diagnose
basal heating contributions.

\begin{figure}[!t]
\begin{center}
%\epsscale{0.9}
%\plotone{IBIS_scan.eps}
\includegraphics[width=0.9\textwidth,clip=]{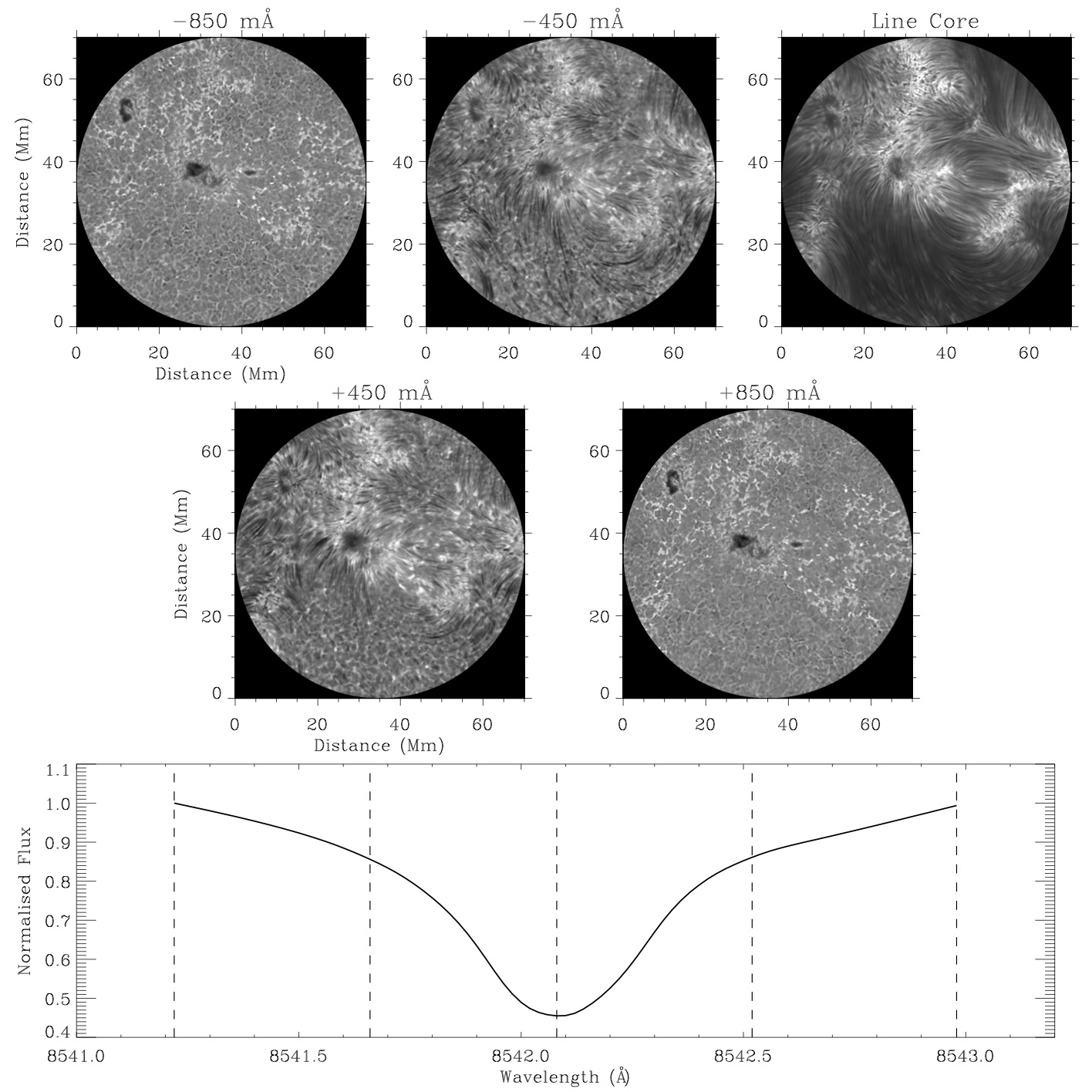}
\caption{A spectral imaging scan, taken by the IBIS instrument on
2011~December~10 across the Ca~{\sc{ii}} absorption
line at $8542${\,}{\AA}, revealing a collection of magnetic pore structures.
The panels display the corresponding intensity images
at specific wavelength positions corresponding to the Ca~{\sc{ii}} line core
$-850${\,}m{\AA} (upper-left), $-450${\,}m{\AA} (upper-middle),
$\pm0${\,}m{\AA} (upper-right), $+450${\,}m{\AA} (middle-left) and
$+850${\,}m{\AA} (middle-right), respectively. The lower panel shows an
`at rest' Ca~{\sc{ii}} profile where the vertical dashed lines indicate the
wavelength positions used to capture the sequence of images displayed
in the upper panels. Note how the chromosphere reveals itself as the
imaging wavelength approaches the deepest part of the absorption profile.
Images based on the data presented by \citet{Jes14}.
\label{Jess_IBIS}}
\end{center}
\end{figure}

\vspace{3mm}
The solar chromosphere also introduces observational difficulties through its
collection of incredibly diverse, rapidly evolving structures covering spatial scales ranging
from sub-arcsecond (e.g., spicules, mottles, fibrils, etc.) through to those in
excess of many hundreds of arcseconds (e.g., filaments; Figure~{\ref{Jess_sCMOS}}). Not only does the
chromosphere comprise of structures covering a vast spread of spatial scales, but it also displays
signatures of supersonic motion and high-frequency oscillatory phenomena in the
forms of evaporated material \citep[e.g.,][]{Act82, Ant84, Ant85, Key11a} and magnetically
guided compressible and incompressible waves
\citep[][to name but a few of the more recent high-impact articles]{DeP04, DeP07, DeP11, Erd07a, Jes09, McI11, Mor12}.
Our present fleet of telescopes able to observe the solar chromosphere includes
the $0.5${\,}m Solar Optical Telescope \citep[SOT;][]{Tsu08, Sue08} onboard the Hinode spacecraft \citep{Kos07},
the $0.76${\,}m Dunn Solar Telescope \citep[DST; formerly the Vacuum Tower Telescope;][]{Dun69} in New Mexico, USA,
the $1${\,}m Swedish Solar Telescope \citep[SST;][]{Sch03} on the island of La Palma,
the $1${\,}m New Vacuum Solar Telescope \citep[NVST;][]{Liu214} at the Fuxian Solar Observatory, China,
the $1.5${\,}m GREGOR telescope \citep{Sch12} at the Teide Observatory, Tenerife, and
the $1.6${\,}m New Solar Telescope \citep[NST;][]{Cao10} at the Big Bear Solar Observatory
\citep[BBSO;][]{Zir70} in California, USA. Each facility has its own unique merits, some of which include
high-cadence multiwavelength imaging, spectropolarimetric imaging,
high resolution spectrographic instrumentation, high-order adaptive optics, and those in
locations with excellent year-round observing conditions. The current suite of solar telescopes capable
of observing the chromosphere have revolutionised our understanding of small-scale
dynamic processes occurring within the interface between the relatively cool photosphere and
the super-heated multi-million degree corona. It is not uncommon for these facilities to be able to
obtain full spectral imaging scans of chromospheric absorption profiles (e.g., Ca~{\sc{ii}},
H$\alpha$, etc.) in as little as a few seconds (Figure~{\ref{Jess_IBIS}}), diffraction-limited narrowband imaging
of deep absorption line cores at frame rates exceeding $40${\,}s$^{-1}$, and
spectral resolutions ($\frac{\lambda}{\delta\lambda}$) exceeding $500{\,}000$ at wavelengths
covering the optical through to the near-infrared. However, even with these powerful telescopes
employing modern detectors and instrumentation, there is clear evidence to suggest that
there are still lower-atmospheric phenomena manifesting below our currently imposed
resolution limits \citep{von95, Lag07, Jes08, Cau08, Cau09, Soc09, Vou10, And13}. Thus, for
the last number of years there has been an impetus placed on further developing the
spatial, temporal and spectral resolutions of our ground- and space-based solar facilities.
The solar physics community
eagerly awaits the arrival of the first next-generation high-resolution facilities, including the
$2${\,}m National Large Solar Telescope \citep[NLST;][]{Has10} in Ladakh, India, and the
$4${\,}m Daniel K. Inouye Solar Telescope
\citep[DKIST, formerly the Advanced Technology Solar Telescope, ATST;][]{Kei03, Rim10}
atop the Haleakal{\={a}} volcano on the Pacific island of Maui, due to receive first light in
2018 and 2019, respectively. Through drastically increased aperture sizes, chromospheric
structures down to $\sim20$~km in size will be able to be detected, tracked and studied
in unprecedented detail.

\vspace{3mm}
While we await the final stages of construction on these
revolutionary facilities we can continue pushing the boundaries of scientific
understanding by employing current generation of telescopes in novel ways.
In this review we will detail recent observations and theoretical interpretations of
oscillatory phenomena found to be propagating through the solar chromosphere. Due to
the cutting-edge research being undertaken around the world in an attempt to address the
long-standing question of how energy and heat manages to pass through the chromosphere on its
way to the corona, often the observations and interpretations put forward by solar physicists
can be anecdotal and fraught with overzealous assumptions. Nevertheless, without
somewhat speculative conclusions the research field would not be advancing at the rate
it is today as researchers attempt to verify or refute the hypotheses put forward. In this
review we will attempt an objective overview of recent observational and theoretical
wave developments, and try to place each scientific result in the context of
atmospheric heating constraints. In the following section
we will summarise the most important theoretical results that form the foundation
knowledge upon which we can start interpreting observed chromospheric waves.

%%%%%%%%%%%%%%%%%%%%
%%%%%%%%%%%%%%%%%%%%
%%%%%%%%%%%%%%%%%%%%
%%%%%%%%%%%%%%%%%%%%
%%%%%%%%%%%%%%%%%%%%
%%%%%%%%%%%%%%%%%%%%
%%%%%%%%%%%%%%%%%%%%
%%%%%%%%%%%%%%%%%%%%
%%%%%%%%%%%%%%%%%%%%
%%%%%%%%%%%%%%%%%%%%

\clearpage
\newpage
\section{Theory of Linear MHD Waves}
\label{theory_section}
\subsection{Linearising the Ideal MHD Equations}\label{sec:linearized:mhd}
In the absence of a magnetic field, the supported plasma eigenmodes are
sound waves which are isotropic (i.e., their speed is independent of the
direction of propagation) and non-dispersive. However,
in the presence of a magnetic field the number of supported waves is dramatically increased.
Importantly, although some of these waves have similarities with sound waves,
they can be highly anisotropic. This is because their characteristics depend on the
degree of alignment of the wavevector ($\bm{k}$) with the direction of the
background magnetic field ($\Beq$), and the ratio of the kinetic pressure
($\peq$) to the magnetic pressure ($\Beq^2 / 2 \mu_0$). This ratio is the
plasma-$\beta$, defined as $\beta = 2 \mu_0 \peq/ \Beq^2$, where $\mu_0$
is the magnetic permeability of free space. A commonly used method to explore
the properties of waves in magnetised plasmas is to consider the linearised
ideal MHD equations. Let us consider small perturbations
about a static equilibrium (i.e., no background flow) where $\bm{v}_0 = \bm{0}$,
\begin{dgroup}\label{eqn:linearized:mhd}
\begin{dmath}\label{eqn:charge:conservation}
\pderiv[1]{}{t}\rhop = -\nabla \cdot (\rhoeq \vp),
\end{dmath}
\begin{dmath}\label{eqn:linearized:momentum}
\rho_{0} \pderiv[1]{}{t}\vp = - \nabla \pp + \frac{1}{\mu_0} \left[ (\nabla \times \Beq)\times \Bp + (\nabla \times \Bp) \times \Beq \right],
\end{dmath}
\begin{dmath}\label{eqn:linearized:adiabatic}
\nabla \pp = \vs^2 \nabla \rhop,
\end{dmath}
\begin{dmath}\label{eqn:linearized:bpert}
\pderiv[1]{}{t} \Bp = \nabla \times (\vp \times \Beq),
\end{dmath}
\end{dgroup}
where $\rhoeq$, $\peq$ and $\Beq$ are the density, kinetic pressure and
magnetic field quantities at equilibrium, with all being functions of the spatial
coordinates. Furthermore, $\rhop$, $\pp$ and $\Bp$ are the
corresponding perturbed quantities, while $\vp$ is the velocity perturbation
and $\vs = \sqrt{\gamma \peq / \rhoeq}$ is the adiabatic sound speed, and, $\gamma$ is the ratio of specific heats.

\subsection{Wave Modes in a Uniform Unbounded Magnetized Plasma}
\label{sec:uum}
Now let us explore the equations in Equation~\ref{eqn:linearized:mhd} in a very simple
setting to illustrate the identifying characteristics of MHD wave modes. For an unbounded, homogeneous and
magnetised plasma, $\peq$, $\rhoeq$ and $\Beq$ are constant, resulting in
Equation~\ref{eqn:linearized:mhd} being rewritten as {\citep{priest2014magnetohydrodynamics}},
\begin{dgroup}\label{eqn:linearized:mhd:uniform}
\begin{dmath}
\pderiv[1]{}{t}\rhop = - \rhoeq \nabla \cdot (\vp),
\end{dmath}
\begin{dmath}\label{eqn:linearized:momentum:uniform}
\rho_0 \pderiv[1]{}{t}\vp = - \nabla\left( \pp + \frac{\Beq \cdot \Bp}{\mu_0} \right) + \frac{1}{\mu_0} \nabla \cdot \left( \Beq \Bp \right),
\end{dmath}
\begin{dmath}
\nabla \pp = \vs^2 \nabla \rhop,
\end{dmath}
\begin{dmath}
\pderiv[1]{}{t} \Bp = (\Beq \cdot \nabla)\vp - \Beq(\nabla \cdot \vp),
\end{dmath}
\end{dgroup}
where the two terms in the right hand side of the momentum equation Equation~\ref{eqn:linearized:momentum:uniform}
are the total pressure perturbation,
\begin{dmath}\label{eqn:total:pressure}
\pTp = \pp + \Beq \cdot \Bp / \mu_0,
\end{dmath}
comprised of the perturbation of the kinetic pressure, $\pp$, and the magnetic pressure
perturbation, $\Beq \cdot \Bp / \mu_0$. The second term in right hand side of
Equation~\ref{eqn:linearized:momentum:uniform} is the magnetic tension. Considering plane wave
solutions for the perturbed quantities,
\begin{align}
\vp,\Bp,\pp,\rhop \propto \planeExp,
\end{align}
where $\bm{x}$ is the position vector and $\bm{k}$ is the wavevector, the equations in
Equation~\ref{eqn:linearized:mhd:uniform} can be combined to produce a dispersion relation.
Specifically, there exist two possibilities: ({\it{i}}) $\bm{k} \cdot \vp = 0$ which corresponds to
the incompressible case, and, ({\it{ii}}) $\bm{k} \cdot \vp \neq 0$ that corresponds to the
compressible case. Using Equation~\ref{eqn:linearized:mhd:uniform} and $\bm{k} \cdot \vp = 0$,
we arrive at the following dispersion relation in terms of the phase speed,
$\vph = \omega / k$ (where $k = |\bm{k}|$), and the angle, $\theta$, 
between the wavevector, $\bm{k}$, and the background magnetic field, $\Beq$, 
\begin{align}\label{eqn:dispersion:uniform:incompressible}
\vph^2 &= \frac{\Beq^2}{\mu_0 \rhoeq} \cos^2{\theta} \\
       &= \va^2 \cos^2{\theta}, \nonumber
\end{align}
which is an anisotropic, non-dispersive wave whose only restoring force is the
magnetic tension. The phase speed in Equation~\ref{eqn:dispersion:uniform:incompressible}
corresponds to phase speed of the Alfv\'en wave \citep{alfven1942existence},
where $\va = |\Beq| / \sqrt{\mu_0 \rhoeq}$. In the compressible case
($\bm{k} \cdot \vp \neq 0$) the system of equations in
Equation~\ref{eqn:linearized:mhd:uniform} can be combined producing the
following dispersion equation,
\begin{dmath}\label{eqn:dispersion:uniform:compressible}
\vph^4 - (\vs^2 + \va^2)\vph^2 + \vs^2 \va^2 \cos^2{\theta} = 0 \ .
\end{dmath}
Equation
Equation~\ref{eqn:dispersion:uniform:compressible} has two roots in terms of the
square of the phase speed, i.e.,
\begin{dgroup}\label{eqn:dispersion:uniform:compressible:solutions}
\begin{dmath}\label{eqn:dispersion:uniform:compressible:solution:fast}
\vph^2 = \frac{1}{2}\left( \vs^2 + \va^2 \right) + \frac{1}{2}\left( \vs^4 + \va^4 - 2\vs^2 \va^2 \cos{2 \theta} \right)^{1/2},
\end{dmath}
\begin{dmath}\label{eqn:dispersion:uniform:compressible:solution:slow}
\vph^2 = \frac{1}{2}\left( \vs^2 + \va^2 \right) - \frac{1}{2}\left( \vs^4 + \va^4 - 2\vs^2 \va^2 \cos{2 \theta} \right)^{1/2}.
\end{dmath}
\end{dgroup}

\begin{table}[!t]
\begin{center}
\footnotesize
\caption{Phase speeds of the slow, fast and Alfv\'{e}n waves for a uniform unbounded magnetised plasma.
\label{tab:uniform:unbounded:medium:summary}}
\begin{tabular}{lccc}
~&~&~ \\
\hline
\vspace{-3mm} \\
					&			& $\beta \gg 1$, $\va \ll \vs$	&   $\beta \ll 1$, $\va \gg \vs$ 	 	\\
\vspace{-3mm} \\
\hline
\hline
\vspace{-2mm} \\
					& $\bm{k} || \Beq$		& Alfv\'{e}n wave, $\vph^2 \sim \va^2$		& Alfv\'{e}n wave, $\vph^2 \sim \va^2$	\\
$\bm{k} \cdot \vp = 0$	&					&									&								\\
					& $\bm{k} \perp \Beq$	& Alfv\'{e}n wave -- does not propagate 		& Alfv\'{e}n wave -- does not propagate	\\
\vspace{2mm} \\
					& 					& Fast wave, $\vph^2 \sim \vs^2$	 		& Fast wave, $\vph^2 \sim \va^2$		\\
 					& 					& approximately isotropic	 			& approximately isotropic		\\
					& $\bm{k} || \Beq$		& magnetic and kinetic pressure in phase	& magnetic and kinetic pressure in phase	\\
					& 					& Slow wave, $\vph^2 \sim \va^2$	 		& Slow wave, $\vph^2 \sim \vs^2$		\\
					& 					& magnetic and kinetic pressure out of phase	& magnetic and kinetic pressure out of phase	\\
$\bm{k} \cdot \vp \neq 0$&					&									&							\\
					& 					& Fast wave, $\vph^2 \sim \vs^2$	 		& Fast wave, $\vph^2 \sim \va^2$		\\
 					& $\bm{k} \perp \Beq$	& approximately isotropic	 			& approximately isotropic		\\
					& 					& magnetic and kinetic pressure in phase	& magnetic and kinetic pressure in phase	\\
					& 					& Slow wave -- does not propagate	 		& Slow wave -- does not propagate		\\
\hline
~&~&~ \\
\end{tabular}
\normalsize
\end{center}
\end{table}

\vspace{3mm}
The solutions in Equation~\ref{eqn:dispersion:uniform:compressible:solutions} correspond
to two magneto-acoustic modes: the fast mode
($\vfm = |\vph|$; Equation~\ref{eqn:dispersion:uniform:compressible:solution:fast}) and
the slow mode ($\vsm = |\vph|$;
Equation~\ref{eqn:dispersion:uniform:compressible:solution:slow}). In summary,
there are three MHD modes, the Alfv\'en mode Equation~\ref{eqn:dispersion:uniform:incompressible},
whose restoring force is only magnetic tension, and the two magneto-acoustic modes
whose restoring force is a combination of the magnetic tension and the total pressure
Equation~\ref{eqn:total:pressure}. The phase speed in Equation~\ref{eqn:dispersion:uniform:compressible}
depends on the angle, $\theta$, and the ratio of the sound speed versus the Alfv\'en
speed. This quantity is proportional to the plasma-$\beta$, which can be rewritten in 
the form of $\beta = (2/\gamma) \vs^2 / \va^2$.
First let us explore the two extremes of the plasma-$\beta$, namely $\beta \gg 1$
and $\beta \ll 1$. Notice that $\beta \gg 1$ means that $\vs \gg \va$, while
$\beta \ll 1$ is equivalent to $\va \gg \vs$. In the limit where $\beta \gg 1$,
Equation~\ref{eqn:dispersion:uniform:compressible:solutions} is reduced to,
\begin{dmath}\label{eqn:limit:beta:gg:1}
\vph^2 \sim
\twopartdef{\vs^2}{Equation~\ref{eqn:dispersion:uniform:compressible:solution:fast},}{ \va^2 \cos^2{\theta} }{Equation~\ref{eqn:dispersion:uniform:compressible:solution:slow},}
\end{dmath}
where the solution corresponding to Equation~\ref{eqn:dispersion:uniform:compressible:solution:fast}
is the dominant mode,
while the solution to Equation~\ref{eqn:dispersion:uniform:compressible:solution:slow}
is a second order correction. As a result, for $\beta \gg 1$ the Alfv\'en and slow modes
vanish and the fast mode, now the only mode, converges to the sound speed, $\vs$. This result
is quite intuitive considering that a high plasma-$\beta$ (i.e., $\beta \gg 1$) implies that the
kinetic pressure dominates the magnetic field, thus the magnetic pressure and tension
in Equation~\ref{eqn:linearized:momentum:uniform} can be neglected. This reduces
Equation~\ref{eqn:linearized:momentum:uniform} to the linearised Navier-Stokes equation.
For a low plasma-$\beta$ scenario (i.e., $\beta \ll 1$),
Equation~\ref{eqn:dispersion:uniform:compressible:solutions} reduces to,
\begin{dmath}\label{eqn:limit:beta:ll:1}
\vph^2 \sim
\twopartdef{\va^2}{Equation~\ref{eqn:dispersion:uniform:compressible:solution:fast},}{ \vs^2 \cos^2{\theta} }{Equation~\ref{eqn:dispersion:uniform:compressible:solution:slow},}
\end{dmath}
which indicates that the fast magneto-acoustic wave converges in
magnitude to the Alfv\'en speed and propagates isotropically.

%Additionally it is interesting to note that the fast magnetoacoustic mode is very
%weakly dependent on pressure and the main restoring forces are magnetic pressure
%and magnetic tension. Furthermore, for a wavevector $\bm{k}$ parallel to $\Beq$ the
%restoring force of the fast magnetoacoustic mode is solely the magnetic tension while
%for increasing $\theta$ magnetic pressure becomes increasingly more important
%until $\theta = \pi / 2$ where the restoring force is only magnetic pressure. In
%contrast, notice that the slow magnetoacoustic mode is highly dependent on
%$\theta$ and admits no waves for $\theta = \pi / 2$.

\vspace{3mm}
Next, we consider $\bm{k} \parallel \Beq$ and $\bm{k} \perp \Beq$. The first case
naturally corresponds the $\theta = 0$, and so
Equation~\ref{eqn:dispersion:uniform:compressible:solutions} is reduced to
\begin{dmath}\label{eqn:k:parallel}
\vph^2 \sim
\twopartdef{\va^2}{Equation~\ref{eqn:dispersion:uniform:compressible:solution:fast},}{ \vs^2 }{Equation~\ref{eqn:dispersion:uniform:compressible:solution:slow},}
\end{dmath}
while for $\bm{k} \perp \Beq$ (i.e., $\theta = \pi/2$) Equation~\ref{eqn:dispersion:uniform:compressible:solutions} reduces to,
\begin{dmath}\label{eqn:k:perpendicular}
\vph^2 \sim
\twopartdef{\va^2 + \vs^2}{Equation~\ref{eqn:dispersion:uniform:compressible:solution:fast},}{ 0 }{Equation~\ref{eqn:dispersion:uniform:compressible:solution:slow}.}
\end{dmath}
The first observation in this case is that the fast magneto-acoustic mode
is no longer isotropic since its phase speed varies from $\va$ to
$\sqrt{\va^2 + \vs^2}$. Note that in the limit $\va \gg \vs$, the phase
speed of the fast mode becomes $\sim \va$ which is in agreement
with Equation~\ref{eqn:limit:beta:ll:1}. Under these conditions the fast mode can be considered
to be approximately isotropic.

\begin{figure}[!t]
\centerline{\includegraphics[width=0.97\textwidth]{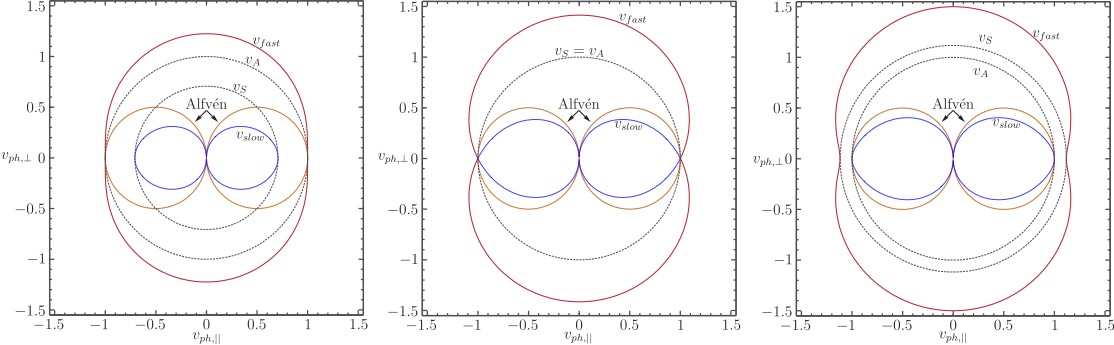}}
\caption{Friedrichs diagrams for $\vs < \va$ with $\beta = 1/ \gamma$ (left),
$\vs = \va$ and $\beta = 2 / \gamma$ (middle) and $\vs > \va$ with
$\beta = 2.5 / \gamma$ (right). The phase speed perturbation of the slow
magnetoacoustic wave is illustrated in blue, the Alfv\'{e}n wave in orange and
the fast magnetoacoustic wave in red. The dotted lines correspond to the sound
and Alfv\'{e}n speed. The horizontal and vertical axes labelled as $v_{ph,||}$ and
$v_{ph,\perp}$ respectively represent the velocity perturbation components along
and perpendicular to the equilibrium magnetic field, $\Beq$.
\label{fig:friedrichs}}
\end{figure}

\vspace{3mm}
An important relation between the slow and fast magneto-acoustic modes
is revealed if we combine Equation~\ref{eqn:charge:conservation},
Equation~\ref{eqn:linearized:adiabatic} and Equation~\ref{eqn:linearized:bpert} to obtain the
following relation between magnetic pressure and kinetic pressure,
\begin{dmath}\label{eqn:magnetic:pressure:to:kinetic:pressure}
\frac{1}{\mu_0} \Beq \cdot \Bp = \frac{\va^2}{\vs^2}\left(1 - \frac{\vs^2}{\vph^2} \cos^2{\theta} \right) \pp .
\end{dmath}
Therefore, according to Equation~\ref{eqn:magnetic:pressure:to:kinetic:pressure},
when $\vph < \vs \cos{\theta}$ the kinetic and magnetic pressures are
out of phase by $\pi$ and so these restoring forces oppose each other.
From Equation~\ref{eqn:dispersion:uniform:compressible:solution:slow} it follows
immediately that this condition holds for the slow magneto-acoustic waves, and
is clearly illustrated in Figure~\ref{fig:friedrichs}. In the case where
$\vph > \vs \cos{\theta}$, the magnetic and kinetic pressure perturbations are
in phase with one another. This condition holds for the fast
magneto-acoustic wave (see Equation~\ref{eqn:dispersion:uniform:compressible:solution:fast})
depicted in Figure~\ref{fig:friedrichs}. For situations where $\vph = \vs \cos{\theta}$, the
magnetic pressure tends to zero, and apart from the trivial solution, this condition is satisfied
when: ({\it{i}}) $\va > \vs$ for $\theta = 0,\pi$ corresponding to the slow magneto-acoustic
wave, ({\it{ii}}) $\va = \vs$ and is satisfied by the the Alfv\'{e}n wave, and
lastly, ({\it{iii}}) $\vs > \va$ which is satisfied by the fast magneto-acoustic wave at
$\theta = 0, \pi$. Also notice that for $\va \gg \vs$ the magnetic pressure is dominant,
while for $\vs \gg \va$ the plasma pressure dominates.

\vspace{3mm}
In summary, in linearised ideal MHD for a homogeneous plasma there are three
distinct waves: the slow and fast magneto-acoustic and the Alfv\'en. The
phase speeds of these waves are well ordered: $0 \leq \vsm \leq \va \leq \vfm$,
and also their velocities are mutually perpendicular, $\vsmv \perp \vav \perp \vfmv$
\citep{goedbloed2004principles}. The Alfv\'en mode is incompressible and is
supported purely by the magnetic tension, while the restoring forces for the two
magneto-acoustic modes is a combination of the total pressure and magnetic
tension. In Table~\ref{tab:uniform:unbounded:medium:summary} we provide a brief summary
of the results in this section. For $\beta \ll 1$, which is valid in magnetically dominated 
regions of the Sun's atmosphere, the fast mode is approximately isotropic while the
slow and Alfv\'en modes exhibit strong anisotropies, with both components
having {\it{preferred}} propagation directions along the magnetic field. It must be 
stressed that the Alfv{\'{e}}n and slow modes do not propagate in directions 
perpendicular to the magnetic field.

\begin{figure}%[!t]
\begin{center}
\includegraphics[width=0.8\textwidth,clip=]{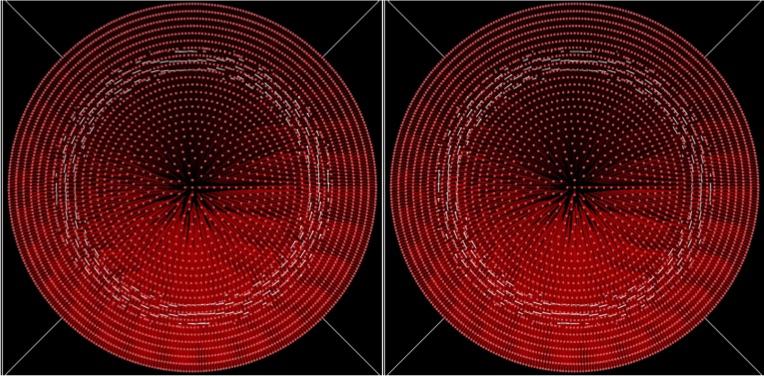}
\includegraphics[width=0.8\textwidth,clip=]{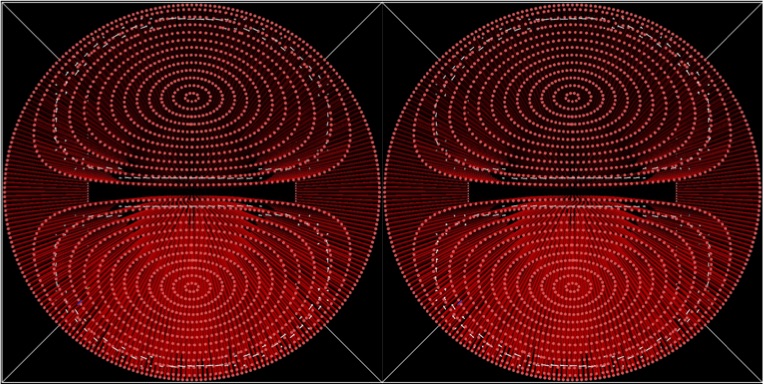}
\caption{{\it{Upper panels:}} The two extrema of the Alfv\'en mode for $n=0$. This
mode is also referred to as torsional Alfv\'en mode. In this figure (and in
Figures~\ref{fig:movie:alfven}--\ref{fig:movie:kink}), the red dotted ropes 
represent the magnetic field lines, while the white arrows describe the velocity field. 
{\it{Lower panels:}} The Alfv\'en mode
for $n=1$. Notice that the magnetic surfaces are decoupled, but that they are
more intricately configured when compared with the torsional ($n=0$) Alfv\'en 
mode displayed in the upper panels.
The movie associated with this figure is available from {\it{http://swat.group.shef.ac.uk/fluxtube.html}}.
\label{fig:movie:alfven}}
\end{center}
\end{figure}

\begin{figure}%[!t]
\begin{center}
\includegraphics[width=0.8\textwidth]{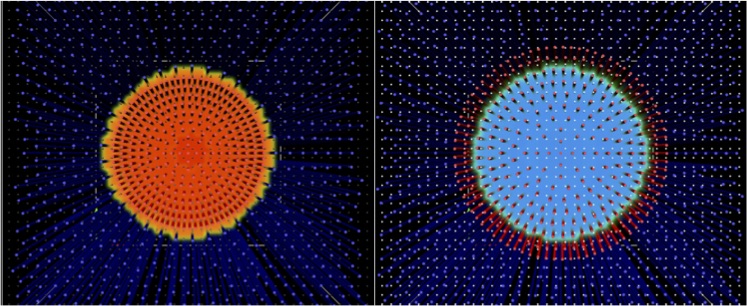}
\includegraphics[width=0.8\textwidth]{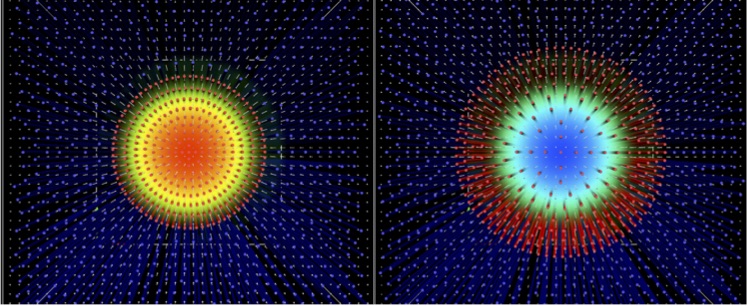}
\caption{The upper panels represent the slow sausage mode ($n=0$), while the 
lower panels describe the fast sausage mode ($n=0$).
The density perturbation ($\rhop$) above the equilibrium background, $\rho_{i}$, 
is highlighted using warmer colours with red denoting the maximum perturbation. 
Conversely, density perturbations below the equilibrium are illustrated with cooler 
colours, with blue representing the minimum. The blue dotted ropes represent the magnetic field outside the flux tube.
Notice that the dominant velocity component for the slow sausage mode is in the direction
along the flux tube, while for the fast sausage mode ($n=0$) this component is zero.
The density perturbation and external magnetic field are represented in similar fashion in 
Figure~\ref{fig:movie:kink}. 
The movie associated with this figure is available from {\it{http://swat.group.shef.ac.uk/fluxtube.html}}.
\label{fig:movie:sausage}}
\end{center}
\end{figure}

\begin{figure}
\begin{center}
\includegraphics[width=0.8\textwidth]{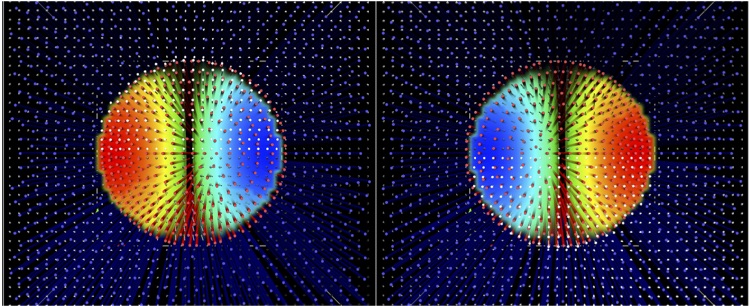}
\includegraphics[width=0.8\textwidth]{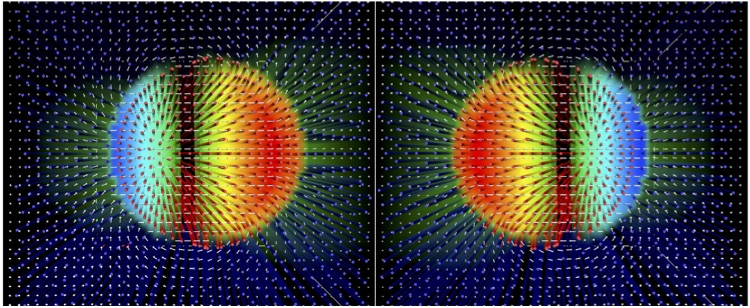}
\caption{{\it{Upper panels:}}
The slow kink mode ($n=1$). As with the
slow sausage mode, the dominant component of the velocity field
is along the direction of the magnetic field. {\it{Lower panels:}}
The fast kink mode ($n=1$). Note that the velocity
component along the magnetic field for this mode is zero, as it
is for the fast sausage mode in Figure~\ref{fig:movie:sausage}.
Another notable feature of this mode is that the divergence of the
velocity inside the flux tube is zero, which suggests that this mode is
(nearly) incompressible.
The movie associated with this figure is available from {\it{http://swat.group.shef.ac.uk/fluxtube.html}} 
\label{fig:movie:kink}}
\end{center}
\end{figure}

\subsection{Wave Modes in a Magnetic Flux Tube}\label{sec:mfluxtube}
To help us understand the much richer variety of MHD waves modes that can supported in more complex 
magnetic geometries, a useful first step is to consider a simple straight magnetic cylinder. \citet{Edw83} 
chose the particular case of a constant
magnetic field inside, $B_i \hat{\bm{z}}$, and outside, $B_e \hat{\bm{z}}$,
the flux tube with a discontinuity at the tube boundary $r = r_a$, where $r_a$
is the tube radius. Similarly the equilibrium density and pressure inside and
outside the tube are taken to be $\rho_i, p_i$ and $\rho_e, p_e$ respectively.
The resulting dispersion relations, assuming no energy propagation towards
or away from the flux tube (thus we allow only $\Me^2 >0$) are the following
\citep{Edw83},
\begin{align}\label{eqn:edwin:roberts:dispersion}
\Mi \rho_e (k_z^2 \vAe^2 - \omega^2) \frac{K_n(\Me r_a)}{K_n^{\prime}(\Me r_a)} = \Me \rho_i (k_z^2 \vAi^2 - \omega^2) \frac{I_n(\Mi r_a)}{I_n^{\prime}(\Mi r_a)}, & \text{ for } \Mi^2 > 0, \\
n_0 \rho_e (k_z^2 \vAe^2 - \omega^2) \frac{K_n(\Me r_a)}{K_n^{\prime}(\Me r_a)} = \Me \rho_i (k_z^2 \vAi^2 - \omega^2) \frac{J_n(n_0 r_a)}{J_n^{\prime}(n_0 r_a)}, & \text{ for } -\Mi^2 = n_0^2 > 0,
\end{align}
where,
\begin{align}
\Mi^2 = \frac{(k_z^2 \vSi^2 - \omega^2)(k_z^2 \vAi^2 - \omega^2)}{(\vAi^2 + \vSi^2)(k_z^2 \vTi^2 - \omega^2)}, \\ \Me^2 = \frac{(k_z^2 \vSe^2 - \omega^2)(k_z^2 \vAe^2 - \omega^2)}{(\vAe^2 + \vSe^2)(k_z^2 \vTe^2 - \omega^2)},
\end{align}
are the internal and external radial wavenumbers, $n$ is the azimuthal
wavenumber and $k_z$ is the longitudinal wavenumber which in the
present work is along the $\hat{\bm{z}}$ direction. For the case where
$\Mi^2 > 0$ (see Equation~\ref{eqn:edwin:roberts:dispersion}), the amplitude of the resulting eigenmodes is heavily localised
near the boundary of the flux tube
and so these are referred to as surface modes. When $\Mi^2 < 0$ the behaviour
of the solutions inside the flux tube is oscillatory, and since only evanescent
solutions are permitted outside the flux tube, the largest wave amplitudes are
observed inside (and in the vicinity of) the flux tube. These modes are referred to as
body waves. The schematic diagram in Figure~\ref{fig:chromosphere:cartoon} depicts velocity 
and density perturbations characteristic to the fast and slow magnetoacoustic modes for 
$n=0$ (sausage mode) and $n=1$ (kink mode). Surface and body waves exhibit similar 
characteristics associated with the slow/fast magnetoacoustic and Alfv\'en modes. 
However, these modes have a substantially different behaviours when compared with the 
eigenmodes studied in \S{\,}\ref{sec:uum}. The parallel component of the wavevector, $\bm{k}$, to the
magnetic field, $\Beq$, is here defined as $k_z$. The azimuthal wavevector, $n$, and
the radial wavevectors, $\Mi$ or $\Me$, form the perpendicular component
to the magnetic field.

\vspace{3mm}
With that in mind, let us explore the similarities and
differences of the corresponding eigenmodes in \S{\,}\ref{sec:uum} and the
modes present in a magnetic flux tube. First notice that the Alfv\'en mode,
shown in Figure~\ref{fig:movie:alfven}, and the
slow mode (see Figures~\ref{fig:movie:sausage} \& \ref{fig:movie:kink}),
when present, exist even when the wavevector (in cylindrical
coordinates in this case) is perpendicular to the magnetic field. This is
not the case in \S{\,}\ref{sec:uum} (see Equation~\ref{eqn:k:perpendicular}) for the slow mode
considering $\bm{k} \cdot \Beq = 0$ when $\bm{k} \perp \vav$.
Additionally, the fast magneto-acoustic mode in \S{\,}\ref{sec:uum} (for the case
where $\beta \ll 1$) was approximately isotropic, while the fast mode in the
magnetic cylinder is highly anisotropic and does not exist for some
azimuthal wavenumbers. Also, the behaviour of
radial harmonics for the fast mode is entirely different.
For instance, for the fast sausage mode ($n=0$) the main restoring force
is the total pressure, while magnetic tension has only a minor role, while the
fast kink mode ($n=1$) appears to be nearly incompressible with the main
restoring force being magnetic tension (e.g., see Figures~\ref{fig:movie:kink} \&
\ref{fig:chromosphere:cartoon}). Nevertheless, despite the differences between
the eigenmodes for the uniform medium and the magnetic flux tube, the velocities
of the three modes present within a magnetic flux tube are still mutually
perpendicular to one another. The practical implication of this is that the slow
and Alfv\'en modes are incredibly difficult to detect in chromospheric flux tubes,
while the fast magneto-acoustic mode is the most prominent.
However, even for fast magneto-acoustic waves we have only successfully detected the
sausage (azimuthal wavenumber $n=0$)
and kink ($n=1$) modes, while modes with $n>1$ are yet to be
observed, mainly as a result of limitations in the spatial and temporal resolutions
of the current generation of telescopes and instrumentation \citep{Zhu00}.

In the following section we will discuss the overwhelming evidence 
that demonstrates the ubiquitous existence of compressible magneto-acoustic 
waves in the solar chromosphere. We will overview the observational 
characteristics which led to the various scientific interpretations, with 
a particular emphasis placed on the energetics of the detected waveforms. 
Importantly, we will show that such oscillatory motion can be readily generated and 
driven at the photospheric layers, with the resulting upwardly propagating 
waves acting as potentially important conduits for supplying 
continual energy to the upper regions of the solar atmosphere.

%Few comments on magnetic interface geometry.
%\cite{wilson1978stratatm} considered magnetic interface.
%\cite{wilson1979waveprop} considered magnetic interface.
%\cite{wilson1979hydromagnetic} magnetic interface.
%
%Few comments on the slab geometry.
%\cite{roberts1981bwave} Slab : Roberts 1981
%
%Discuss extensions of the cylindrical geometry.
%\cite{spruit1982propagation} magnetic flux tube radiating (leaky)
%\cite{Edw83} the classic magnetic cylinder.
%\cite{cally1985magnetohydrodynamic} leaky modes.
%\cite{cally1986leaky} leaky modes in magnetic cylinders.
%\cite{roberts2000waves} Invited review on oscillations in the corona
%\cite{cally2003coronal} Overview of leaky modes.
%\cite{ruderman2006leaky} Leaky modes - Ruderman.
%\cite{erdelyi2006sausage} Twist incompressible magnetic flux tube 2006
%\cite{ruderman2007nonaxisymmetric}
%\cite{erdelyi2007linear} Twist compressible magnetic flux tube $n=0$ 2007
%\cite{erdelyi2010magneto} Twist compressible magnetic flux tube $n>0$ 2010

%% Perhaps its relevant from the chromosphere but as the paper is structured at the moment it would be a stretch to include expanding flux tubes.
%\cite{roberts1978vertical} considered expanding magnetic flux tube considering dependence upon height of the boyancy, compressibility and magnetic forces.

%%%%%%%%%%%%%%%%%%%%
%%%%%%%%%%%%%%%%%%%%
%%%%%%%%%%%%%%%%%%%%
%%%%%%%%%%%%%%%%%%%%
%%%%%%%%%%%%%%%%%%%%
%%%%%%%%%%%%%%%%%%%%
%%%%%%%%%%%%%%%%%%%%
%%%%%%%%%%%%%%%%%%%%
%%%%%%%%%%%%%%%%%%%%
%%%%%%%%%%%%%%%%%%%%

\clearpage
\newpage
\section{Compressible Waves}
\label{compressible_section}
As mentioned in \S{\,}\ref{sec:uum}, compressible waves are characterised by 
$\nabla \cdot \vp \neq 0$. Physically, this means that these waves have the ability to perturb the local plasma
density. As a result, such perturbations give rise to periodic intensity fluctuations since the plasma emission is
modulated by the induced compressions/rarefactions.
The manifestation of such waves in the solar atmosphere has been
well-documented since the 1960s when researchers first 
identified periodic fluctuations in
both the intensity and velocity fields of solar optical 
observations \citep{Lei60, Lei62, Noy63}.  
At first, these intensity and velocity oscillations were 
interpreted as purely acoustic 
waves. This was to be expected since acoustic wave 
based heating theories had already 
been proposed earlier by, e.g., \citet{Sch48} and \citet{Bie48}. 

\subsection{Magnetoacoustic Waves}
From an MHD perspective, acoustically dominated magnetoacoustic wave modes should naturally occur 
in the Sun's atmosphere under all plasma-$\beta$ regimes (see \S{\,}\ref{sec:linearized:mhd} for the definition 
of plasma-$\beta$). One does not have to assume that waves of an acoustic character can only occur in 
regions that have little or no magnetic field. The important caveat to add is that in regions of strong magnetic 
field (i.e., low plasma-$\beta$), the acoustically dominated MHD wave modes are very anisotropic, with their 
direction of propagation significantly aligned to the magnetic field direction (see, e.g., the magnetoacoustic 
slow mode of a $\beta \ll 1$ homogeneous plasma in Table~\ref{tab:uniform:unbounded:medium:summary} 
and the left panel of Figure~\ref{fig:friedrichs}). In the following sections we review studies of these waves in 
different representative plasma-$\beta$ regimes of the Sun's chromosphere, i.e., quiet Sun, network 
locations and active regions.

\subsubsection{Quiet Sun and network locations}
After intensity oscillations in the solar atmosphere were 
interpreted as the signatures of acoustically dominated 
waves, the next logical step was to attempt to track 
these wave motions higher up in the solar atmosphere. Initial
work by \citet{Deu71} was able to follow velocity and 
intensity fluctuations through to the upper-photospheric
layers by employing narrowband {\nad} and 
{\mgb} filters.
%However, these measurements were 
%obtained from the relatively low (by modern standards) 
%resolution $17$~cm vectormagnetograph at the Anacapri
%Observatory in Napoli, Italy \citep{Deu69}, and as a 
%result were only able to indicate wave propagation on
%spatial scales of ~$8$--$100''$, corresponding to 
%near-supergranular -- active-region sizes. 
Then, utilising the vacuum tower telescope 
(now the Dunn Solar Telescope) at the National Solar Observatory,
New Mexico, alongside dedicated chromospheric 
H$\alpha$ measurements, \citet{Deu74, Deu75} was able to
detect propagating waves down to spatial scales 
on the order of a few arcseconds. However, these
measurements were designed to shed light on the 
geometrical formation heights of the spectral lines used
in the study, and therefore made no estimation of the 
energetics carried by these waves. Subsequent work
revealed that the upward phase velocity of the waves was 
too large to be explained by traditional acoustic
phenomenon, and instead the embedded magnetic field 
must also be considered to better
understand the wave energies and dynamics 
\citep{Ost61, Mei76, Ulm76}. 
%Over the following decades the examination
%of these oscillations from a magnetoacoustic perspective 
%rose to the forefront of observational solar physics, with 
%\citet{Ulm76} aptly asking whether magnetoacoustic 
%oscillations were an important heating mechanism for the
%chromosphere and the corona. 
Many publications 
followed which addressed the energetics of
upwardly propagating magnetoacoustic waves 
\citep{Ath78, Ath79a, Ath79b, Whi79a, Whi79b, Lit79, Mei80, Sch80},
but none were able to find sufficient mechanical 
flux to balance the heavy radiative losses 
experienced in the solar chromosphere.
Of course, the main purpose of this review article 
is not to provide an in-depth overview of historical 
results, but instead review recent advances in the 
field of propagating wave phenomena. 
Therefore, for a more detailed overview of the initial pioneering 
work undertaken in relation to wave studies in the 
lower solar atmosphere, we 
refer the reader to a series of early review articles by 
\citet{Fri72}, \citet{Bon81} and \citet{Nar90, Nar96}, and 
of course, the references contained therein.

\begin{figure}[!t]
\begin{center}
%\epsscale{0.6}
%\plotone{Fossum_Carlsson_2006_ApJ.png}
\includegraphics[width=0.7\textwidth,clip=]{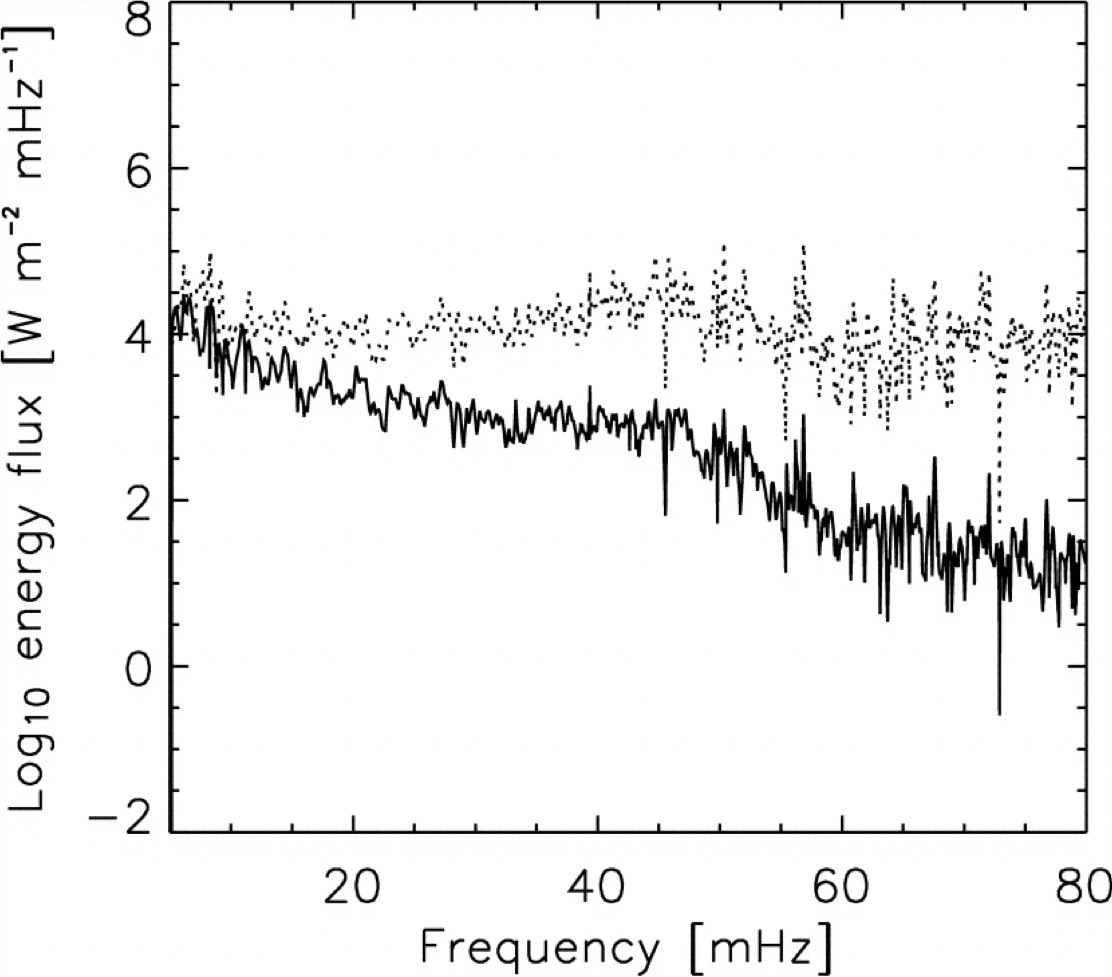}
\caption{Observed acoustic energy flux available for chromospheric heating (solid line) and the
corresponding acoustic energy flux $86$~km below at the $\tau_{500} = 1$ layer (dotted line). Since the
radiative damping in the solar photosphere increases with frequency, the flattening of the acoustic spectrum
at higher heights is believed to correspond to an increase in the overall high-frequency photospheric acoustic
power. However, here the behaviour above $30$~mHz ($\lesssim30$~s) is critically dependent on the subtraction
of inherent noise characteristics, as well as the instrumental temporal and spatial resolutions. It has subsequently
been shown that the TRACE instrument may not have sufficient spatial {\it{and}} temporal resolutions to
accurately define and characterise the acoustic power available for chromospheric heating.
Image adapted from \citet{Fos06}.
\label{FosCarl}}
\end{center}
\end{figure}

\vspace{3mm}
In more recent times, and following the analysis of Transition Region and Coronal Explorer \citep[TRACE;][]{Han99} data,
\citet{Fos05a, Fos06} were unable to detect sufficient power in high-frequency ($5$--$50$~mHz; $20$--$200$~s)
magnetoacoustic oscillations and concluded that these waves cannot constitute the dominant heating mechanism of the
solar chromosphere. However, this study was limited by the cadence TRACE can achieve ($\sim13$~s), its
coarse spatial sampling ($\approx0{\,}.{\!\!}''5$~pixel$^{-1}$) and the onboard filter transmissions \citep{Fos05b}.
Consequently, physically small oscillation sites with short coherence lengths may be smeared out by the
coarse spatial and temporal resolutions. Furthermore, it was suggested by \citet{Jef06} and \citet{Wed07}
that such methods will overlook dynamic patterns created on sub-resolution scales, and as a direct
result severely underestimates the actual mechanical flux \citep{Kal07, Kal08}. In a follow-up article,
\citet{Cun07} employed similar UV TRACE observations alongside revised $1$D simulations, detailed
by \citet{Ram03}, to reveal that the complex small-scale time-dependent topologies that manifest within the
solar chromosphere produce a network of hot filaments embedded into broad cool regions. The
authors suggest that the hot chromospheric components of solar emission consist of small pockets
embedded in much cooler material that is unrelated to the Ca~{\sc{ii}} emission previously used as a temperature
diagnostic. As a result, the limited spatial resolution of the TRACE instrument makes a direct comparison
between the measured radiative fluxes and the implied wave energy fluxes difficult, if not impossible using
purely $1$D simulations. Employing higher resolution observations from the G{\"{o}}ttingen
spectrometer/polarimeter \citep{Pus06, Bel08}, \citet{Bel09, Bel10a} were able to find significant energy flux
($\sim2000$~W{\,}m$^{-2}$) for magnetoacoustic periodicities as low as $40$~s at
lower chromospheric heights. Then, utilising the Imaging Magnetograph eXperiment \citep[IMaX;][]{Mar11}
two-dimensional spectropolarimeter onboard the Sunrise balloon-based observatory \citep{Sol10, Bar11},
\citet{Bel10b} uncovered yet more evidence to support the hypothesis that the lower chromosphere is
bombarded with high-energy magnetoacoustic waves with energies on the order of
$6400-7700$~W{\,}m$^{-2}$. The work of \citet{Bel09, Bel10a, Bel10b} strengthened the support
for atmospheric heating through magnetoacoustic wave dissipation, and coupled with the opposing
findings of \citet[Figure~{\ref{FosCarl}}]{Fos05a, Fos06} inspired many groups to push the examination of
magnetoacoustic waves to even smaller spatial extents, especially with kG strength magnetic bright points
\citep[MBPs;][]{Dun73, Ste85, Sol93, San94, Ber01, Ste01, Cro09, Cro10, Jes10a, Key11, Key13, Utz13}
offering potentially efficient waveguides on sub-arcsecond scales.

\begin{figure}[!t]
%\epsscale{0.8}
%\plotone{Kontogiannis_2010_AandA1.pdf}
\begin{center}
\includegraphics[width=0.9\textwidth,clip=]{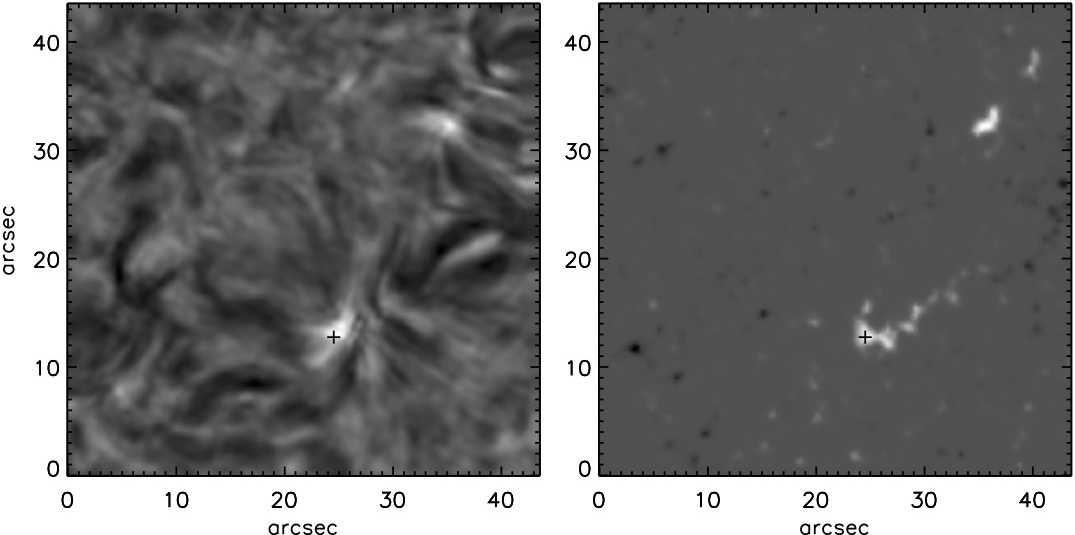}
\end{center}
\caption{A chromospheric H$\alpha$ core image (left) and co-spatial line-of-sight
magnetogram (right). The black cross indicates the location of concentrated magnetic fields in the
solar photosphere, with field strengths exceeding $1000$~G, which can connect upwards through
various layers of the solar atmosphere, thus providing an efficient channel for the propagation of
compressible waves. Image adapted from \citet{Kon10}.
\label{Kontogiannis}}
\end{figure}

\vspace{3mm}
\citet{McA02} studied MBPs in network locations with high resolution ground-based observations and
found a multitude of magnetoacoustic wave power spanning the deep photosphere through to the
upper chromosphere. Follow-up work incorporating phase analysis routines allowed the authors
to characterise the waves as upwardly propagating, with their magnetoacoustic nature potentially
offering a means for energy deposition on small spatial scales \citep{McA03}.
\citet{Kon10} employed period-mapping techniques to investigate the linkage between small-scale
concentrated photospheric magnetic flux elements to oscillations found in simultaneous
chromospheric H$\alpha$ time series (Figure~{\ref{Kontogiannis}}).
The authors uncovered a complex relationship depending on both the
strength and orientation of the encompassing magnetic fields, but ultimately found evidence for waves tracing
the path of small-scale magnetic fields through to chromospheric heights, indicative of acoustically
dominated waves in a low plasma-$\beta$ regime. Using detailed cross-correlation methods
on Solar Optical Telescope \citep[SOT;][]{Tsu08, Sue08} data from the Hinode spacecraft \citep{Kos07},
\citet{Law12} demonstrated a direct linkage between upwardly propagating magnetoacoustic modes and aureoles
of enhanced oscillatory power at chromospheric heights, suggesting how powerful photospheric motions at
periodicities nearing the acoustic cut-off may be able to produce shock-wave heating of the localised chromospheric
plasma \citep{Car92, The97, Kri01, Blo04, Vec09}. However, most shock phenomena arises at the interface with
downflowing material in the mid-chromosphere, and as a consequence little of the resulting heat and motion can be
found within the upper regions of the solar chromosphere \citep{Car97}. While shock waves may not be a dominant
heating mechanism for the magnetised solar chromosphere, current research is investigating their possible role in the
generation of incompressible wave modes at higher atmospheric heights \citep[e.g.,][]{DeP04, Rou07, Cau08, Rut08, Kur13}.

\begin{figure}[!t]
%\epsscale{1.0}
%\plotone{Liu_2014.png}
\includegraphics[width=0.97\textwidth]{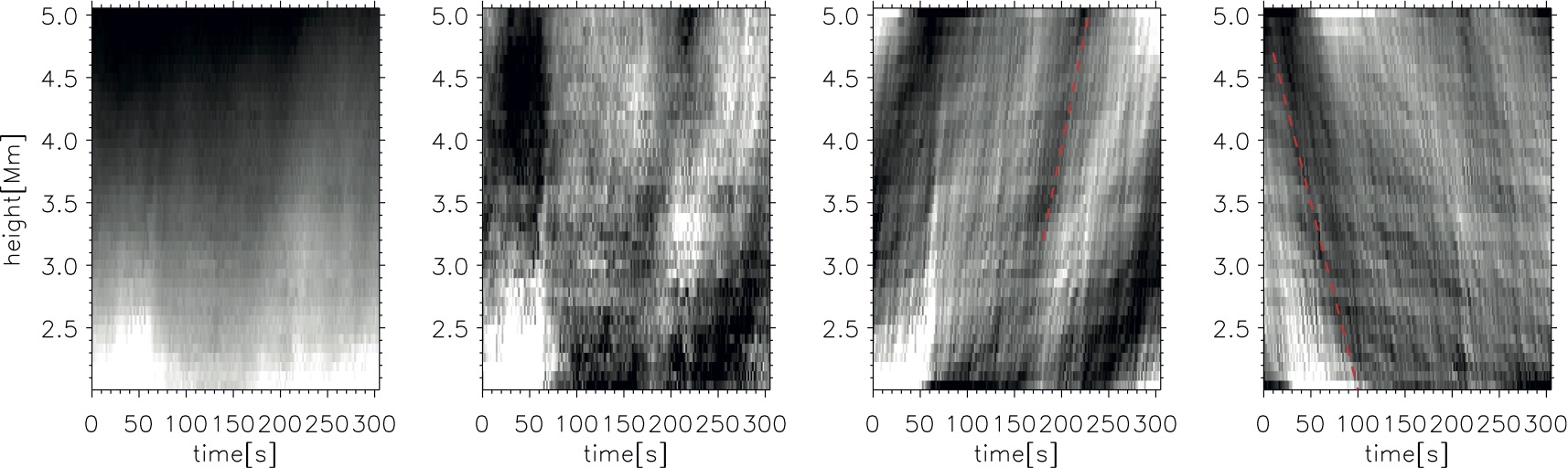}
\caption{Time--distance intensity maps of off-limb spicules, from left to right,
of original Hinode/SOT {\cah} data, background-subtracted data,
and those filtered for upwardly and downwardly propagating magnetoacoustic modes, respectively. Here, the `height'
measurement is indicative of the atmospheric height above the solar photosphere, and therefore represents
magnetoacoustic wave modes reaching chromospheric heights. The presence of both upwardly and downwardly
propagating waves (as indicated by the dashed red lines orientated in different directions) suggests the
superposition of such phenomena may cause previous evaluations of magnetoacoustic
energy flux to be underestimated. Image adapted from \citet{Liu14}.
\label{Liu}}
\end{figure}

\vspace{3mm}
In a series of papers, \citet{Bec09, Bec12, Bec13a, Bec13b} employed sub-arcsecond resolution
observations from the POlarimetric LIttrow Spectrograph \citep[POLIS;][]{Bec05} on the Vacuum
Tower Telescope, Tenerife, to analyse and track the velocity and intensity perturbations of small-scale magnetic
elements reaching chromospheric
heights. The authors found that, even for small-scale magnetic elements, the embodied magnetoacoustic
energy was simply insufficient to maintain the chromospheric temperature rise by a factor of about five.
However, \citet{Liu14} have recently shown that many structures (including omnipresent spicules, mottles and
fibrils) in the lower solar atmosphere demonstrate signatures
of the superposition of both upwardly and downwardly propagating magnetoacoustic wave modes, indicating
that while the upwardly propagating signatures dominate, the mere presence of downwardly propagating
waves may artificially reduce the derived magnetoacoustic energy flux (Figure~{\ref{Liu}}).

\vspace{3mm}
Over the last $50$ years there has been an abundance of studies attempting to quantify the role magnetoacoustic
waves play in the heating of the outer solar atmosphere. As time progressed and new high resolution facilities became
commissioned (TRACE, Hinode/SOT, etc.), researchers attempted to probe the energetics of magnetoacoustic waves
further still by harnessing the improved spatial and/or temporal resolutions on offer. However, each time
\citep{Fos05a, Fos06, Bec09, Bel09, Bel10a, Bel10b, Bec12, Bec13a, Bec13b} the authors were
unable to conclusively verify that these wave modes carry
sufficient energy to play a dominant role in atmospheric heating. Perhaps, as highlighted by the work of
\citet{Jef06}, \citet{Wed07}, \citet{Kal07, Kal08}, to name but a few, we still require yet higher spatial and temporal resolutions
to be able to accurately constrain the rapid fluctuating dynamics synonymous with propagating
magnetoacoustic wave modes in small-scale magnetic elements. With the upcoming
National Large Solar Telescope \citep[NLST;][]{Has10}, Daniel K. Inouye Solar
Telescope \citep[DKIST, formerly the Advanced Technology Solar Telescope, ATST;][]{Kei03, Rim10}, Solar
Orbiter \citep{Mul13}, Solar--C \citep{Shi11} and European Solar Telescope \citep[EST;][]{Col10} facilities all
offering unprecedented views of the Sun, it is only a matter of time until we are able to accurately quantify the
contribution of magnetoacoustic waves to plasma heating.

\subsubsection{Active regions}
\label{compressive_ar}
Active regions are typically large-scale structures extending throughout the solar atmosphere and
visible as an intense manifestation of magnetic fields. 
Sunspots, pores and plage are the usual
constituent representation of the magnetic field topology in the solar chromosphere, with overall
sizes in the range of $30$--$10{\,}000$ million square km \citep[$10$--$3000$ micro solar
hemispheres;][]{Kop53, How92, Mar93, Bau05} and 
field strengths regularly exceeding 
$1000${\,}G at the photospheric level, with over $6000${\,}G 
documented in extreme cases \citep{Liv06}. 
Thus, solar active regions provide
idealised conduits for wave and energy transportation into the outer
solar atmosphere. Indeed, a wide variety of wave phenomena has been
observed in active region structures for over $40$~years \citep{Bec69, Bogdan2006}.
Initial work on oscillatory phenomena in sunspots helped validate the detection
of long-period magnetoacoustic oscillations, which are generated by the response of the
convectively-inhibiting sunspot to the 5-minute global $p$-mode oscillations \citep{Tho82, Lit92}.
While oscillations in solar active regions are dominated by periodicities
intrinsically linked to the global $p$-mode spectrum \citep[on the order of
$3$--$5$~minutes;][and the references therein]{Gol77a, Gol77b, Gab92, Bau96, Laz97},
an extended range of alternative wave periods can also be identified in the active region locality,
spanning three orders-of-magnitude from in-excess of one hour \citep{Dem85}
through to less than several seconds \citep{Jes07}.

\begin{figure}[!t]
\begin{center}
%\epsscale{0.85}
%\plotone{Zirin_Stein_1972.png}
\includegraphics[width=0.85\textwidth,clip=]{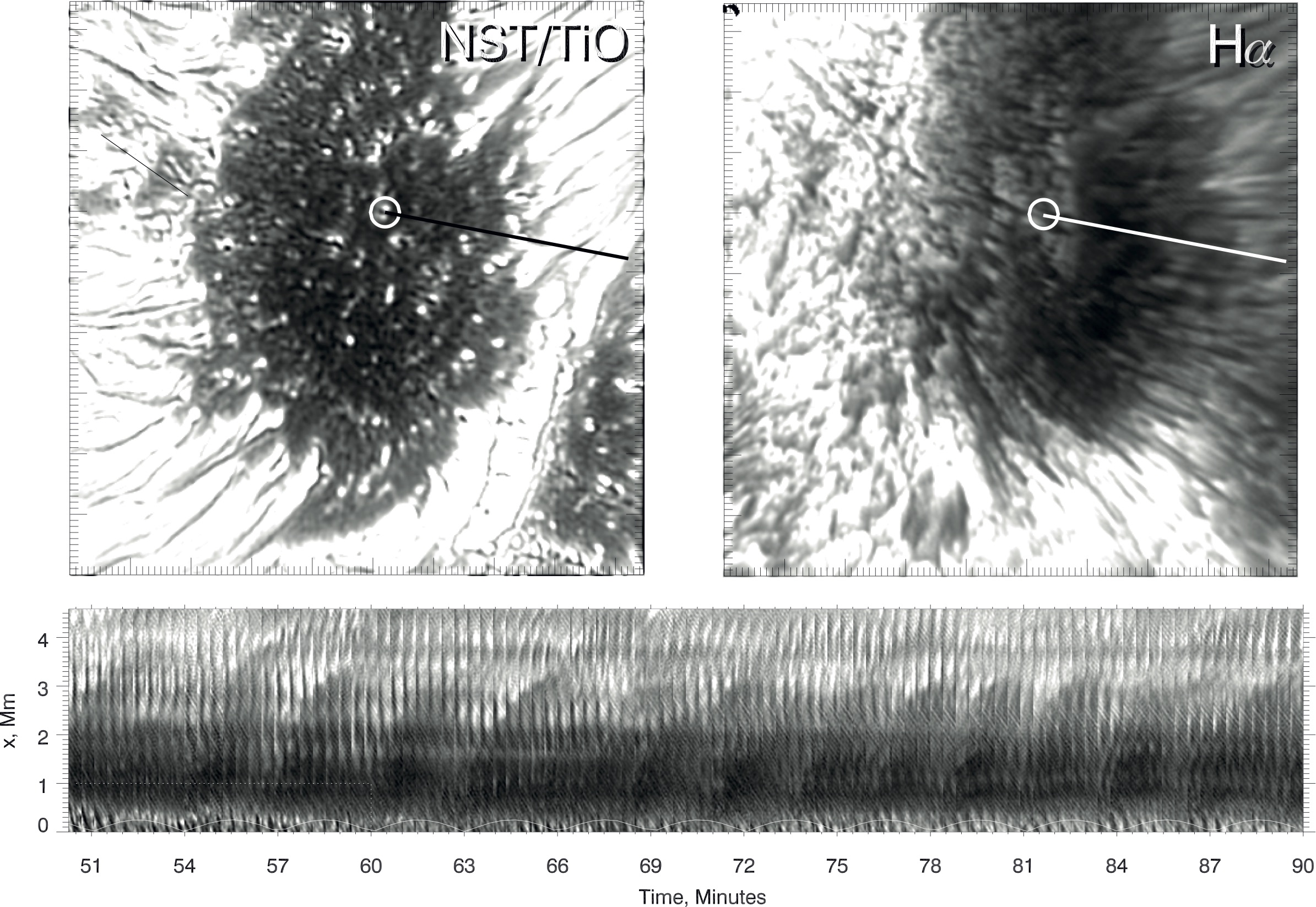}
\caption{A photospheric TiO image (upper left) and a simultaneous 
and co-spatial H$\alpha$ 
core snapshot (upper right) of a sunspot 
acquired by the high-resolution NST. Both 
images have been unsharp masked to better reveal fine-scale 
details, while the long tick marks 
on the axes represent 1000~km intervals. The black and white lines 
indicate the location of the time--distance cut displayed in the lower 
panel, while the large white circles highlight the position of a photospheric 
umbral dot that forms the starting point of the time--distance cut. The 
propagation of RPWs is clearly evidenced by the diagonal trends 
present in the time--distance diagram, where curved features either 
represent the acceleration of wave activity or the superposition of 
differing wave periodicities along the observational line-of-sight. 
The white curve at the bottom of the lower panel displays a 
constant 3~minute periodicity to highlight the repetitive and 
ubiquitous nature of all RPW phenomena.
Images adapted from \citet{Yur14}.
\label{Yurchyshyn}}
\end{center}
\end{figure}

\vspace{3mm}
Running penumbral waves (RPWs) are a common sight in the chromospheric layer of
sunspots \citep{Nye74}. \citet{Giov72} and \citet{Zir72} provided the first observational evidence of this
phenomenon when they
detected concentric wave fronts propagating outwards through the penumbra of the sunspot in
narrowband H$\alpha$ images (Figure~{\ref{Yurchyshyn}}). Interpreted as magnetoacoustic modes,
\citet{Bris97} and \citet{Kob04} demonstrated how the wave signatures are actually comprised
of the superposition of many individual wave periods, each propagating with independent
phase speeds. \citet{Kob06} examined the relationship between propagating intensity and
velocity disturbances to conclude that the frequencies and phase speeds of RPWs are largest
($>3$~mHz or $<300$~s, 40~{\kms}) at the inner penumbral boundary,
decreasing to their lowest values ($<1$~mHz or $>1000$~s, 10~km{\,}s$^{-1}$)
at the outer penumbral edge. Additionally, \citet{Kob00} has shown
evidence that the RPWs can propagate to distances exceeding
$\sim15''$ ($\sim10{\,}000$~km) from the outer edge of the
penumbral boundary, suggesting the waves have considerable energy to overpower the
signatures of ubiquitous quiet-Sun $p$-mode oscillations. However, while RPWs
are readily observed in chromospheric imaging and spectroscopic sequences, their
origin has been under intense debate ever since their discovery. Some consider
RPWs to be a purely chromospheric phenomenon driven by trans-sunspot waves,
while others believe they are the observational signature of upwardly propagating
magnetoacoustic waves guided along the intense magnetic fields of the underlying sunspot
\citep{Chri00, Chri01, Geor00, Cen06, Tzio06, Tzio07}. The recent work of
\citet{Blo07} has added momentum to the interpretation that RPWs are
a chromospheric visualisation of upwardly propagating magnetoacoustic oscillations through
use of high-resolution spectroscopic measurements, obtained with the Tenerife
Infrared Polarimeter \citep[TIP;][]{Mar99} attached to the German Vacuum Tower
Telescope in Tenerife, Canary Islands. Here, the authors suggested that RPWs can
readily propagate along magnetic field lines in a low plasma-$\beta$ regime (i.e., dominated by magnetic
pressure) environment, and therefore most likely explained as a signature of the
channelling of magnetoacoustic waves through to chromospheric heights. Indeed, in
the lower solar atmosphere the strong magnetic field strengths associated with sunspot
structures \citep[up to $\sim6000$~G;][]{Liv06} results in extremely large magnetic
pressures when compared to the localised gas pressure
\citep[i.e., plasma-$\beta \ll 1$;][]{Mat04}. Under these conditions MHD wave modes
can exist that have field-aligned perturbations which are much larger than
the cross-field perturbations. These compressive and acoustically dominated
MHD wave modes are strongly guided by the magnetic field and hence are
very anisotropic, thus producing stable waveguides for oscillations to propagate along
\citep{Rob78, Nak95, Erd08, Hin08, Lun12, Wil14}. While \citet{Blo07} provided
tentative evidence that RPWs are in-fact upwardly propagating magnetoacoustic
waves, their evidence relied on single-slit spectroscopic measurements with a spatial
resolution of $0{\,}.{\!\!}''8$. As a result, all locations within the vicinity of the sunspot
were not examined with a high degree of precision, and the conclusive proof as to the origin
of RPWs remained elusive.

\begin{figure}[!t]
%\epsscale{0.6}
%\plotone{Jess_2013.jpg}
\begin{center}
\includegraphics[width=0.7\textwidth,clip=]{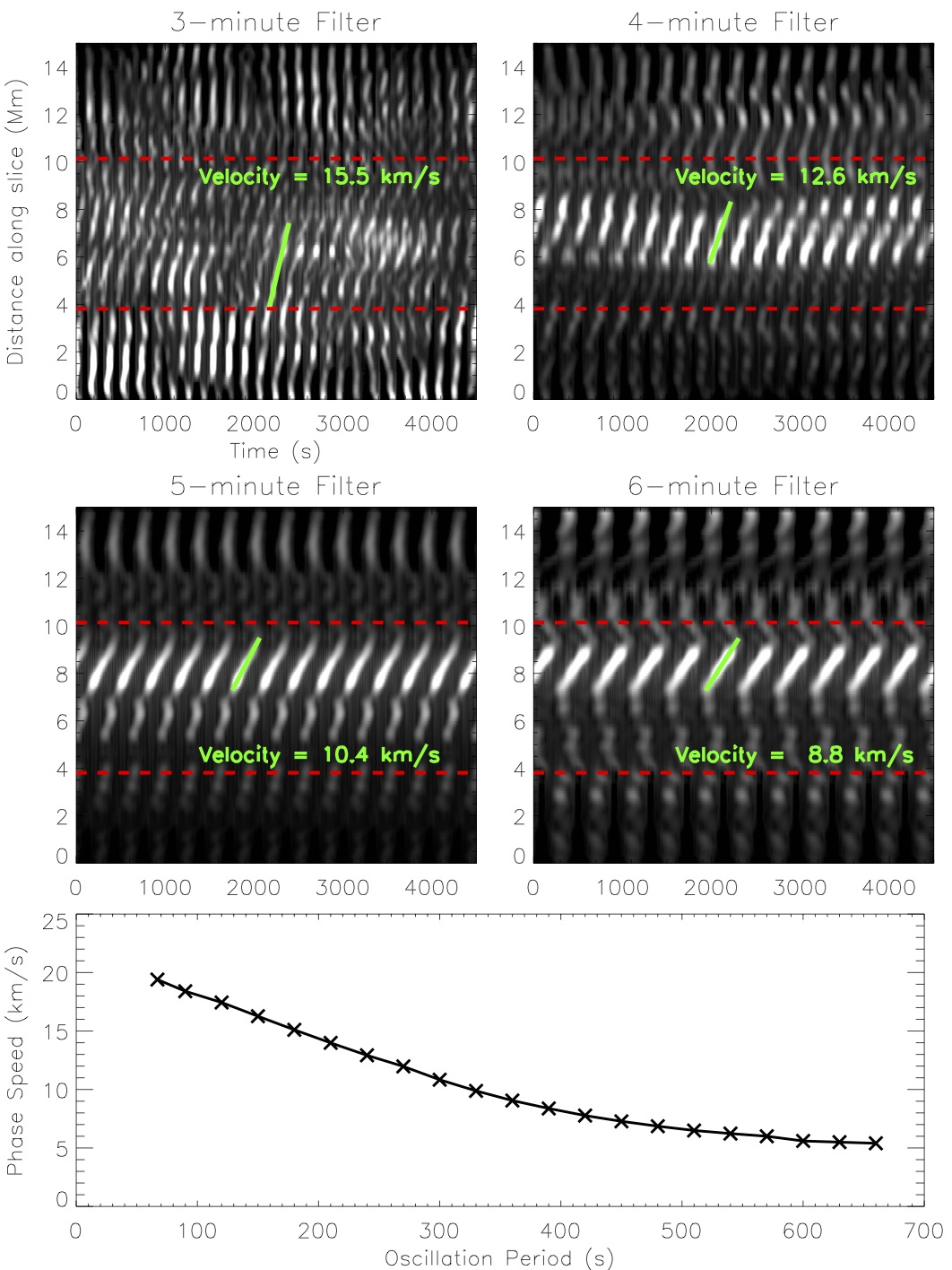}
\caption{Time--distance diagrams of chromospheric sunspot data (upper and
middle panels), where $0$~Mm indicates the centre of the underlying umbra. Each
time--distance slice was extracted using identical solar coordinates, but with
Fourier filtering techniques previously employed to isolate particular periodicities
corresponding to $3$ (upper-left),
$4$ (upper-right), $5$ (middle-left), and $6$~(middle-right)~minutes. Red
horizontal dashed lines highlight the inner- and outer-penumbral
boundaries at $\approx3.8$ and $\approx10.1$~Mm, respectively,
from the centre of the umbra. Solid green lines indicate the lines-of-best-fit
used to calculate the gradient in each of the time--distance diagrams, and thus represents
a measure of the period-dependent phase speeds. The lower panel displays
the RPW phase speed (in {\kms}) as a function of oscillatory period.
Image reproduced from \citet{Jess13}.
\label{Jess}}
\end{center}
\end{figure}

\vspace{3mm}
Pushing the boundaries yet further by employing high-resolution images obtained with
the Rapid Oscillations in the Solar Atmosphere \citep[ROSA;][]{Jes10b} and
Hydrogen-Alpha Rapid Dynamics camera \citep[HARDcam;][]{Jess12} instruments on the
Dunn Solar Telescope,
\citet{Jess13} compared the dynamics of RPWs with magnetic field extrapolations obtained
with the Helioseismic and Magnetic Imager \citep[HMI;][]{Sch12HMI} onboard the
Solar Dynamics Observatory \citep[SDO;][]{Pes12}. The authors found that the composition
of both the observed frequencies and the spatial locations at which they were present
conclusively agreed with the predicted cut-off frequencies imposed by
the geometry of the magnetic fields at chromospheric heights, something which was
initially proposed by \citet{Rez12} who used lower resolution UV data from the
Atmospheric Imaging Assembly \citep[AIA;][]{Lem12} onboard SDO. Furthermore,
\citet{Jess13} were able to determine the phase speeds of the propagating
magnetoacoustic RPWs as a function of their oscillatory period by decomposing the
original time series into narrow frequency bands through implementation of $3$--dimensional
Fourier filtering techniques (Figure~{\ref{Jess}}). The authors were able to corroborate the
results of \citet{Kob06}, but place more stringent thresholds on the periodicities and
phase speeds of the RPWs as a direct result of the high temporal and spatial resolutions offered
by the ROSA and HARDcam instruments. Consequently, the linkage between the photospheric
$p$-mode spectrum, the geometry changes of the magnetic field lines with atmospheric
height, and the resulting wave signatures visible in simultaneous chromospheric observations
directly supports the interpretation that RPW phenomena are the chromospheric signature of
upwardly-propagating magnetoacoustic waves generated in the photosphere. This
has since been further substantiated by \citet{Yuan14}, who utilised chromospheric UV
AIA images to reveal how the distribution of oscillatory power varied in the vicinity
of a sunspot as a function of spatial location, and further suggested how such wave
characteristics may reflect on the localised magnetic and thermal composition. However, while it has been
demonstrated through multiwavelength chromospheric observations that upwardly
propagating magnetoacoustic oscillations are rife within sunspot penumbrae,
the energetics associated with these waves are negligible with regards to the
overall radiative losses experienced in chromospheric active regions \citep{Nye76, Gall78}.

\begin{figure}[!t]
%\epsscale{0.5}
%\plotone{Nagashima_2007.jpg}
\begin{center}
\includegraphics[width=0.6\textwidth,clip=]{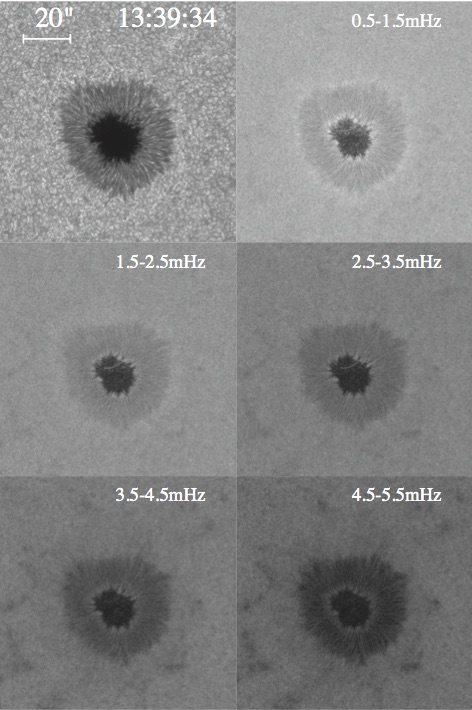}
\caption{A photospheric G-band image (upper-left) of an active region
observed by the SOT instrument onboard Hinode on 2007 January 8. The remaining
panels depict the spatial mapping of Fourier power of magnetoacoustic oscillations in narrow frequency
bins corresponding to $0.5$--$1.5$~mHz ($667$--$2000$~s), $1.5$--$2.5$~mHz
($400$--$667$~s), $2.5$--$3.5$~mHz ($285$--$400$~s), $3.5$--$4.5$~mHz
($222$--$285$~s) and $4.5$--$5.5$~mHz ($180$--$222$~s). It is clear that highly magnetic
locations surpress magnetoacoustic power over all frequency ranges.
Image reproduced from \citet{Nag07}.
\label{Nagashima}}
\end{center}
\end{figure}

\vspace{3mm}
Oscillations manifesting in the near-vertical magnetic field configurations of sunspot
umbrae have recently began to attract the attention of the solar physics community again.
Magnetic fan and plume structures are commonly observed to extend outwards from sunspots
into the solar corona, often with lengths exceeding many hundreds of Mm \citep{Cur08, Kri12b, Rao14}.
One of the first studies which uncovered propagating wave phenomena in such coronal
structures was by \citet{Def98}, who used the Extreme-ultraviolet Imaging Telescope
\citep[EIT;][]{Del95} onboard the Solar and Heliospheric Observatory \citep[SoHO;][]{Dom95} to
identify quasi-periodic perturbations in the brightness of EUV image sequences. More recent
studies, incorporating higher resolution telescopes such as TRACE, interpreted
these oscillations as the signatures of upwardly propagating magnetoacoustic waves
with velocities in the range of $70$--$165$~{\kms} and periodicities of $180$--$420$~s
\citep{Ofm99, Ofm02, DeM03, DeM04, Men04, Kri11, Kri12a, Kob14, Liu2014}. However, while the
coronal characteristics of magnetic fan and plume oscillations were well documented,
the origin of these waves remained elusive. \citet{DeM02} suggested that the
most likely explanation would be a photospheric driver directly exciting the magnetic
footpoints of the fan and plume structures. This scenario requires the magnetoacoustic
wave trains to be able to propagate from the photosphere,
through the chromosphere and transition region, and into the corona.
\citet{Kho06} produced numerical simulations of the lower solar atmosphere and revealed how
$\sim3$~minute magnetoacoustic oscillations generated at the base of a sunspot umbra can
readily propagate upwards through the lower layers and into the transition region. Thus, a
key science goal was now not only to detect these waves, but to track them through the
chromosphere to coronal heights. However, three minute umbral oscillations are
notoriously difficult to detect at photospheric heights. Both \citet{Lit85} and \citet{Bal87} were
unable to detect photospheric signatures of $3$~minute umbral oscillations, claiming they
may be swamped by noise due to their very low amplitudes. Employing the higher sensitivity
SOT instrument onboard Hinode, \citet{Nag07} revealed how
all oscillatory power within sunspot umbrae is drastically reduced (Figure~{\ref{Nagashima}});
a common phenomenon now referred to as `acoustic power suppression'
\citep{Woods81, Tho82, Tit92, Par07, Cho09, Ilon11, Cou13}.
Even more recently, \citet{Kob08, Kob11a, Kob11b} were not only able to detect
photospheric $3$~minute oscillations, but the authors tentatively claimed that the
locations of minimum photospheric power also corresponded to heightened power in
co-spatial chromospheric observations. Unfortunately
the spatial resolution obtained by the Horizontal Solar Telescope \citep{Kob09} at the
Sayan Solar Observatory, Russia, was on the order of $1''$, thus preventing
precise characterisation of the exact umbral structures displaying the $3$~minute
periodicities.

\begin{figure}[!t]
\begin{center}
%\epsscale{0.8}
%\plotone{Jess_2012.png}
\includegraphics[width=0.85\textwidth,clip=]{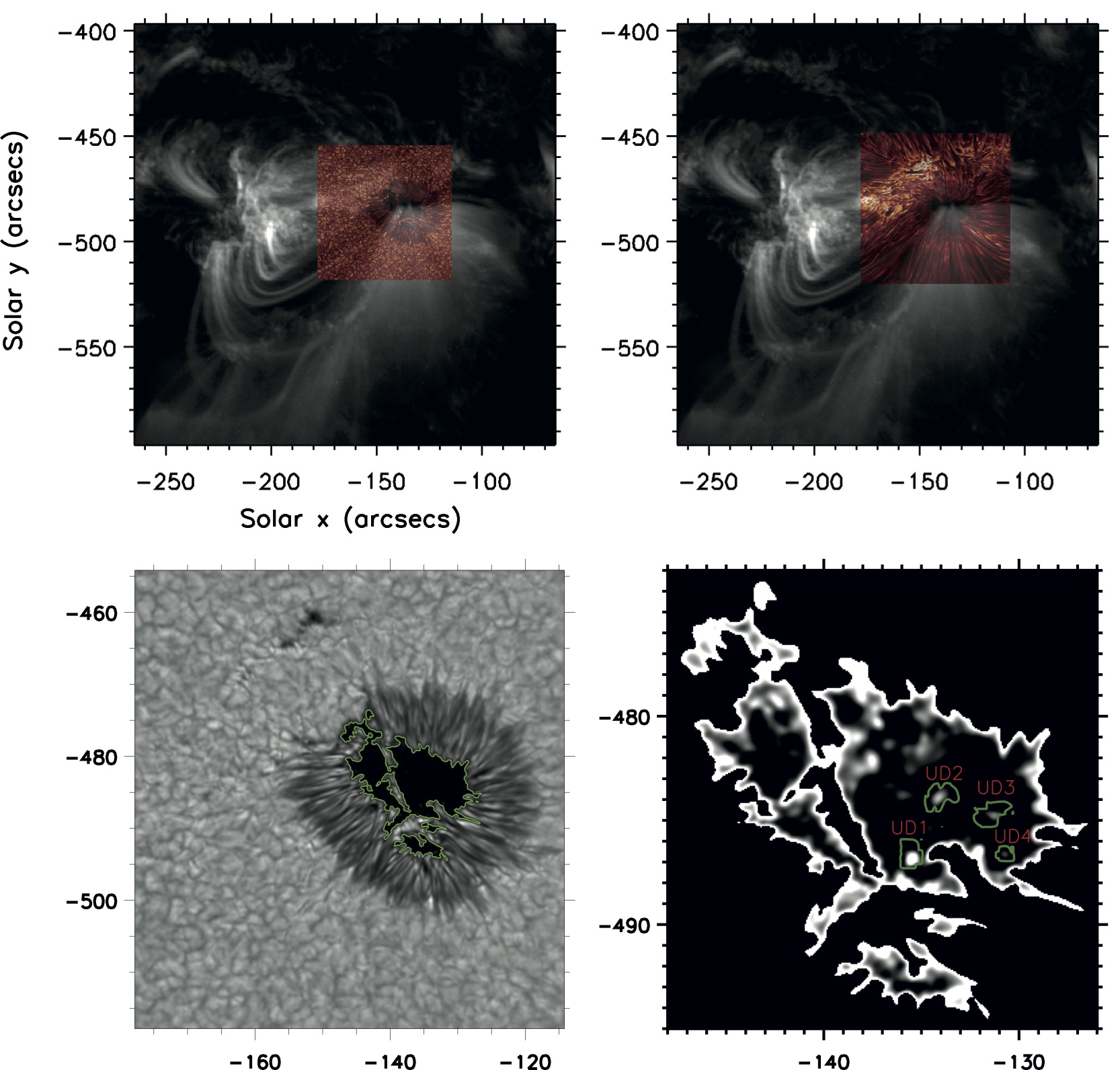}
\caption{The top panels display coronal EUV 
($171${\,}{\AA}) images taken by
the AIA instrument onboard SDO on 2011 July 13 and 
interlaced with co-spatial and co-temporal ROSA 
4170{\,}{\AA} continuum (upper left) and 
H$\alpha$ (upper right) snapshots. It is clear
the high degree of co-alignment precision now 
possible between multiwavelength
{\it{and}} multi-instrument image sequences.
The lower left panel shows the full ROSA 
4170{\,}{\AA} continuum field-of-view, where solid 
green contours highlight the perimeter of the sunspot 
umbra. The lower right panel displays a magnification of 
the umbra itself, and reveals a number of small-scale 
intensity enhancements within the dark umbral background. 
Such umbral dots, labelled UD1, UD2, UD3 and UD4, display 
$3$~minute oscillatory power several orders-of-magnitude
higher than in the surrounding sunspot umbra (green contours) 
and are believed 
to be the locations where the intense coronal fans, each 
displaying prominent slow magnetoacoustic wave phenomena, 
are anchored. Images adapted from \citet{Jess12}.
\label{Jess2}}
\end{center}
\end{figure}

\vspace{3mm}
Undertaking a multiwavelength study spanning the near infrared through to the EUV,
\citet{Jess12} were able to provide evidence that small-scale photospheric umbral dots
directly contribute to the presence of propagating magnetoacoustic waves observed
in coronal fan structures. First, it was noted that umbral dots, visible as concentrated
brightenings in the sunspot umbra with diameters $\sim 0{\,}.{\!\!}''5$, displayed
$3$~minute oscillatory power several orders of magnitude higher than the surrounding
umbral locations. Regions of heightened and localised power were also co-spatial in
simultaneous chromospheric H$\alpha$ and Ca~{\sc{ii}} image sequences. Employing
spectral imaging techniques with the Interferometric BIdimensional Spectrometer
\citep[IBIS;][]{Cav06}, \citet{Jess12} were able to compare the phase relationship
between intensity and line-of-sight velocity measurements to categorise the wave
signatures as upwardly propagating slow magnetoacoustic modes. Then, by interlacing
co-temporal EUV images from the AIA instrument onboard SDO with the chromospheric
observations, it was found that the footpoints of the coronal fans lay directly
on top of the umbral dot structures displaying heightened oscillatory power
(Figure~{\ref{Jess2}}). Almost unbelievably, it appears that photospheric
structures with diameters $\sim 0{\,}.{\!\!}''5$ ($360$~km)
can drive propagating magnetoacoustic oscillations in coronal structures
not only several thousand km above their position, but on structures
which have expanded into the local plasma to diameters often
exceeding $10''$ ($7000$~km). Estimations of the energy carried
by such propagating disturbances has been performed by \citet{Def98} and
\citet{DeM00}, producing an incredible span of values in the range of
$0.1$--$100$~W{\,}m$^{-2}$, thus opening up possibilities for such
magnetoacoustic waves to contribute significantly to the heating of the
lower corona through compressive dissipation. Recent work by \citet{Kid12}
and \citet{Uri13} have verified the temperature-dependent nature of the
propagation speeds of disturbances in fan/plume structures, suggesting
the magnetic field topology from the photosphere upwards may play an important
role in the observed dynamics; similar to the frequency filtering observed in
RPW phenomena \citep{Rez12, Jess13, Yuan14}. However, in contrast to
coronal fan and plume structures observed directly above sunspots,
those positioned within active regions, yet in non-sunspot locations,
appear to display vastly different characteristics. Often the non-sunspot
structures display wave periodicities longer than $10$~minutes
\citep{Ber99, Mar09, Wan09}, and as a result
cannot be interpreted in terms of upwardly propagating $p$-mode oscillations
\citep{Wan13}. Another outstanding issue is how such low frequency
waves actually penetrate into the corona since acoustic-based cutoff
theories cannot explain this. Subsequently, it has been suggested that small-scale
nanoflare activity in the solar chromosphere may be able to trigger such
low-frequency wave phenomena \citep{Ofm12}, although conclusive evidence
for such a distinctly different driver has not yet been observed.

\vspace{3mm}
Solar pores are often described as the first evolutionary stage of a typical
sunspot structure. Their defining characteristics at lower atmospheric heights
are a relatively small (compared to fully developed sunspots) umbral core without
the presence of surrounding penumbrae \citep{Sob99}. \citet{Cho10, Cho13} have
recently provided observational evidence to support numerous
numerical studies \citep[e.g.,][]{Kno88, Cam07} that suggest how rapid
cooling within pore umbrae, through the inhibition of convective motion, drives strong downflows
which collide with the dense lower layers below the photosphere, producing
reflected upflows that can assist with the transportation of significant energy flux
to chromospheric heights. Furthermore, the general properties
associated with pores are often similar to those found in fully developed sunspot
umbrae, including their temperatures and magnetic field inclination angles
\citep{Kopp92, Mug94, Sut96, Cri12}. As with sunspots, the highly magnetic nature of
pores allow them to act as efficient waveguides for magnetoacoustic
oscillations. Using the TRACE satellite, \citet{Bal00} studied the upward
propagation of magnetoacoustic waves in near-circular pores to chromospheric heights.
The authors found that the
observational signatures best fitted the `whispering gallery' mode first put forward
by \citet{Zhu00}, whereby the detected magnetoacoustic waves induced larger-amplitude
magnetic field oscillations than for physically larger magnetic structures (e.g., sunspots).
These oscillations rapidly diminish in amplitude towards the edge of the pore's magnetic
radius \citep[thus defining a discontinuity boundary of the magnetic field;][]{Hir03}, which can be
substantially larger than its visible radius when observed in optical wavelengths \citep{Kep96}.
However, due to poor seeing conditions and the resulting inability to accurately co-align
their ground-based data with that from the TRACE satellite, \citet{Bal00} were unable
to evaluate the energy flux of magnetoacoustic waves reaching chromospheric heights, but
instead pointed out the need for higher resolution (both temporally and spatially)
observations to better isolate propagating wave trains.

\begin{figure}[!t]
%\epsscale{0.4}
%\plotone{Sobotka_2013.eps}
\begin{center}
\includegraphics[width=0.5\textwidth,clip=]{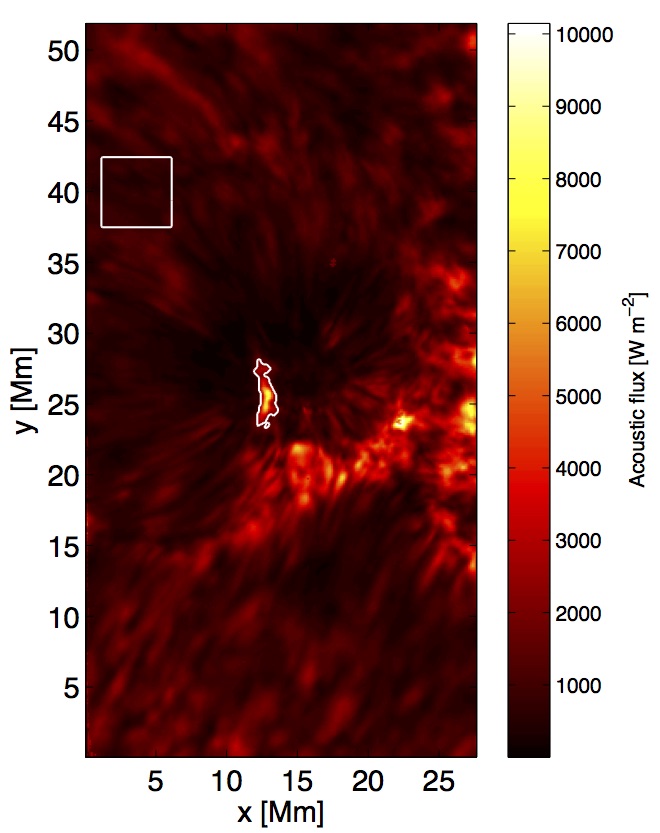}
\caption{Map of the total magnetoacoustic power flux, measured from a series
of Dopplergrams acquired with the IBIS instrument at the DST, and summed over
all magnetoacoustic wave periods in the range of $100-1000$~s.
The solid white line highlights the extreme localised energy flux, often exceeding
$10{\,}000$~W{\,}m$^{-2}$ in the chromosphere, found in a light-bridge region
separating two distinct solar pores. A white box outlines a region of the quiet
chromosphere, which still displays heightened magnetoacoustic flux on the order
of $1000$~W{\,}m$^{-2}$. Image reproduced from \citet{Sob13}.
\label{Sobotka}}
\end{center}
\end{figure}

\vspace{3mm}
Employing the IBIS instrument on the DST, \citet{Sta11, Sta12} and \citet{Sob12, Sob13}
found evidence for magnetoacoustic waves, with periodicities in the range of
$100-1000$~s, leaking upwards into the chromosphere along the pore's inclined
magnetic fields. The authors claimed that the energy flux of the upwardly propagating
waves was sufficient to balance the entire radiative losses of the pore's chromosphere
structure, deemed to be $\sim 3400$~W{\,}m$^{-2}$ averaged over the surface area
of the pore, with localised peaks reaching in excess of $10{\,}000$~W{\,}m$^{-2}$ for
particularly bright regions of the pore's chromosphere (Figure~{\ref{Sobotka}}). It
appears from the recent literature that the small-scale, yet highly magnetic nature of
solar pores provide idealised wave conduits to efficiently transport energy into higher
layers of the solar atmosphere. 
Interestingly, however, \citet{Sob13} uncovered distinct wave 
characteristics in a solar pore that also incorporated a 
light bridge. The authors found that the three minute oscillations 
dominated the pore umbra, while significant five minute periodicities 
were observed above the light bridge. Recently, 
\citet{Yuan14a} were able to identify identical 
wave characteristics in a large-scale sunspot that also displayed 
a prominent light bridge. Here, the authors suggested that the 
presence of significant five minute oscillations above the 
light bridge may be the result of the creation of 
standing magnetoacoustic waves along the thin edge of the 
light bridge. Ultimately, such findings pose challenges to the 
connectivity and traditional suppression of five minute $p$-mode 
oscillations typically observed within the highly magnetic 
vicinity of pore and sunspot features. Thus, as suggested by 
\citet{Yuan14a}, modelling the $p$-mode interaction with a 
pore and/or sunspot that has a prominent light bridge will be 
an interesting topic for future theoretical consideration. 

A limiting factor in the quest for a
global heating mechanism is the fact that solar pores, just like their larger
sunspot counterparts, are not sufficiently common to provide continual energy
flux to the outer layers of the Sun's atmosphere. Furthermore, due to their
limited size, and thus their inability to efficiently inhibit the surrounding convective motions
on long-term time scales, solar pores often display minimal signatures at higher
atmospheric heights (transition region and corona). \citet{Sut98solo} and
\citet{Sut98} employed full-Stokes analysis of pore structures and found that they
displayed a vertical magnetic field gradient of $\sim 5$~G{\,}km$^{-1}$, marginally
inflated when compared to large-scale sunspots
\citep[$1$--$3$~G{\,}km$^{-1}$;][]{Pah90, Bru95, Rue95, Ber06}, thus depleting
their observational (and magnetic) signatures rapidly as one moves away from
the photospheric layer. Nevertheless, pore structures provide observers with one
of the most idealised wave conduits in the lower solar atmosphere: a nearly circularly
symmetric waveguide that is heavily susceptible to external motions, buffeting and driving forces.
As a result, it is foreseen that a multitude of focussed efforts will be undertaken on solar
pores in the near future in an attempt to compare observations more readily with MHD
wave theory, thus opening up possibilities of being able to refine and/or refute theoretical
wave flux predictions.

\subsection{Sausage Waves}
\label{sausages}
Sausage oscillations have proven to be one of the most difficult of the compressible wave modes to
identify observationally. These waves are typically identified through simultaneous periodic intensity {\it{and}} area
fluctuations in magnetic solar structures including pores, spicules and coronal loops. The high spatial resolution
necessary to identify the fractional area changes meant that early attempts were limited to
studying oscillations in the radio emission of coronal loops
\citep[e.g.,][]{Drooge1967}. However, more recently \citet{Nakariakov2003} demonstrated that incorrect
theoretical interpretation of the dispersion relations had been applied to previous
radio observations. It was found that the long-wavelength cutoff and the highly dispersive nature
of the phase speeds were not considered, and therefore the earlier
results needed to be revisited to apply these corrections.
\citet{Asc04} subsequently catalogued the relevant radio
observations and derived refined properties for these waves through the new
theoretical understanding. The oscillations presented were
shown to be fast sausage-mode oscillations which were confined to small segments of the
magnetic loop that corresponded to higher harmonic modes. The radio frequencies of these
waves were shown to be able to constrain the plasma density since the oscillations
could only exist at atmospheric heights greater than $\sim 40$~Mm, representing the
apex of the loop where the density contrast with respect to the background is greatest.
This work also confirmed the
observations of global fast sausage modes by \citet{Asai2001}, whereby oscillatory
behaviour was evident throughout the entire magnetic loop.
These observations were made using microwave images from
the Nobeyama Radioheliograph \citep[NoRH;][]{Nakajima1994} and images
from the Yohkoh soft X-ray telescope \citep[SXT;][]{Tsuneta1991},
and provided the first observational evidence that sausage-mode waves may be able to
propagate through the lower solar atmosphere providing the magnetic field guidance
was sufficiently strong \citep[$B \simeq 40$~G;][]{Asai2001}.
In the years since, there have been
numerous studies conducted on sausage-mode oscillations at coronal heights
\citep[e.g.,][]{Srivastava2008}, but evidence to support their existence within
the solar chromosphere has proved to be much more elusive.

\vspace{3mm}
The first lower atmospheric observations of sausage-mode waves were by
\citet{Dorotovic2008}. White light observations of a magnetic pore were
taken with the Swedish Vacuum Solar Telescope
\citep[SVST, now renamed the SST;][]{Scharmer1985}.
Periodic area changes in the photospheric pore were observed by analysing the area time
series using the wavelet analysis techniques of \citet{Torrence1998} . This
analysis identified area oscillations with periods on the order of $20-70$ minutes,
and it was suggested that the long periods present were the signature of
magnetoacoustic gravity modes, although the existence of
these waves have yet to be directly confirmed observationally. This work verified
the existence of sausage modes at photospheric heights and showed that highly magnetic
pore structures were viable conduits for these waves.

\vspace{3mm}
With the development of sensitive high-cadence camera systems (e.g., ROSA), it has
become possible in recent years to study sausage-mode oscillatory phenomena at
unprecedentedly high spatial and temporal resolutions.
\citet{Mor11} imaged a group of magnetic pores using a blue
continuum ($4170${\,}{\AA}) channel with ROSA, thus maximising the diffraction-limited
spatial resolution and allowing highly sensitive measurements of any area changes to
be undertaken. In this study, \citet{Mor11} employed Empirical Mode Decomposition
\citep[EMD;][]{Huang1998} techniques to identify simultaneous oscillations in the pore
area and intensity, with periodicities in the range of $50-600$~s detected. The
shorter periods, when compared to the results of \citet{Dorotovic2008}, were
thought to be a result of the waves being driven by the global solar $p$-mode
spectrum instead of magnetoacoustic gravity modes. However, the majority
of the sausage oscillations were only observed
in the area time series, without simultaneous intensity perturbations,
indicating that they did not possess a large quantity of wave energy. For
those oscillations that were concurrently observed in both the area and intensity,
it was determined that the intensity fluctuations were
$180^{\circ}$ out-of-phase with the area changes. Although this characteristic
was not interpreted at the time, such a phase relationship was later shown
to be evidence that these oscillations were fast sausage mode
waves \citep{Moreels2013}.
Further study of photospheric sausage modes was conducted by
\citep{Dorotovic2014}, with the aim of distinctly identifying features of fast and slow
modes. Employing active region observations acquired with the SVST and
the Dutch Open Telescope \citep[DOT;][]{Rutten2004}, wavelet
analysis was performed on pore and sunspot features
to identify fluctuations in the areas and intensities of these structures, with
the resulting phase relationships studied using EMD. Standing photospheric oscillations
were identified with periods ranging from $4-32$ minutes, with the
observed modes interpreted as a combination of fast surface and slow sausage
modes. Such photospheric work has shown that sausage modes can form in
magnetically active structures such as pores and sunspots at photospheric
heights, and that both fast and slow modes can exist in the photosphere. This
supports the possibility that these waves can carry energy from the solar
surface to higher heights to aid atmospheric heating, although the search for
such waves within the chromosphere is in it's infancy.

\begin{figure}[!t]
\begin{center}
\includegraphics[width=0.8\textwidth,clip=]{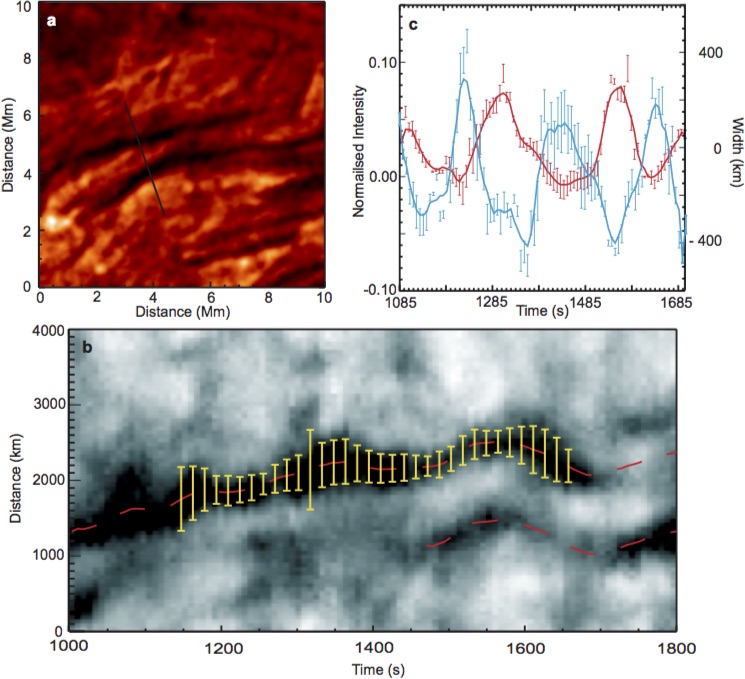}
\caption{Panel (a) depicts a cropped ROSA H$\alpha$ snapshot containing
a pair of relatively large dark chromospheric flux tubes. Using the cross-cut
(black line) to extract intensity information, panel (b) displays the resulting
time--distance diagram revealing the dynamic motion of the waveguides.
Times are given in seconds from the start of the data set, while the overplots
are the results from a Gaussian fitting routine to show the non-linear
fast MHD kink wave (red line shows the central axis of the structure) and
the fast MHD sausage mode (yellow bars show the measured width of structure).
The transverse motion has a period of $232 \pm 8$ s and bi-directional
phase speeds equal to
$71 \pm 22$~{\kms} upwards and $87 \pm 26$~{\kms} downwards.
The typical velocity
amplitudes are $5$~{\kms}. The fast MHD sausage mode has a period of
$197 \pm 8$~s, a phase speed of $67 \pm 15$~{\kms} and apparent velocity
amplitudes of $1 - 2$~{\kms}. Panel (c) displays a comparison between the
detected intensity
(blue) and width (red) perturbations resulting from the Gaussian fitting.
The data points have been fitted with a smoothed $3$-point boxcar function.
The observed out-of-phase behaviour is typical of fast MHD sausage waves.
The error bars plotted are the one-sigma errors on each value calculated
from the Gaussian fitting. Image reproduced from \citet{Mor12}.
\label{Morton2012}}
\end{center}
\end{figure}

\vspace{3mm}
A major piece of work that has initiated our improved understanding of sausage-mode waves
in the solar chromosphere was that by \citet{Mor12}. In this study,
H$\alpha$ observations were acquired using ROSA on the DST, with the field-of-view
cropped in order to observe a $34\times34$~Mm$^{2}$ region of the quiet chromosphere.
The imaged region contained hundreds of fine-scale structures,
composed of elongated fibrils and short mottles, which accurately mimic a
theoretical flux tube. Alongside the observed incompressible kink
modes (see \S{\,}\ref{incompressible_section}), periodic
intensity fluctuations were also seen to exist alongside the expansions and contractions of the
visible cross-sections of these chromospheric structures. A difficulty arose when attempting
to isolate multiple wave periods as a result of the short lifetimes of the waveguides. Instead,
intensity perturbations in a series of time--distance diagrams which lay along
the axis of the structure were used (Figure~\ref{Morton2012}). The extracted
intensity perturbations, alongside simultaneous area oscillations, identified the presence
of sausage-mode waves that were seen in numerous structures across the entire
field-of-view.
Traditionally, intensity fluctuations observed through narrowband 
filters are often considered synonymous with density perturbations of 
the plasma \citep[e.g.,][]{KLIBRA2014}. However, additional 
circumstances can manifest that may introduce alternative 
interpretations for observed intensity fluctuations. 
As mentioned in \S~\ref{difficulties_section}, 
line-of-sight Doppler shifts can result in a narrowband filter 
sampling a different part of an absorption profile (i.e., the blue/red 
wings instead of the line core), thus causing a shift in observed 
intensity \citep[see, e.g.,][]{Jes07}. Also, when 
employing a broadband filter, the observed intensities can be 
thought of as following a simple Planck function under the assumption 
of local thermodynamic equilibrium. Thus, any perturbations in 
intensity can be interpreted as a signature of local temperature 
fluctuations. However, this interpretation hinges upon the accuracy of 
the local thermodynamic equilibrium approximation.
Propagation speeds were deduced by \citet{Mor12}, and were shown to be
on the order of the Alfv{\'{e}}n speed, thus indicating that they are most likely
fast sausage modes. It was also inferred that some of these waves are
propagating upwards through the atmosphere. This is due to many of the chromospheric
structures being inclined with respect to the solar surface, allowing upwardly propagating
waves to be identified within single-channel images. Analysis of the energetic properties of these waves was
conducted to ascertain their potential suitability as conduits for atmospheric heating.
An important parameter to calculate when undertaking energy analyses is the dimensionless
variable, `$ka$', the product of the wavenumber, $k$, and waveguide
half-width, $a$. Waves are defined as `trapped', whereby they retain energy as they
propagate in the absence of external damping, if $ka$ is greater than a constant
dependent on the internal and external Alfv{\'{e}}n speeds \citep{Cal86}. 
In terms of external damping, there is a rich variety of viable 
dissipation mechanisms identified for various wave modes 
manifesting throughout the solar atmosphere. These include resonant 
absorption and phase mixing 
\citep[e.g.,][]{Goossens2001, Ruderman2002, Goossens2002}, 
turbulent mixing \citep{VanB2011}, in addition to examples of 
mode conversion \citep[e.g.,][]{Ulm1991, Kalkofen1997} 
and thermal conduction \citep{Ofman2002,  Men04, Erdelyi2008}.
Furthermore, the waves may be defined as `leaky' should they readily dissipate
their embodied energy without first being prompted by external effects.
\citet{Mor12} estimated the internal and external Alfv{\'{e}}n speeds based
on the known properties of a cold, dense chromosphere and defined the
$ka$ value at which the waves transferred across the trapped/leaky divide
as $ka\simeq0.2$. The wave activity displayed in Figure~\ref{Morton2012}
was used to estimate $ka \sim 0.09\pm0.03$, with the results suggesting
that the observed sausage-mode waves lie within the leaky
regime and will therefore radiate energy away from the magnetic structure
without the need for external damping. This is clearly an advantageous
characteristic to promote localised atmospheric heating, and estimates
of the individual wave energies produced values on the order of
$11700 \pm 3800$~W{\,}m$^{-2}$. This is a significant amount of energy, and
\citet{Mor12} highlighted the fact that if only $5$\% of the chromosphere was
connected to the corona via flux tubes then a total energy flux of
$460 \pm 150$~W{\,}m$^{-2}$ would be available to the corona for atmospheric
heating. Despite contemporaneous SDO imaging,
it was unclear how the observed sausage-mode waves interacted with the corona.
However, these early results indicate that sausage oscillations may play an important
part in supplying both the chromosphere and the corona with the necessary
energy flux to maintain their elevated temperatures.

%Resonant absorption, conversion into leaky mode and thermal conduction

\begin{figure}[!t]
\begin{center}
\includegraphics[width=0.8\textwidth,clip=]{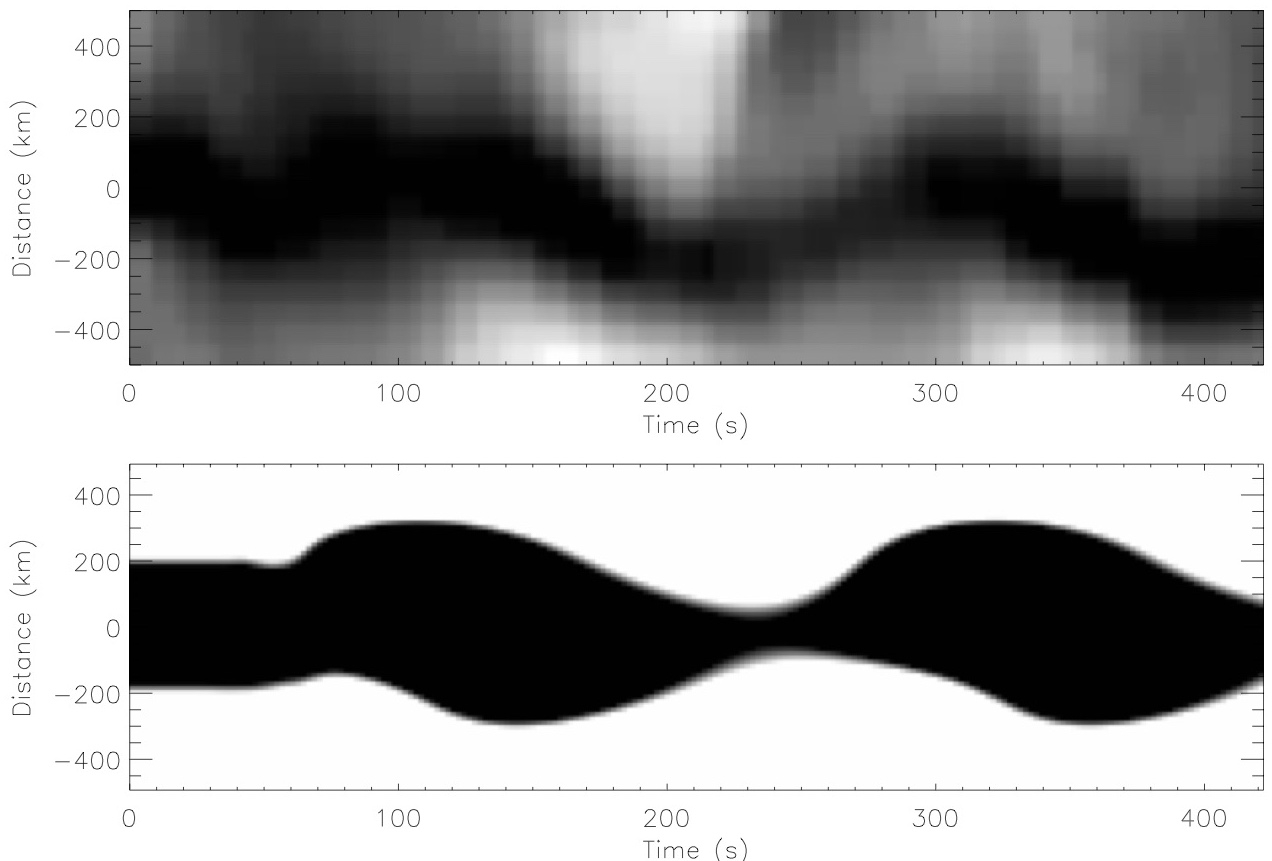}
\caption{The top panel displays a time--distance diagram of high-cadence chromospheric H$\alpha$
observations cut perpendicularly through the central axis of a spicule. The bottom panel displays
a comparitive time--distance diagram of simulated
chromospheric spicule densities having first been driven by out-of-phase compressive oscillations
at the solar surface. There is a remarkable degree of similarity between the
two panels, with both kink (transverse displacement of
spicule axis) and sausage (periodic compressions and rarefactions)
modes visible. This clearly shows how the velocity gradients generated as a result of
out-of-phase compressive oscillations at the footpoints of spicule structures can create
both compressible {\it{and}} incompressible wave modes at higher atmospheric heights.
Image adapted from \citet{Jess2012}.
\label{Jess2012}}
\end{center}
\end{figure}

\vspace{3mm}
Supplementary images of the lower solar atmosphere highlighted
that MBPs may be the photospheric anchor points of the chromospheric
waveguides undergoing sausage-mode oscillations \citep{Mor12}. Although
no information is provided by the authors regarding whether
simultaneous oscillations are present at lower atmospheric heights,
this suggestion is consistent with the previous work of \citet{Jess2012} who
examined the connectivity between photospheric MBPs and chromospheric
spicules. The primary aim of this work was not to study sausage-mode oscillations,
but instead to examine the mode-coupling between compressible and incompressible
waves found in MBPs and their chromospheric spicule counterparts. Through
use of G-band, {\cak} and H$\alpha$ filtergrams obtained with ROSA,
\citet{Jess2012} found compressive fluctuations across the body of an isolated
photospheric MBP that coupled into incompressible transverse oscillations
in the lower chromosphere. Importantly, the compressive magnetoacoustic
oscillations were found to be $90^{\circ}$ out-of-phase at opposite sides of the
MBP. Employing the Lare2D numerical code \citep{Arber2001} and modelling
a spicule as a thin magnetic flux tube, the authors found
that a $90^{\circ}$ out-of-phase behaviour at the photospheric level not only
produced velocity gradients that caused the spicule axis to displace transversally,
but the motions also induced compressions and expansions in the waveguide,
thus promoting the manifestation of both compressible sausage modes and
incompressible transverse waves at chromospheric heights (Figure~{\ref{Jess2012}}).
The similarity between the observed and simulated spicule dynamics
clearly shows how thin, magnetic structures omnipresent throughout the solar
chromosphere can readily support sausage-mode wave generation and propagation.
While no analyses of the sausage-mode energetics was performed by \citet{Jess2012},
the work of \citet{Mor12} highlights the impressive energy these waves can carry; more than
sufficient to balance the extreme localised radiative losses experienced in the chromosphere
and corona.

\vspace{3mm}
The study of sausage mode waves in the chromosphere is a new and
developing field of research. Despite the small volume of published material,
their importance in terms of energy transport through the dynamic chromosphere
is becoming more clear. It has been shown that these waves can be generated
in the photosphere through a variety of mechanisms, including the mode-conversion
of fundamental $p$-mode magnetoacoustic oscillations. It has also
been established that sausage-mode oscillations can carry a significant energy flux,
leading to the conclusion that these waves may act as an energy conduit for supplying
higher atmospheric heights with the necessary energy to balance radiative losses.
Many of these outstanding questions can be answered by employing new and existing
technology. For instance, a key goal for sausage mode
research is to utilise multiwavelength imaging to identify propagation from
the photosphere to the chromosphere and beyond in an attempt validate the efficiency at which
they transport energy. In this regard, the high resolution imaging of ROSA will
continue to be vital. Complimentary approaches would involve imaging spectroscopy techniques
(through use of the IBIS and CRISP instruments) to examine the line-of-sight velocities,
thermal and non-thermal spectral widths, and the manifestation of spectral line asymmetries in
order to more accurately categorise the wave modes, phase speeds, oscillation amplitudes
and energetics through prominent phase relationships intrinsic to each particular mode
\citep{Moreels2013a}.

\vspace{3mm}
While the contribution of slow magnetoacoustic waves to energy transportation needs to be 
re-assessed in light of these recent results, incompressible (or Alfv{\'{e}}nic) waves 
have consistently been a more-favoured mechanism for efficient energy transport. However, 
it wasn't until 2007 that direct evidence for ubiquitous incompressible waves was first presented. 
In the following section, we draw upon these post-2007 results and review the publications related to 
the direct observations of chromospheric incompressible waves. We demonstrate that 
their ubiquity has allowed for the typical properties of these waves to become relatively 
well constrained, although there are still some outstanding questions. We also discuss 
investigations that have provided clues as to how these waves are generated and 
how they evolve as they propagate through the chromosphere.

%%%%%%%%%%%%%%%%%%%%
%%%%%%%%%%%%%%%%%%%%
%%%%%%%%%%%%%%%%%%%%
%%%%%%%%%%%%%%%%%%%%
%%%%%%%%%%%%%%%%%%%%
%%%%%%%%%%%%%%%%%%%%
%%%%%%%%%%%%%%%%%%%%
%%%%%%%%%%%%%%%%%%%%
%%%%%%%%%%%%%%%%%%%%
%%%%%%%%%%%%%%%%%%%%
\clearpage
\newpage
\section{Incompressible Waves}
\label{incompressible_section}
Incompressible waves are characterised by $\nabla \cdot \vp = 0$ (see \S{\,}\ref{sec:uum}). 
In practice, waves that nearly satisfy this condition are labelled as incompressible. 
Therefore, from a physical point of view, incompressible waves can exhibit small 
pressure perturbations while the dominant restoring force is magnetic tension. The lack of
compression makes it particularly difficult to dissipate the wave energy
unless large gradients in the Alfv{\'{e}}n speed exist (e.g., \citealp{HEYPRI1983}).
This has made incompressible waves a favourable mechanism for
transferring energy from the convective motions in the photosphere
up into the upper chromosphere and coronae, playing the role as the
dominant energy carrier in many simulations that appear to be able to
generate hot corona \citep[e.g.,][]{CRAVAN2005, SUZINU2005, VERVEL2007, MATSHI2010}.

\vspace{3mm}
In a plasma medium composed of fine-scale magnetic structures, the incompressible
(or nearly incompressible) motions can be split into two main categories,
those of MHD fast kink motions (in the long wavelength limit, i.e.,
$\lambda{\gg}a$, where $\lambda$ is the wavelength and $a$ is the
radius of the waveguide) and the torsional Alfv{\'{e}}n mode
\citep[for theoretical discussions of the individual mode properties see, e.g.,][]{SPR1982, Edw83, BENetal1999, GOOetal2009}.
Over the last eight or so years, periodic motions of fine-scale
structures in the magnetically dominated upper chromosphere, in both
imaging and spectroscopic observations, have been associated with the presence
of both types of incompressible wave. These recent observations have
built upon numerous historic reports of periodic
variations in Doppler signals and filtergrams of various chromospheric
spectral lines, which were often interpreted in terms of Alfv{\'{e}}n waves, although
the exact nature of the signals remains ambiguous (see, e.g.,
\citealp{NIKSAZ1967}; \citealp{PASetal1968}; \citealp{SAW1974}; or
\citealp{ZAQERD2009} and \citealp{Mat13} for recent reviews).
The advantage of recent observations is that they possess the ability to
observe at high-spatial and temporal resolutions, allowing fine-scale
structures in the upper chromosphere, alongside its intrinsic dynamics, to be resolved.
This has only been made possible through;
\textit{(i)} seeing-free, space-based chromospheric observations
provided by Hinode/SOT;
\textit{(ii)} advances in reducing atmospheric
distortion through both instrumental (e.g., adaptive optics) and image reconstruction
techniques such as speckle \citep{WOEetal2008} and Multi-Object Multi-Frame
Blind-Deconvolution \citep[MOMFBD;][]{VANNetal2005} suitable for ground-based
observations, e.g., ROSA at the DST
\citep{Jes10b} and CRISP at the SST \citep{Sch08};
\textit{(iii)} increased spectral resolution, e.g., IBIS at the DST \citep{Cav06}
and the TRI-Port Polarimetric Echelle-Littrow spectrograph at the SST
\citep[TRIPPEL;][]{KISetal2011}.

\begin{figure}[!t]
\begin{center}
\includegraphics[width=0.95\textwidth,clip=]{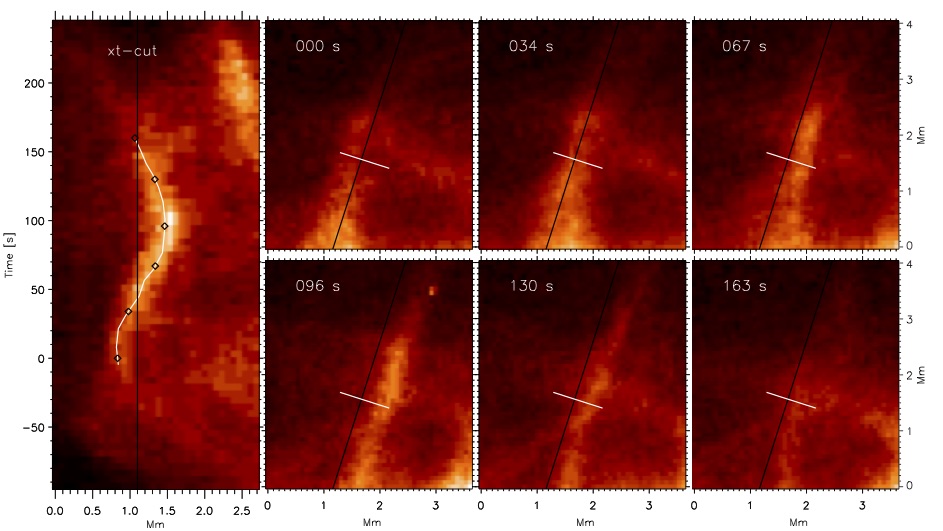}
\caption{Examples of fast kink wave motion observed in chromospheric spicule structures.
The panels demonstrate the
larger amplitude motions observed in spicules, with the largest (left-hand) panel showing a
time--distance diagram, while the smaller panels reveal sequential snapshots of
the spicule. This is in contrast to the typically smaller-displacement
fast kink waves present in fibrils (see Figure~{\ref{Morton2012}}).
Image reproduced from \citet{DeP07}.
\label{fig:wave_examp}}
\end{center}
\end{figure}

\vspace{3mm}
The interpretation of signatures pertaining to the torsional Alfv{\'{e}}n wave is
still contentious \citep[for a detailed overview see][]{Mat13}.
On the other hand, the interpretation of
the observed motions of fast kink waves is fairly straight-forward, with the
displacement of the central axis of the magnetic structure unambiguous in
images, e.g., Figures~{\ref{fig:movie:kink}},{\ref{Morton2012}} \& {\ref{fig:wave_examp}}.

\subsection{Observations and measurements}
In this section, we will review the observations of both types of
incompressible motions in the chromosphere. However, we believe it
necessary to split the observations into separate categories based on the nature of
the chromospheric structure. The reason for this will become obvious after
consideration of the different observations. Ultimately, these
categories essentially pertain to whether the features are thought to be
closed within the chromosphere, or open and connected to the corona, which
would undoubtedly lead to differing plasma properties, something
apparently reflected in the measured properties of the waves.

\vspace{3mm}
Before discussing the observations, we bring the readers attention to a
particular point of potential interest. The measurements of incompressible wave
phenomena have been performed using data from a variety of spectral
lines that are typically associated with the chromosphere, e.g., H$\alpha$,
Ca~{\sc{ii}}~H/K and the Ca~{\sc{ii}} infrared triplet at $8542${\,}{\AA}
(see \S{\,}\ref{difficulties_section} for a more detailed overview).
These lines have different properties with respect to opacities and
formation regions in the atmosphere (e.g., \citealp{RUT2007}; \citealp{LEEetal2009, Lee12}),
hence the local plasma properties could differ for chromospheric features
observed in various spectral lines, potentially leading
to subtle variations in measured wave parameters. The differing behaviour
of the related chromospheric phenomena is highlighted in
\citet{ROUetal2009}, who report higher velocities in Rapid Blue-shifted Events
(RBEs) observed in H$\alpha$ than in Ca {\sc{ii}} $8542${\,}{\AA} and suggest the
larger opacity in H$\alpha$ allows the sampling of higher layers. While the
effects of this are easier to avoid in limb observations, where the apparent
height in the atmosphere of a wave measurement can be deduced, this is not the
case for on-disk measurements. At present, there is no clear evidence
for any variation in wave properties measured using different lines --
although this may be due to the fact such an investigation has not yet been
undertaken. With this said, we give the wavelength used during each
observation discussed in the following but do not give any significance to this when
comparing results.

\subsubsection{Spicules}
\label{sec:wave_spic}
It is generally well known that spicules are jets of chromospheric
material that outline almost vertical magnetic field lines and
penetrate into the upper layers of solar atmosphere
(e.g., \citealp{BEC1968}). Spicules are predominantly observed
at network boundaries, appearing as a dense forest at the
limb and best seen on-disk in H$\alpha$ wing images after the
more `static' component of the chromosphere is removed
(\citealp{ZIR1988,RUT2007}).
More recently, there has been the sub-classification of spicules
into Type-I and Type-II varieties (\citealp{DEPetal2007a}). The observation
of the second type of spicule is said to be only possible with
high cadence, seeing free observations, such
as those provided by Hinode/SOT. Type-II spicules are apparently faster
moving than the traditional Type-I spicule and the material that
composes them is not seen to fall back towards the surface,
suggesting the plasma may be heated to coronal temperatures
as it rises and these spicules may play an important role in supplying heated
mass to the corona (\citealp{DeP11}). However, this idea
is currently contentious with vigorous opposition to the classification
of the spicules (e.g., \citealp{ZHAetal2012}) and their contribution to
coronal mass supply (\citealp{MADetal2011}; \citealp{KLI2012};
\citealp{GOO2014}; \citealp{PATetal2014}; \citealp{KLIBRA2014};
\citealp{PETetal2014}).

\begin{figure}[!t]
\begin{center}
\includegraphics[width=0.7\textwidth]{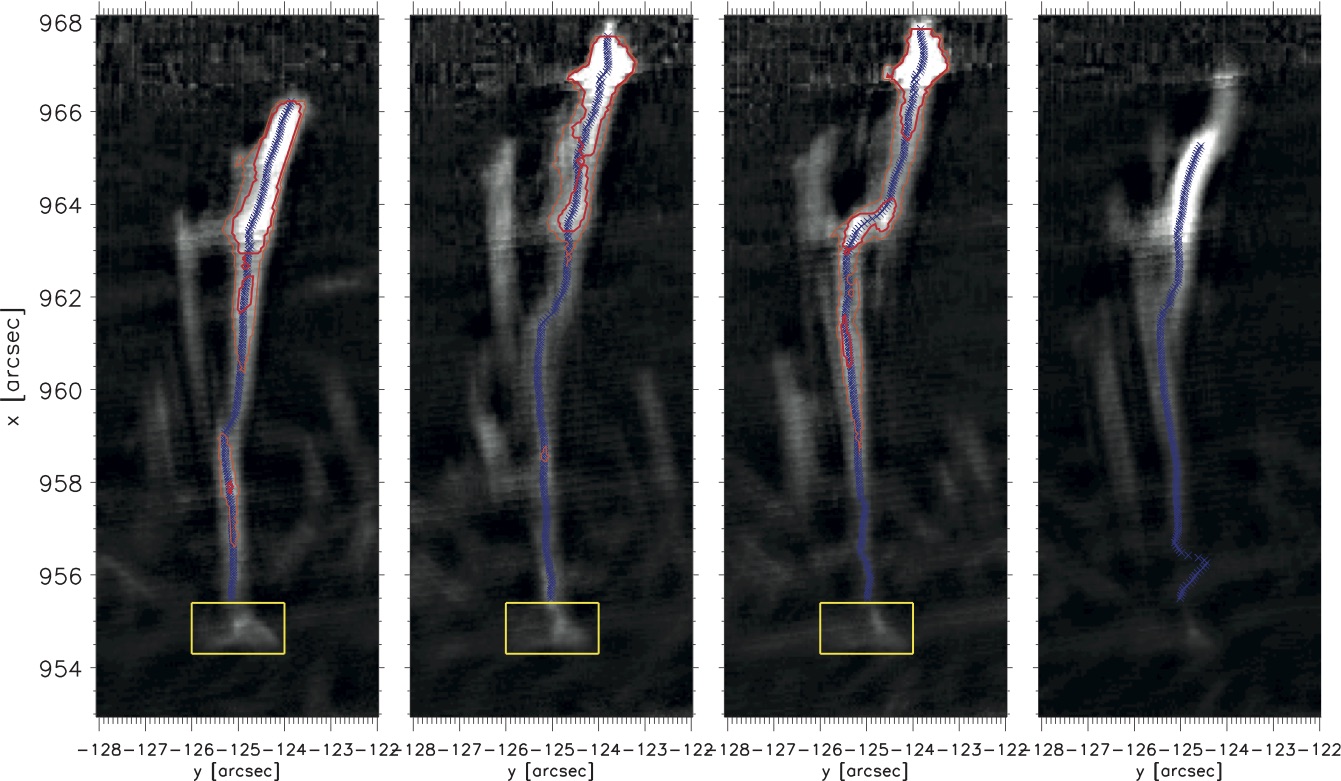} \\
\vspace{0.2cm}
\includegraphics[width=0.7\textwidth]{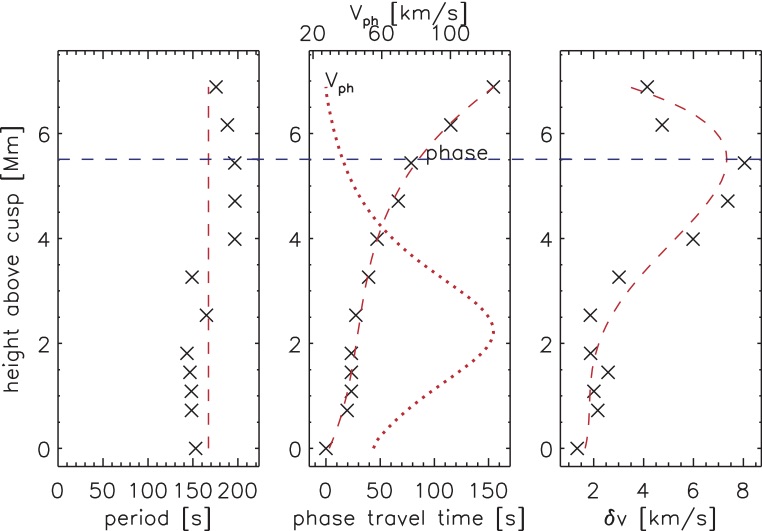}
\caption{A spicule oscillation observed with Hinode/SOT, where the upper panels
display the spicule structure and highlight the propagation of a wave front along the
feature. The lower panels show the measurements
of the kink wave properties as a function of atmospheric height. Here
$\delta v$ is the velocity amplitude and $v_{ph}$ is the propagation velocity.
Images adapted from \citet{HEetal2009b}.
\label{fig:wave_he}}
\end{center}
\end{figure}

\vspace{3mm}
Hinode/SOT observations suggested the need for refined
spicule classifications, and also revealed that spicules undergo pronounced
transverse displacements \citep{DeP07}. Limb observations in {\cah}
of a coronal hole found that the majority of spicules underwent
transverse displacements on the order of $500-1000$~km with
time-scales of $10$ to $300$~s, and had typical velocity
amplitudes of $10-20$~{\kms}. It appears that the authors primarily measured
uni-directional displacements, i.e., no sign of periodicity, however,
they did report that the longer lived spicules demonstrated
signatures of oscillatory motion (e.g., Figure~\ref{fig:wave_examp}).
Moreover, by comparing the observations to Monte Carlo
simulations, the authors estimated that the typical period of
oscillations had to lie between $150-300$~s. The subsequent interpretation
of the authors was that these observations could be explained by
Alfv{\'{e}}n waves, which led to intense debate
\citep{Erd07a, VANetal2008b}.
As mentioned in the introduction to this section, in a highly
structured atmosphere -- particularly one that exhibits structuring of the density
perpendicular to the direction of the magnetic field -- the pure Alfv{\'{e}}n
wave is a torsional motion. The transverse displacements of the waveguide
central axis represents the fast kink mode, which is Alfv{\'{e}}nic in the sense that
it is highly incompressible in the observable limit and that magnetic tension is
the dominant restoring force \citep{GOOetal2009}. While this may seem like a technical
detail, serious discrepancies can occur in estimates of the wave energy flux
depending on whether one assumes the observed waves are Alfv{\'{e}}n or
kink waves (\citealp{GOOetal2013}; \citealp{VANetal2014}). Assuming the waves
were Alfv{\'{e}}n waves, \citet{DeP07} calculated that they transported
around $4000-7000$~W{\,}m$^{-2}$. Recently, \citet{VANetal2014} re-evaluated the
estimates of \citet{DeP07}, and for typical filling factors of $5-15$\% the
energy flux is greatly reduced to $200-700$~W{\,}m$^{-2}$. The physical reason
for this reduction in energy flux is that for a kink wave the energy is strongly
localised in the neighbourhood of the flux tube density enhancement. This
is not the case for the more idealised bulk Alfv{\'{e}}n wave scenario, where the
waves are assumed to travel through a homogeneous plasma, resulting in a
spatially uniform energy flux. The bulk Alfv{\'{e}}n wave model is therefore
especially unsuited to thin, overdense magnetic structures such as spicules and
fibrils. Additionally, \citet{DeP07} provide an estimate for the typical
Alfv{\'{e}}n speeds in spicules by using previous measurements of spicular
magnetic fields ($B\sim10$~G) and densities ($\rho\sim10^{-11}-10^{-10}$~\kgm),
estimating $v_A=B/\sqrt{\mu_0\rho}\sim45-200$~{\kms}.

\vspace{3mm}
Before continuing, it is worth noting that the values for the propagation speed of the
fast kink wave is actually a weighted average of the internal and
external Alfv{\'{e}}n speeds. The subsequent propagation velocity will then
be greater than the internal Alfv{\'{e}}n speed of the spicule plasma, i.e. that quoted
by \citet{DeP07}. Nonetheless, it is clear that the fast kink waves, if
propagating, will traverse a typical spicule length in a matter of tens of seconds to
minutes. Additionally, the value of magnetic field used is conservative, with
spectropolarimetric inversions suggesting field strengths up to $\sim 50$~G may
be present \citep{TRUetal2005, LOPCAS2005, CENetal2010}, hence,
potentially providing larger values of the Alfv{\'{e}}n
speed than those given. Combining the fast propagation speeds with the
 long wavelengths of kink waves, e.g., $20{\,}000$~km for a wave with a
period of $100$~s, this makes it extremely difficult to observe and measure the
propagation of the waves along spicules unless high cadence data is used
and rigorous measurement techniques are employed.

\vspace{3mm}
The presence of transverse motions of spicules was also reported in
\citet{SUEetal2008b}, giving similar values for amplitudes but, interestingly,
noting that the lateral motions and oscillations become more prominent as height
increases. This would suggest that the amplitude of the waves
increases with height in the atmosphere, and would imply a decrease in the
average density with height as one might expect (see
\S{\,}\ref{sec:wave_seis}). Furthermore, \citet{SUEetal2008b}
highlight that some spicules show evidence for rotational motions.
A number of authors also investigated spicules with Hinode/SOT, analysing a
few individual spicule oscillations in more detail. \citet{KIMetal2008}
studied three spicule oscillations in what appears to be a coronal hole
region. Using time--distance diagrams, they saw wave motion at numerous heights
along the spicules and reported that there was no evidence of phase shifts
between the differing heights, hence the authors gave an estimate for the
phase speeds of the perturbations as $260-460$~{\kms}. However, the data
used has a cadence of 16~s and it is unclear which techniques were used to
measure spicule displacement and phase shifts, adding to the uncertainty
in the given values.

A thorough analysis of four spicule oscillations in a coronal hole was
undertaken by \citet{HEetal2009}. Notably, they found evidence for waves with
periods $< 50$~s, significantly less than that suggested by the Monte Carlo
comparisons of \citet{DeP07}. Additionally, \citet{HEetal2009} provided
the first measurements of propagation speeds of the kink waves
using cross-correlation of signals from time-distance diagrams that were generated at various positions along the
spicules. The measured propagation speeds ranged between $59-150$~{\kms}
and all waves were upwardly propagating. They also show evidence for an
increase in wave amplitude with height along two of the spicules analysed,
supporting the reports of \citet{SUEetal2008b}.
Another investigation into fast kink waves in spicules was carried out
by \citet{HEetal2009b}, although this time only one event was studied in
detail. The spicule was located above plage in an active region. In this
event, the amplitude and phase speed were able to be measured at
12 separate positions along the spicule (see Figure~\ref{fig:wave_he}).
The measurements clearly showed initial increases in velocity
amplitude and phase speed, followed by decreases in both quantities.
Perhaps surprisingly, the measured phase speed of the wave is as
little as $25$~{\kms} in the upper sections of the spicule, suggesting a
weakening of magnetic field with height. The observed variations of the
velocity amplitude and phase speed are not simultaneous, suggesting
a complex variation in plasma parameters (see \S{\,}\ref{sec:wave_seis}
for further details).

\begin{figure}[!t]
%\center
\includegraphics[width=1\textwidth, clip=true]{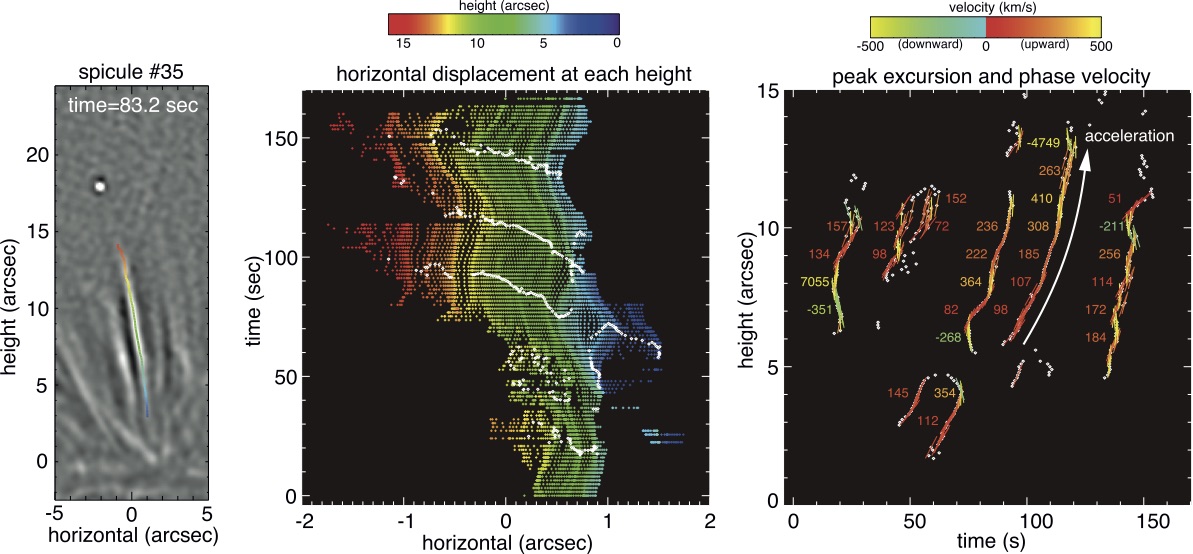}
\caption{Examples of wave propagation along an off-limb spicule.
The left panel shows the spicule that has been highlighted with coloured line,
where the colour variation corresponds to increasing atmospheric height. The middle
panel shows the results from following the spicule over time. The colours
correspond to those in the first panel and show the transverse displacement at each
height along spicule (horizontal axis) with time (vertical axis). The displacement
is observed to travel along the spicule, which is highlighted by the white lines.
The right panel shows the propagation speeds calculated from the gradient
of the white lines. Image reproduced from \citet{OKADEP2011}.
\label{fig:wave_prop}}
\end{figure}

\vspace{3mm}
A significant investigation into the wave properties of spicules was
performed by \citet{OKADEP2011}. Again, the focus was on the properties of
coronal hole spicules, although a unique automated technique to track the spicules
was developed, extracting spicules from images, locating the central axis of
the spicules along their length and following them over
time. The technique is subject to various conditions, first
removing short-lived and small-scale structures. A total of 89 suitable spicules
were identified and used for further study, with the authors suggesting that they
are likely isolating Type-II spicules. In Figure~\ref{fig:wave_prop} we show
an example of their results. The middle panel in the figure contains a
significant amount of information and we will try to provide a brief summary here,
however, it is strongly suggested an interested reader refers to
\citet{OKADEP2011} for a full explanation. The middle panel shows the
horizontal displacements of a spicule as a function of time, for each
position along the spicule length, where each position is given a
different colour. The maximum displacements of the spicule are
highlighted by the white lines. It can be seen that the position of the
maximum displacements moves upwards or downwards along the
spicule as time increases, suggesting the maximum displacement is
propagating along the spicule. Using the gradient of these white lines,
propagation velocities of the waves can be obtained and are shown
in the right hand panel. The average properties of the waves observed
gave typical periods of $45\pm30$~s and velocity amplitudes of $7.4\pm3.7$~{\kms}.
Interestingly, this is almost half the value suggested by the measurement of
predominantly uni-directional motions in \citet{DeP07}. This has
implications for the estimated energy flux -- if the amplitudes are half those of
previously reported it means the spicules may only carry a quarter of the
energy flux previously estimated by \citet{DeP07}. The authors find that
a majority of the waves are upwardly propagating, with approximately one third
downwardly propagating. The results also suggest that, on average, the
phase speeds increase with height. The authors additionally refer to the
presence of standing waves being present, however, the observational
evidence is unconvincing. The authors suggest that when 
upwardly and downwardly propagating waves pass each other a standing
wave is present (these features are seen in Figure~\ref{fig:wave_prop} as
anomalously high values of phase speed). A superposition of counter
propagating waves is not, however, a standing wave. By definition a standing wave
has fixed nodes which do not oscillate. What appears to be observed is
just the temporary superposition of counter propagating waves, which
would cause the apparent anomalous high phase speeds. Further objections
to the interpretation of standing waves are given in \citet{LIPetal2014}.

\vspace{3mm}
Another significant and thorough analysis of spicule properties was carried
out by \citet{PERetal2012}, and the results also included statistics on spicule
displacements from coronal holes, quiet Sun locations and active regions. The
study provides statistically significant measurements for both transverse
displacements and velocity amplitudes. The authors measure both uni-directional
and sinusoidal motions, further splitting the results between spicules that show
either parabolic or linear trajectories, which essentially tries to
distinguish between Type-I and Type-II varieties. They suggest that
`linear spicules' are dominant in both coronal holes and
quiescent regions, with parabolic profiles rarely occurring in these regions.
A few other papers have also reported the transverse motions in
spicules, and we briefly summarise them here.
\citet{TAVetal2011} and \citet{EBAetal2012}
both demonstrate examples of spicules that undergo transverse
displacements but do not provide any solid analysis of the events.
\citet{Jess2012} provide a unique study of spicules with an on-disk
observation in H$\alpha$ and provide evidence regarding the potential excitation
mechanism (see \S{\,}\ref{sec:wave_gen} for further discussion).
\citet{YURetal2012} also note the presence of periodic and linear
transverse motions in H$\alpha$ observations of on-disk Type-II spicules. However,
these observations appear to be of RBEs rather than Type-II spicules
(although there is the suggestion that these two phenomenon are one in the same
-- \citealp{ROUetal2009}).

\vspace{3mm}
Aside from the transverse displacement of the spicules, evidence for
torsional motions in spicules has been provide by \citet{DEPetal2012}. Using both
H$\alpha$ and {\cah}, the authors are able to resolve oppositely directed Doppler
shifts on either side of the spicule suggesting motion in opposite
directions. Using Monte-Carlo simulations, the authors suggested that
amplitudes of $\sim 30$~{\kms} and periodicities of $100-300$~s represent the
observed Doppler signatures well. However, it has been demonstrated by
\citet{GOOetal2014} that it also possible to interpret the observed Doppler
velocity in terms of a kink motion (see, \S{\,}{\ref{theory_section}} for more details).
\citet{RUT2013} also reports evidence for torsional motions in spicules
using Dopplergrams, showing examples of
spicules with red and blue shifts apparently on opposite sides of the feature.
However, the author compares red and blue wing H$\alpha$ ($\pm 600$~m{\AA})
images that are taken 1~minute apart so it is unclear whether this is
torsional behaviour or just transverse displacement of the spicule along the
observer's line-of-sight. As the torsional motions are likely to propagate at the Alfv{\'{e}}n
speed, they will traverse the spicule relatively quickly. Hence, the blue-red
asymmetry across the spicule will undergo a relatively rapid
evolution, fading and then reappearing with the asymmetry on opposite
sides due to the periodicity of the waves.

\subsubsection{Fibrils}
\label{sec:wave_fib}
Chromospheric fibrils are elongated structures that span
supergranular cells, lying almost horizontally in the
chromosphere \citep{FOU1971, ZIR1972}. The fibrils are typically
associated with strong concentrations of photospheric flux, i.e.,
network boundaries or plage regions. They spread out from these regions,
showing a greater degree of topological organisation in active regions
compared to quiet regions. The other footpoint of fibrils is assumed to
lie within opposite polarity flux, but this is not always evident (\citealp{REAetal2011}).
Fibrils appear as dark features in the line cores of chromospheric absorption lines as
a result of them being a local density enhancement that leads to increased
scattering of the photospheric radiation (e.g., \citealp{Lee12}).
The first resolved observation of kink waves in fibrils appears to be
that of \citet{PIEetal2011}, who measured a single oscillating feature in Ca~{\sc{ii}}
8542~{\AA} -- although the authors mention that there is evidence for further
transverse displacements in other features. It was found that the velocity
amplitude of the wave was on the order of $1$~{\kms}, significantly less than
that typically associated with the fast kink waves measured in spicules.
Additionally, an attempt was made to measure the speed of
propagation from phase analysis, where small phase shifts were measured
giving a value of $190$~{\kms}. However, the authors suggest that the
measurement is subject to large uncertainties and are not convinced
by the estimated value, stating that the actual speed of propagation
may be either too fast (leading to small phase shifts) to be measured
robustly or the wave is simply not propagating.

\begin{figure}[!t]
\begin{center}
\includegraphics[width=1\textwidth, clip=]{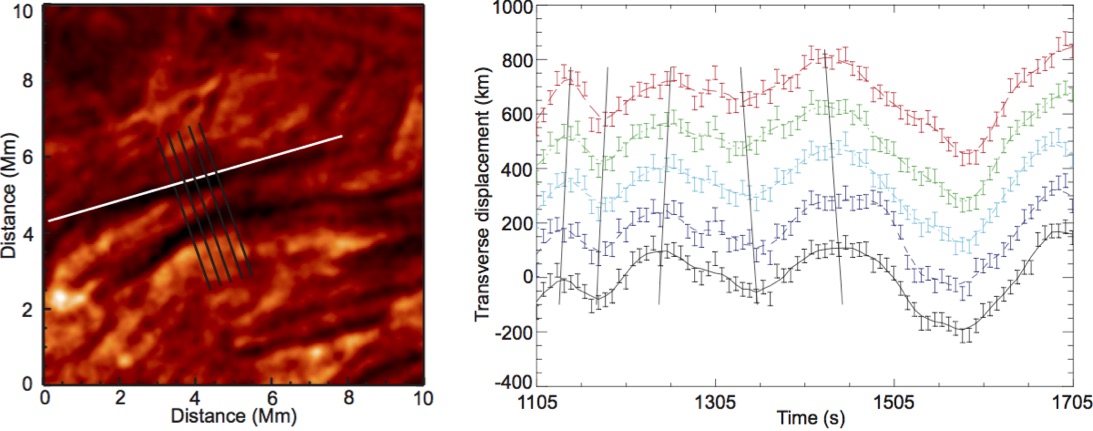}
\caption{H$\alpha$ observations of chromospheric fibrils. The left panel displays a
cropped field-of-view containing elongated dark fibril structures, with one axis highlighted by the
solid white line (same fibrilar structures as displayed in Figure~{\ref{Morton2012}}).
Perpendicular black lines indicate locations where cross-cuts were made, with the
right panel displaying the displacement of the fibrils central axis from
each of the cross-cuts. The straight lines connect the peaks and troughs of the
sinusoidal displacement and highlight the propagation of waves along the fibril.
Image reproduced from \citet{Mor12}.
\label{fig:wave_prop_fib}}
\end{center}
\end{figure}

\vspace{3mm}
A more general look at transverse waves in quiet Sun fibrils in
H$\alpha$ is given in \citet{Mor12}, who use data from the
ROSA instrument. The authors observe
evidence for ubiquitous transverse wave motions of the fibrils,
along with evidence for fast propagating compressional modes
(see, \S{\,}{\ref{sausages}} for further details). They measure the
uni-directional displacements of over 50 dark fibrils using
time--distance diagrams, but also measure and report a number
of sinusoidal displacements. The observed displacements have a
mean values of $315\pm130$~km and velocity amplitudes of
$6.4\pm2.8$~{\kms}, suggesting the waves had significantly smaller
amplitudes than those observed at the limb in a similar manner
(i.e., \citealp{DeP07}). Additionally, measurements of
phase speeds of some of the observed transverse displacements
reveal evidence for counter propagating waves travelling with
speeds in the region of $50-250$~{\kms} (Figure~\ref{fig:wave_prop_fib}).
Subsequent studies of the fibrils in H$\alpha$ ROSA data were
given in \citet{MORetal2013, MORetal2013b}. Here, an advanced feature
tracking routine was exploited to
examine the periodic motions of the fibrils in time--distance diagrams
and $\sim 740$ and $\sim 840$ individual measurements were made
in active and quiet Sun regions, respectively. This extended analysis
suggested that typical displacement amplitudes and velocity amplitudes
were smaller than those in \citet{Mor12} (see Table~\ref{tab:A1})
and periods were $120\pm50$~s. As noted in \citet{MORetal2013b},
these results are subject to a series of observational constraints,
with higher frequency waves ($P<50$~s, where $P$ is period)
likely to be underrepresented since they will have displacements on
the order of the spatial resolution, while lower frequency waves ($P>250$~s)
will also be underrepresented as the lifetimes of fibrils are of the same
order. The large number of events measured additionally enabled
the authors to derive the first estimates for the velocity power spectra
of the chromospheric transverse displacements -- this exciting result
will be discussed further in \S{\,}\ref{sec:wave_gen}.

\begin{table}[!t]
\begin{center}
\tiny
\caption{Average (or measured) properties of fast kink waves.
\label{tab:A1}}
\begin{tabular}{llcccccl}
\hline
& & & & & & & \\
%\ra{1.3}
Structure & Region & {$\xi$ (km)} & {$P$ (s)} & {$v$ (km/s)} & {$c_{ph}$ (km/s)} & No. Events & {Reference} \\ [0.3ex]
\hline \\ [0.3ex]
Spicule & CH &  $200-500$ & $150-350$ & $20\pm5$ & -   & 95 &   \citet{DeP07} \\
 &CH &  -  & $60-240$ & $20\pm5$ & -    &  -&   \citet{SUEetal2008b} \\
 & CH &  $1000$ & $130$ & $15$ & 460    &  1 &   \citet{KIMetal2008} \\
&&  $700$ & $180$ & $8$  & 310   &   1 &     \\
&&  $800$ & $170$ & $9$  & 260   &   1 &     \\
 & CH &  $36$ & $48$ & $4.7$ & 75-150  &  1 &     \citet{HEetal2009} \\
 &&  $36$ & $37$ & $6.1$ & 59-117  &   1 &    \\
 &&  $130$ & $45$ & $18.1$ & 73   &   1 &     \\
 &&  $166$ & $50$ & $20.8$ & 109-145  &  1 &    \\
& CH &  $55\pm50$ & $45\pm30$ & $7.4\pm3.7$ & 160-305 &  89 &     \citet{OKADEP2011} \\
& &  $600$ & $180$ & $22$ & - &  1 &     \citet{EBAetal2012} \\
 & QS &  $670$ & $220$ & $19.2$ & - &  1 &    \citet{Jess2012}\\
& &  $630$ & $139$ & $28.3$ & - &  1 &     \\
& &  $160$ & $65$ & $14.8$ & - &  1 &      \\
& &  $410$ & $158$ & $16.2$ & - &  1 &      \\
& &  $380$ & $129$ & $18.5$ & - &  1 &      \\
&&  $200$ & $105$ & $11.8$ & - &  1 &      \\
& &  $190$ & $171$ & $7.2$ & - &  1 &      \\
&AR&  $283\pm218$ & $ - $ & $14\pm112$ & - &  112 &    Type-I -\citet{PERetal2012}\\
&AR&  $463\pm402$ & $ - $ & $18\pm12$ & - &  58 &    Type-II\\
&QS&  $245\pm211$ & $-$ & $16\pm11$ & - &  174 &    \\
&CH&  $342\pm257$ & $-$ & $20\pm12$ & - &  170 &    \\
\hline \\ [0.3ex]
Fibrils  &&  $135$ & $135$ & $1$ & 190  & 1 & \citet{PIEetal2011}   \\
 & QS &  $315\pm130$ & $-$ & $6.4\pm2.8$ & 50-90   & 103  & \citet{Mor12}   \\
 &QS &  $71\pm37$ & $94\pm61$ & $4.5\pm1.8$ & -   &  & \citet{MORetal2013}   \\
& QS &  $94\pm47$ & $116\pm59$ & $5.5\pm2.4$ & -  & 841  & \citet{MORetal2013b}   \\
& AR &  $73\pm36$ & $130\pm92$ & $4.4\pm2.4$ & - & 744   &   \\
\hline \\ [0.3ex]
RBEs & &  $300$ & $-$ & $8$ & -   &  35&   \citet{ROUetal2009} \\
 & CH &  $200$ & $-$ & $4-5$ & -   &  960&   \citet{SEKetal2012} \\
 & QS &  $200$ & $-$ & $8.5$ & -   &  1951 &  average -  \citet{SEKetal2013} \\
 & &  $220$ & $-$ & $11.7$ & -   &  1951 &  maximum  \\
\hline \\ [0.3ex]
Mottles & QS & $200\pm67$  & $165\pm51$ & $8.0\pm3.6$ & -   &  42&   \citet{KURetal2012} \\
 & QS &  $\sim172$ & $120\pm10$ & $\sim9$ & 50   &  1&   \citet{Kur13} \\
 & QS &  $252$ & $180\pm10$ & $8.8\pm31$ & $101\pm14$  &  1&    \\
 & QS &  $327$ & $180\pm10$ & $11.4\pm3.3$ & $79\pm8$   &  1&    \\
\hline
\end{tabular}
\end{center}
\end{table}
\normalsize

\subsubsection{Other features}
\label{sec:wave_other}
In this section we briefly discuss measurements of transverse
displacements in other chromospheric structures. This is not to
belittle the importance of these features, it is simply because it
is unclear how these structures fit into the chromospheric scene.

\vspace{3mm}
The first of these are RBEs, which are apparently fast-moving
plasma flows observed in the blue wings of
chromospheric spectral lines. \citet{ROUetal2009} provides a thorough
study of the phenomenon, and are able to measure 35 examples of the transverse
displacement of the RBEs. The average measured amplitudes are
$0.3$~Mm for displacement and $8$~{\kms} for velocity. Subsequent
studies by \citet{SEKetal2012, SEKetal2013} provide similar
numbers following a larger statistical survey
(see Table~\ref{tab:A1}), and the authors demonstrate that the
distributions are similar for measurements in both
H$\alpha$ and Ca~{\sc{ii}}~8542{\,}{\AA}. They also show that
the average transverse velocity ($4-8$~\kms) is approximately one-third
to a half of that associated with the maximum transverse velocities
($8-11$~\kms). \citet{SEKetal2013} takes an additional
step and classifies the RBEs in relation to the type of transverse
displacement they observe, i.e. uni-directional and periodic.
Subsequently, they found that the maximum velocity amplitudes
for the periodic motions ($7.5$~{\kms}) are less than the
uni-directional motions ($11.8$~{\kms}), similar to measurements in
both spicules and fibrils.
\citet{KURetal2012, Kur13} analysed the transverse displacements
of chromospheric mottles thought to be connected to spicules. A number of
periodic events are analysed and provide displacements, velocity amplitudes
and periods that are in line with those seen in spicules. \citet{KURetal2012}
also measure the time variation of the amplitude, which in a number of
cases appear to decay with time. Although the authors talk about
damping times in the paper, it is likely that the events are propagating wave
packets of finite length rather than damped wave motion.

\vspace{3mm}
Preceding the spicule observations of apparent torsional motions,
\citet{Jes09} also demonstrate evidence for torsional Alfv{\'{e}}n waves in the
chromospheric counterpart to an MBP. Utilising the
H$\alpha$ line, the authors measured periodic variations in the non-thermal line
widths. The indicator that the observed variation was torsional was a
$180^{\circ}$ phase delay between signals on the opposite sides of the
chromospheric MBP, with the resulting chromospheric absorption profile
shifts displayed in \citet{Mat13}.

\begin{figure}[!t]
\begin{center}
\includegraphics[width=0.49\textwidth, clip=]{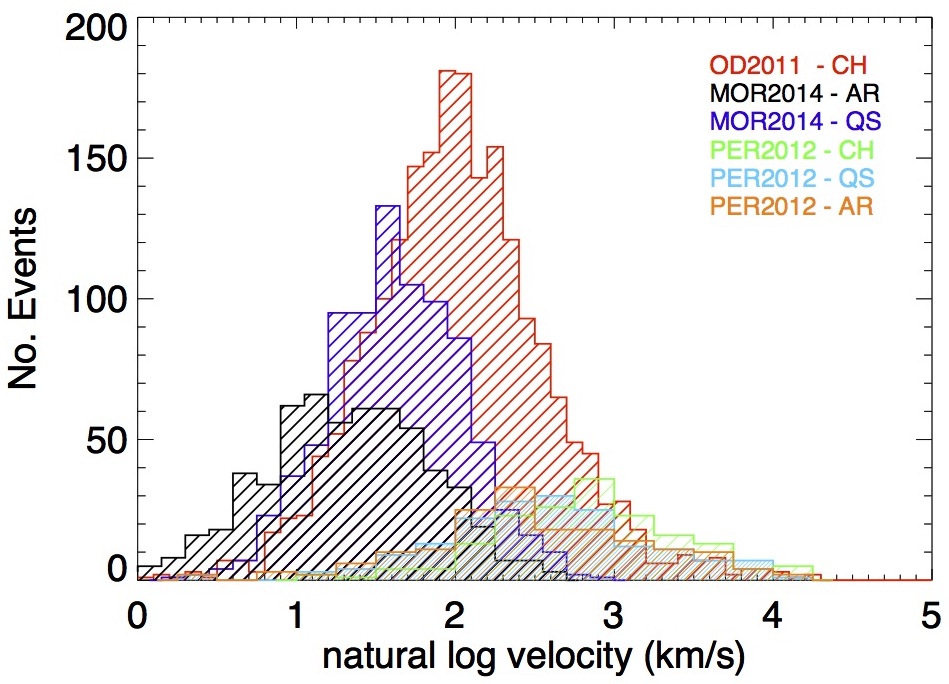}
\includegraphics[width=0.495\textwidth, clip=]{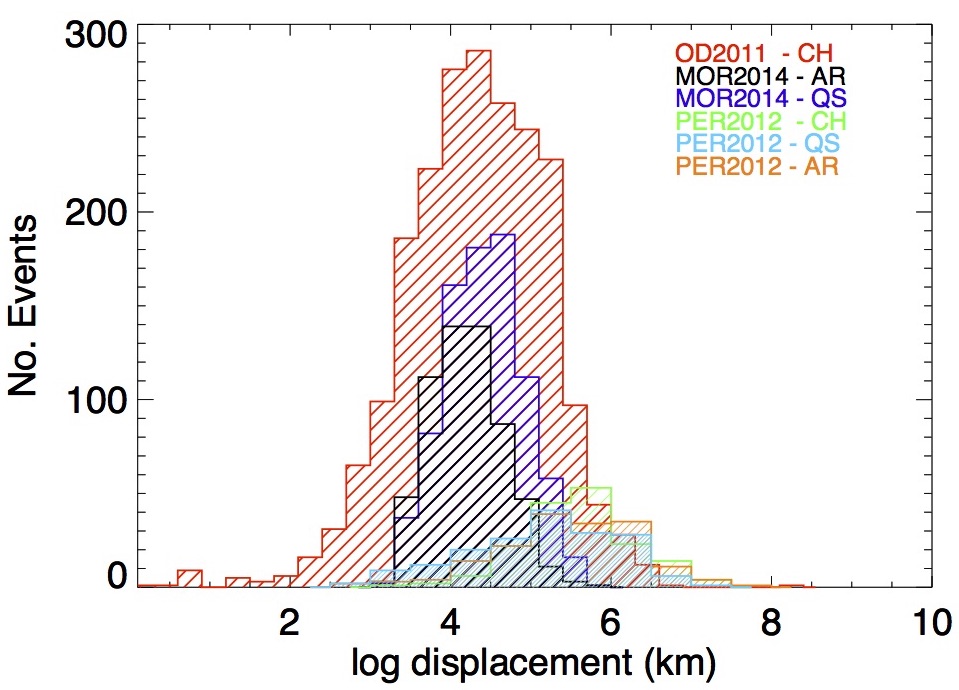}
\end{center}
\caption{Histograms showing the distributions of velocity (left) and displacement (right) amplitudes for
different chromospheric features. The labels correspond to: OD2011 -- \citet{OKADEP2011};
MOR2014 -- \citet{MORetal2013b}; PER2012 -- \citet{PERetal2012}.
\label{fig:wave_hist}}
\end{figure}

\subsubsection{Summary}
It is clear from this growing body of evidence that the fast kink
waves are ubiquitous throughout the chromosphere and present in almost
all chromospheric features. However, the amplitudes of the waves can be
very different for the various features (e.g., Table~\ref{tab:A1},
Figure~\ref{fig:wave_hist}). We highlight this further with two histograms
that show the distributions of wave amplitudes in the different structures.
Although spicules have larger velocity amplitudes, it is unclear
whether they carry a greater energy flux than other chromospheric
features (energy flux is given roughly by $F\sim\rho c_{ph} v^2$,
where $c_{ph}$ is the phase speed of the wave and $v$ is the
velocity amplitude). For example, the density of fibrils appears to be
$100-1000$ times that of spicules
(\citealp{BEC1972}; \citealp{Lee12}), which gives
approximately equal values for the energy flux in both features.
(Note, this number isn't the total energy flux associated with
spicules/fibrils since it doesn't account for differing filling factors of
the structures, and therefore is simply the approximate energy
flux per wave packet.)
The fate of the observed wave energy is likely to be different.
Spicules have a connection to the corona and the
observations of \citet{OKADEP2011} suggest that some of the waves
leave the chromosphere to deposit their energy elsewhere.
Fibrils, on the other hand, appear to be closed to the upper atmosphere,
which means the energy likely stays contained in the chromosphere.
It is then likely that the waves within these two structures play
different roles in energy transport through the solar atmosphere.

\subsection{Magneto-seismology}
\label{sec:wave_seis}
Solar magnetoseismology (SMS) has its origins in exploiting MHD 
oscillations in the corona to determine the physical conditions in the 
local plasma (\citealp{UCH1970}; \citealp{ROBetal1984}). To date,
there have been numerous successful applications in the corona (i.e., coronal seismology) with 
significant focus on fast MHD kink waves (e.g., \citealp{Nak05}; \citealp{RUDERD2009}; \citealp{ANDetal2009}).
However, the associated coronal scale-heights and time-scales for
evolution are typically (much) larger than chromospheric values.
Fortunately, most of the assumptions used to derive the SMS
techniques for coronal applications are still largely applicable to
oscillations in chromospheric structures. However, in the chromosphere
one needs to carefully consider the influence of flows on the SMS
techniques because the effects of such phenomena become important when the flow speed, $U$,
is on the order of the kink speed, $c_k$. This can be seen in the governing
wave equation when flow is included, e.g., \citet{MORERD2009},
\citet{RUD2011}, \citet{SOLetal2011} and \citet{TERetal2011}, where terms
on the order of $(U/c_k)^2$ are present. The full development of SMS
techniques that include the influence of flows should be
the next step for those who are theoretically minded and would
improve the applicability of SMS to a wider selection of situations.

\vspace{3mm}
To date, very few applications of SMS have been made to chromospheric
features. The first attempt was made by \citet{KIMetal2008}, who were able to
measure parameters for a few oscillatory events described in
\S{\,}\ref{sec:wave_spic}. The authors use the following relationship,
\begin{equation}
B_0=\sqrt{\frac{\mu_0}{2}}\frac{\lambda}{P}\sqrt{\rho_i+\rho_e} \ ,
\end{equation}
where $B_0$ is the magnetic field, $\mu_0$ is the magnetic permeability,
$P$ is the period, $\rho$ is the plasma density, $\lambda$ is the wavelength, and the subscripts $i$ and $e$
refer to internal and external values, respectively\footnote{Note, that
the value $B_0$ is the root of the sum of the squares of the external and internal magnetic fields,
i.e., $\langle B \rangle=\sqrt{B_i^2+B_e^2}$.}. The authors use previously measured
spicules densities and
values for the phase speed to derive the wavelength, thus estimating values
of $10-80$~G for the magnetic field strength.
The large range of values is partially due to the fact that density estimates for
spicules vary by an order-of-magnitude. Secondly, the fact that the authors
are not able to measure the phase speed directly also adds to the uncertainty.

\begin{figure}[!t]
%\center
\includegraphics[width=1\textwidth, clip=]{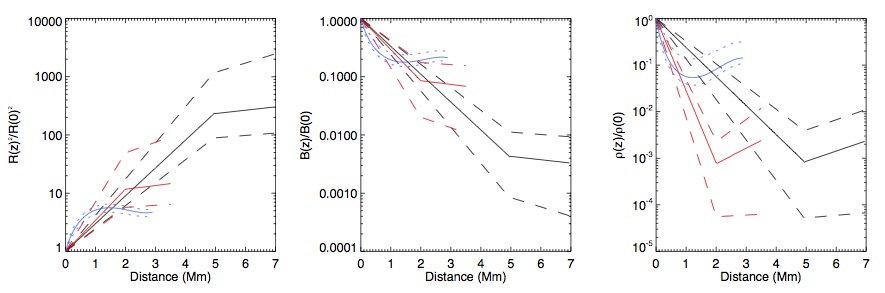}
\caption{Results from magneto-seismological inversions assuming no wave damping.
Displayed are the normalised variations in area (left), magnetic field (middle) and plasma
density (right). The results are from: spicules (black -- \citealp{VERTetal2011},
blue -- \citealp{MOR2014}); and mottles (red -- \citealp{Kur13}). The dashed lines
are the $95${\%} confidence bounds for the seismologically determined results.
Note that each observation does not start at the same atmospheric height,
so the horizontal axis corresponds to the distance from the first
measurement along the spicule. Note that the results here come from either limb
or disk observations, which could mean line-of-sight effects may play a
role in differences between them.
\label{fig:wave_seis}}
\end{figure}

\vspace{3mm}
A more advanced SMS application was given in \citet{VERTetal2011},
where the authors exploited the measurements of a propagating fast kink wave
along an active region spicule from \citet{HEetal2009b}, specifically the
amplitude and phase speed values (see, \S{\,}\ref{sec:wave_spic} and
Figure~\ref{fig:wave_he}). Combining these measured quantities with the
theory that describes fast kink waves in a magnetic flux
tube with longitudinal variations in the magnetic field and plasma density
(e.g., \citealp{VERERD2008}, \citealp{RUDetal2008}), the gradients in plasma
density and magnetic field strength can be estimated. The normalised
variation in each of these quantities is shown in Figure~\ref{fig:wave_seis}.
Note, that the
absolute values of quantities cannot be measured from the observations.
Additional information would be required to do this, i.e., a value for
the magnetic field or density at a particular height.

\vspace{3mm}
The general trend measured for the plasma density gradient is that likely to be expected,
i.e., the density is found to decrease with height. This shows agreement
(within errors) with the gradient in plasma density estimated through other
techniques, e.g., \citet{MAK2003} from
eclipse spectra. There is the suggestion that the density begins to increase
towards the top of the spicule, although a constant or decreasing density
profile is within the error bars. It is also expected the magnetic field
weakens with height as it expands to fill the coronal volume as a result of
magnetic structures increasing in size. This is precisely what is found from the SMS inversions.
The inferred expansion suggests a significant increase in the spicule
radius, a factor $10$, which leads to a factor
of $100$ decrease in magnetic field strength. This may seem large,
but let us assume that the spicule is anchored in a MBP
with an initial field strength $\sim 1000$~G. The value at
the spicule head is then $10$~G, in line with approximate coronal
values of magnetic field strength (e.g., \citealp{VERetal2013}). The
rate of decrease in the magnetic field strength is then $\sim 0.25$~G km$^{-1}$,
which is comparable to the rate of decrease observed in
sunspots and active regions between the photosphere and chromosphere
(e.g., \citealp{LEKMET2003}, and the discussion in \S{\,}\ref{compressive_ar}).
Additionally, the seismologically derived
expansion is less than the estimated upper bound for the expansion of
flux tubes from the photosphere to the corona (\citealp{TSUetal2008}).

\vspace{3mm}
Similar analysis is performed by \citet{Kur13} for a mottle observed
on-disk. The results (Figure~\ref{fig:wave_seis}) suggest a similar variation in
quantities to the inversion of \citet{VERTetal2011}. The increase in radius up to
$2$~Mm is a factor of $\sim 3$ for both observations, and consequently
the variation in magnetic field is also similar. Interestingly, the density gradient
is steeper in the mottle and only decreases by a factor $\sim 10^{-3}$.
This coherent behaviour is not unexpected as the density along the spicules should
drop from chromospheric values ($\sim 10^{-9}$~\kgm) to coronal values
($\sim 10^{-12}$~\kgm).
\citet{MOR2014} recently undertook a study of a fast kink wave along a
spicule that occurred in the penumbra of a sunspot. The inversions revealed
similar variations in spicule expansion and magnetic field gradient to the
previous two studies, although the gradients are steeper. Significantly,
\citet{MOR2014} was also able to directly measure the expansion along the
spicule by fitting a Gaussian to the spicules cross-sectional flux profile. He
found good agreement between the seismologically determined values and the
directly measured value after taking into account optical effects from the telescope.
The density decrease along the spicule was almost an order-of-magnitude less
than that found in \citet{VERTetal2011} and \citet{Kur13}.

\begin{figure}[!t]
\begin{center}
\includegraphics[width=0.8\textwidth, clip=]{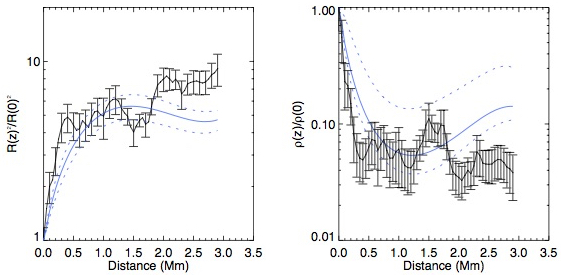}
\includegraphics[width=0.4\textwidth, clip=]{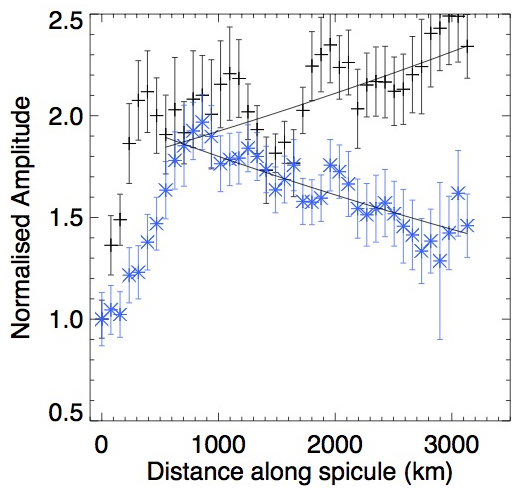}
\end{center}
\caption{Results from magneto-seismological inversions combining spicule width
variations. Displayed are the normalised variations
in area (upper-left) and plasma density (upper-right). The blue lines are the inversions
shown in Figure~\ref{fig:wave_seis}. The black line in the upper-left panel shows
the expansion inferred from the measurements of the Gaussian width
of the spicule with $\sigma$ error bars given. The black line in the upper-right
panel is the density gradient determined using a combination of the measured
width and measured phase speed. The lower panel displays the measured
amplitude of an fast kink wave along a spicule (blue stars) and the amplitude
determined from the measured expansion and the phase speed measurements
(black crosses). The difference between the two amplitudes suggest the wave
is damped as it propagates along the spicule. The black solid lines correspond
to exponential fits to each amplitude profile.
Image reproduced from \citet{MOR2014}.
\label{fig:wave_seis_pt2}}
\end{figure}

\vspace{3mm}
One feature that all these observations have in common is that the rate of change of magnetic
field strength reduces drastically at a certain height, and perhaps surprisingly,
the density appears to begin to increase again with height. The
second of these quirks can easily be explained. Firstly, the amplitude of the
wave is dependent upon the density (i.e., $\xi\propto\langle\rho\rangle^{-1/4}$- \citealp{MOR2014}) and is
independent of the magnetic field. Note that the relationship of the kink wave 
amplitude to density is similar to that found for the Alfv\'{e}n wave.
It is expected that the fast kink{fast kink} waves will undergo some form of
damped motion, with resonant absorption an excellent candidate for such
damping (e.g., \citealp{GOOetal2011}). The relations used for the SMS
inversions do not take into account the possibility of damping. Hence, if any of the observed waves 
are being damped, the standard SMS technique will underestimate the gradient in density,
and consequently gradients in expansion and magnetic field strength
(as pointed out in \citealp{VERTetal2011}). If the evolution of the amplitude is dominated by the damping rather
than the variation of plasma parameters, then it is likely that non-physical results will
be obtained from the inversions. As \citet{MOR2014} was able to measure
the expansion along the spicule via another technique (for a comparison
see Figure~\ref{fig:wave_seis_pt2}), this could be
combined with the measured phase speed to derive the actual variation
in density along the structure. As should be expected, the density decreases
continuously along the spicule (Figure~\ref{fig:wave_seis_pt2}). Using this density
profile, the expected amplitude variation along the spicule for an undamped
wave was calculated, with comparison to the measured amplitude suggesting
the spicule's wave motion was indeed damped (Figure~\ref{fig:wave_seis_pt2}
and see \S{\,}\ref{sec:wave_gen} for further details). The damping is
found to have a quality factor, $\xi_E$, equal to $\tau_D/P=L_D/\lambda\sim0.34$,
where $\tau_D$ is the damping time and $L_D$ is the damping length.

\vspace{3mm}
The initial, rapid change in magnetic field strength inferred from all observations
may also have a relatively straightforward explanation. Spicules
are jets of chromospheric plasma that follow vertical field lines, reach well
into the corona, and as a result may be departures from the traditional gravitationally
stratified atmosphere. However, if it is assumed that the spicules are
superimposed on the traditional atmospheric profile, then the
external values for both density and magnetic field strength will rapidly
decrease at the height at which the spicule crosses
the transition region. This may also be related to the magnetic flux merging
height, which depends on the flux distribution in the photosphere. If the
magnetic flux distribution is of the small scale `salt and pepper' format then the
merging height will be quite low compared to a more simple large scale dipole
source. The important point is that the merging height and the height of the
transition region could vary over all atmospheric locations. This is
suggested by the SMS results in Figure~\ref{fig:wave_seis}. Note, that while
the external values of density and magnetic field may change drastically, the
internal values along the spicule may not. The measured variations are the
average values of these quantities, which will reflect the average
highly localised behaviour.

\vspace{3mm}
Finally, we mention that the fast kink waves are not the only useful
tool for probing the chromosphere, but torsional Alfv{\'{e}}n waves also have
the potential to reveal information about the local plasma conditions.
Inspired by the observations of \citet{Jes09}, \citet{Ver10, VERTetal2011}
and \citet{FEDetal2011} have demonstrated that the torsional motions
can be used to map the magnetic field in the chromosphere. This is
another exciting avenue for SMS and hopefully will be
built upon with future observations.

\subsection{Wave generation and damping}\label{sec:wave_gen}
From the preceding sections it is apparent that the chromosphere is
subject to ubiquitous incompressible motions, with
the body of evidence for this ever increasing. This leaves us with
two very pertinent questions: how are these waves generated and
what is the fate of the energy that they transport?

\vspace{3mm}
The first of these questions is perhaps somewhat easier to provide
answers to. It has been postulated that the horizontal component of the
convective motions is able to excite incompressible motions
(e.g., \citealp{HOL1972}; \citealp{SPR1981}; \citealp{CHOetal1993})
and this forms the basis for many simulations related to the heating of
the solar atmosphere and solar wind acceleration via MHD waves
(e.g., \citealp{CRAVAN2005}; \citealp{SUZINU2005};
\citealp{FEDetal2011}) and spicule formation (e.g, \citealp{MATSHI2010}).
Complementary mechanisms of wave
generation may also be present. For example, $p$-modes
(or more generally slow magnetoacoustic waves) can also excite fast
MHD waves via mode conversion (\citealp{CARBOG2006}), although this
only occurs along inclined magnetic fields. Additionally, it is well known
that magnetic reconnection can also release some of its energy in the
form of MHD waves (e.g., \citealp{YOKSHI1996}), with periodic reconnection
mechanisms also viable (e.g., \citealp{MCLetal2009}).
Observational evidence that demonstrates the excitation mechanisms
of incompressible waves is very limited at present, and is typically restricted to
isolated examples. Both \citet{HEetal2009b} and
\citet{YURetal2012} show examples of oscillating spicules with an
inverted Y-shaped structure, which they suggest shows evidence
of a reconnection event -- following \citet{SHIetal2007}.
Consequently, they put forward the idea that some
of the energy released from the reconnection is used to
generate the kink wave.

\vspace{3mm}
Evidence for mode conversion generating transverse waves in
spicules has been offered by \citet{Jess2012}. The authors use the
multiwavelength capabilities of the ROSA instrument to identify and
examine the photospheric foot-points of the spicules. The
spicules are observed to be rooted in MBPs found in G-band images,
which correspond to strong regions of magnetic field
\citep{Ber01, Jes10b}. The MBPs are found to
display significant intensity oscillations that are upwardly propagating,
and can be interpreted as slow magnetoacoustic waves. Importantly, the
intensity oscillations are $90^{\circ}$ out-of-phase across the bright point,
suggesting a double `piston-like' action. This out-of-phase behaviour
leads to velocity gradients across the spicule and excites the
fast kink wave.
Hints at excitation mechanisms for incompressible waves in
fibrils have also been reported in \citet{MORetal2013, MORetal2013b}.
In \citet{MORetal2013}, the multiwavelength capabilities of the
ROSA instrument are again exploited in an attempt to connect
the dynamics of the photosphere to the chromosphere.
The authors identify an MBP that is associated
with the footpoints of fibrils and the bright point appears to exist
within a photospheric vortex. It has been demonstrated that these
photospheric vortices can generate significant Poynting flux
\citep{SHEetal2011b,SHEetal2012,MOLetal2012,WEDetal2012}
and excite MHD waves \citep{FEDetal2011, VIGetal2012,
SHEetal2013}. The authors
observe quasi-periodic twisting motions of the
chromospheric counterpart of the MBP that can be
identified as torsional Alfv{\'{e}}n waves. Additionally, there
appears to be a coupling between the torsional motions of the
large-scale magnetic structure and the transverse motions of the
fibrils, although the underlying physics is unclear.
While not apparently periodic, magnetic features that show
evidence for uni-directional `swirling' motion have also been
identified in Ca~{\sc{ii}}~8542{\,}{\AA} observations
\citep{WEDROU2009}. \citet{WEDetal2012} related these
motions to photospheric vorticities which had been
observed previously by \citet{BONetal2008}. It was also revealed
that the emission in the upper solar atmosphere simultaneously increased,
suggesting localised plasma heating during the
lifetime of a swirl event.

\begin{figure}[!t]
\begin{center}
\includegraphics[width=0.48\textwidth, height=7.9cm, clip=true]{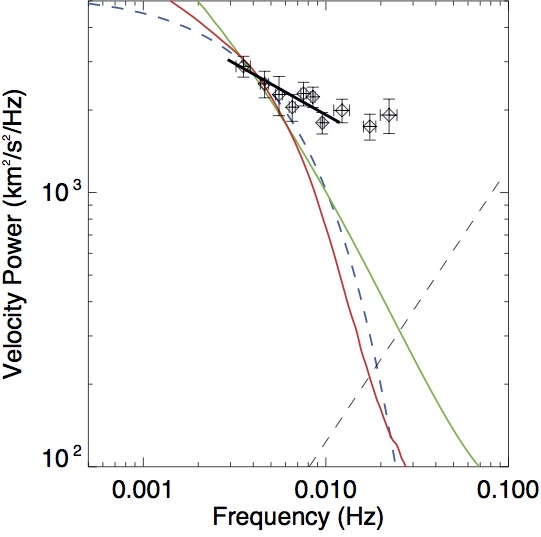}
\includegraphics[width=0.48\textwidth, height=8cm, clip=true]{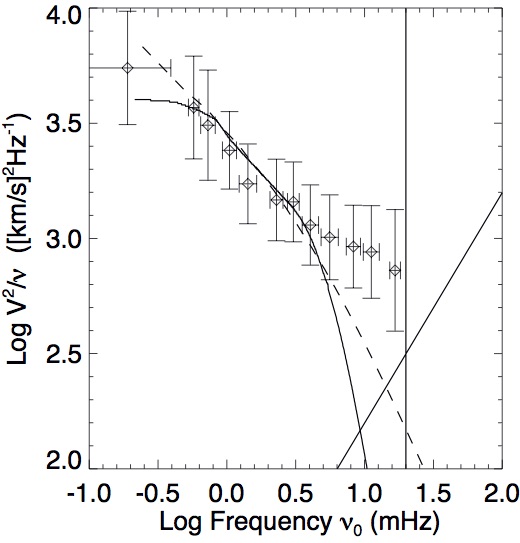}
\end{center}
\caption{Velocity power spectra of incompressible motions. The left panel shows
the velocity power spectra measured from the transverse displacements
of quiescent chromospheric fibrils (data points) in \citet{MORetal2013b}.
The coloured lines correspond to photospheric velocity power spectra from
\citet[][red dash-dot line]{MATSHI2010} and
\citet[][green solid and blue dashed lines]{CHITetal2012}.
The right hand panel is the velocity power spectra of transverse
displacements of prominence threads from \citet[][data points]{HILetal2013}.
Again the over plotted lines show the photospheric power spectra.
\label{fig:wave_spectra}}
\end{figure}

\vspace{3mm}
A more general attempt at identifying the driver of fast kink
waves in fibrils was undertaken in \citet{MORetal2013b}. The authors
measured the velocity power spectrum of the observed chromospheric
waves in order to compare it to spectra measured from granular flows (e.g.,
\citealp{MATKIT2010}; \citealp{STAetal2013}) and motions of
MBPs (\citealp{CHITetal2012}). This involved measuring over 700
sinusoidal transverse displacements in fibrils in order to produce a
statistically significant spectra. Comparison of the photospheric and
chromospheric spectra demonstrated a good correlation between the
gradients at low frequencies
(Figure~\ref{fig:wave_spectra}), hinting that the
granular motions play a dominant role in generating the transverse
waves in fibrils. These conclusions are given support from similar
observations in prominences \citep{HILetal2013}, where the
longer lifetimes of the prominence structures (relative to fibrils)
allows the extension of the `chromospheric' power spectra to even lower
frequencies (Figure~\ref{fig:wave_spectra}).

\vspace{3mm}
As for the fate of the observed waves, there have been few
observational hints. This is partly due the short time-scales of the
chromospheric structures, with both spicules and fibrils having
lifetimes on the order of $100-300$~s
(\citealp{PERetal2012,MORetal2013b}). This does
not mean that the magnetic fields vary on such short time-scales,
but is instead related to the variations in the dense plasma that
defines the structures. The short visible lifetimes mean that it becomes
difficult to track wave packets for an extended period of time,
unfortunately reducing the chance of observing their full evolution.
Additionally, short length-scales, both of the features
($\sim 4-10$~Mm) and the chromospheric scale heights
($500-1000$~km), means that fast propagating waves can
travel along the structures in tens of seconds leading
to large variations in the wave amplitudes
and phase speeds. Hence, as suggested earlier,
careful measurements of high-quality data is needed to
reveal information regarding the details of the waves propagation
over extend distances (see, \S{\,}\ref{sec:wave_seis}
for details on measurements of this type).

\vspace{3mm}
The first clue that incompressible waves suffer from wave
damping at chromospheric heights came from \citet{HEetal2009b}.
As discussed in \S{\,}\ref{sec:wave_seis}, the authors were able
to measure the variation in amplitude of an embedded kink wave over an
extended range of heights. An initial increase in amplitude is observed,
followed by a decrease in amplitude which the authors suggested
may be due to damping. However, it was not possible
to prove this, as changes in plasma density can also lead to variations
in amplitude, although it would seem unlikely that the spicule density
would increase with height. A similar profile for the amplitude of a kink
wave along a mottle was also observed in \citet{Kur13}.
Very recently, a strong observational case was made for wave
damping in spicules. \citet{MOR2014} observed a similar amplitude evolution for a kink
wave in a spicule as outlined by the two previous publications.
However, a key step was that \citet{MOR2014} was also able to
directly measure the variation in the width of the
spicule with height. This additional information was combined
with the phase speed measurements allowing the true density
profile along the spicule to be
derived. Using this information, the expected variation in amplitude
in the absence of wave damping was estimated, and it was
demonstrated that the amplitude should have continued to
increase with height (Figure~\ref{fig:wave_seis_pt2}). In light of
the this additional information, wave damping was suggested
as the cause for the observed amplitude
profile, and allowed for an estimate of the quality factor
($\xi_E=0.34$) and the frequency-independent
$\alpha$ factor ($\alpha=L_{D}/P=0.07$~{\mms}).
These values suggested that
the damping was significantly stronger than that associated with the damping of
propagating coronal waves, where measured values are
on the order of $\xi_E\sim2.69$ and
$\alpha\sim1.6$~{\mms} (\citealp{VERTHetal2010}).

\vspace{3mm}
\citet{MORetal2013b} also provided an insight into the
fate of the kink waves. The authors compared the velocity
power spectrum measured from Coronal Multi-channel Polarimeter
\citep[CoMP;][]{Tom08} observations of quiescent
coronal loops \citep{TOMMCI2009} to the chromospheric
velocity power spectrum of fibrils. The comparison demonstrated
that the coronal power is significantly less than that observed in the
chromosphere (some of which is likely due to the poorly resolved
velocity amplitudes in the CoMP data), and that the coronal spectra has a
much steeper power law. The steeper power law implies a frequency-dependent
damping mechanism is in action between the chromosphere and corona, which dissipates
higher frequency waves more efficiently. They put forward the
idea that the kink waves are mode-converted via resonant
absorption on their journey from the chromosphere to the corona,
finding an estimate for the quality factor of $\xi_E=1.35$ and
$\alpha=0.2$~{\mms}.
These estimates are spatially averaged values, which are averaged
over the distance from the chromosphere to the height of the
CoMP measurements ($\sim 15$~Mm). These results support the
idea of enhanced damping of kink waves in the
chromosphere and transition region compared to that found in the
corona.

%%%%%%%%%%%%%%%%%%%%
%%%%%%%%%%%%%%%%%%%%
%%%%%%%%%%%%%%%%%%%%
%%%%%%%%%%%%%%%%%%%%
%%%%%%%%%%%%%%%%%%%%
%%%%%%%%%%%%%%%%%%%%
%%%%%%%%%%%%%%%%%%%%
%%%%%%%%%%%%%%%%%%%%
%%%%%%%%%%%%%%%%%%%%
%%%%%%%%%%%%%%%%%%%%

\clearpage
\newpage
\section{Future Directions and Concluding Remarks}
The question of what heats the outer solar atmosphere to its multi-million degree temperatures
has remained at the forefront of astrophysical research for well over 50 years now. In order
to conclusively determine the key drivers, transportation processes and dissipation mechanisms
we must strive to answer a number of outstanding fundamental science questions, notably:
Where is the energy generated; is it locally produced in the outer solar atmosphere, or is it
something which manifests below the photospheric layers and is transported outwards through
the chromosphere to the corona? If it is the latter, then what physical processes allows this
energy to propagate upwards against the steep temperature gradient intrinsically embedded
within the solar atmosphere? Is the energy flux carried by magneto-hydrodynamic (MHD)
waves, and if so, which mode(s) of oscillation plays a dominant role in the energy transfer? Then,
ultimately, how does the wave energy flux dissipate in the form of localised heating, and what
physical mechanisms instigate and support this energy conversion?

\vspace{3mm}
Of course, the challenging optically thick, photon starved, rapidly changing and
magnetically complex nature of the chromosphere often deters observers and theorists alike.
However, it is refreshing to see that significant strides are currently
being made to detect and understand MHD wave phenomena in the lower solar atmosphere.
Indeed, in recent years there has
been a multitude of MHD wave observations documenting
localised oscillations with sufficiently high energy densities to balance
the monumental radiative losses experienced in the high-temperature
outer-atmospheric environments. However, the increased radiative losses found in certain
active regions place the dominant role MHD waves play more globally in some doubt. Furthermore,
there is still a significant way to go until we fully understand the underlying physics and mechanics,
since we are yet to physically observe wave dissipation and its subsequent conversion into heat.
On the other hand, it is only more recently with the advent of high-order
adaptive optics, high-efficiency imaging detectors and vastly improved data reduction and
analysis tools that we have begun to probe MHD wave phenomena anywhere close to the
intrinsic spatial and temporal scales it is believed to operate on.
Ground- and space-based facilities, including
IBIS, CRISP, Hinode/SOT, ROSA and
HARDcam, have provided the necessary sensitivity to be able to not only
detect MHD oscillations, but also to track their dynamic evolution as a function of spatial position,
time, and perhaps most importantly, atmospheric height. Thus, we do not have
to wait for future missions to increase our understanding -- the currently available
fleet of instruments can still be exploited to provide crucial pieces to the puzzle.

\vspace{3mm}
Looking towards the future, it is currently
expected that high-cadence imagers (e.g., Hinode/SOT, ROSA and HARDcam) will be
employed simultaneously alongside cutting-edge 2D spectropolarimeters (e.g., IBIS and CRISP)
to obtain multiwavelength time series with the highest spatial, temporal and spectral
resolutions currently achievable. Such datasets will allow key MHD wave parameters to
be extracted with unprecedented accuracy, such as their amplitudes, propagation speeds
and phase relationships. The multiwavelength nature of the data will allow detected MHD phenomena
to be traced as a function of height, through to the uppermost regions of the solar chromosphere.
Then, as these waves bombard the transition region interface, state-of-the-art space-borne satellites,
such as IRIS \citep{DeP14}, will allow the spectroscopic signatures (including thermal widths,
shock waves and rebound characteristics) to be fully investigated at the precise location where
the steepest temperature gradient resides. Only a simultaneous and comprehensive imaging and
spectral catalogue covering a vast array of atmospheric heights will allow observers to uncover
the true extent of MHD-governed energy flow through the Sun's tenuous atmosphere.
In particular, we draw the readers attention to some key questions that may be answered
through the approaches outlined above, notably:
\begin{itemize}
\item It is apparent that chromospheric measurements show distinct
variations in wave amplitudes for different structures (spicules, mottles, fibrils, etc.)
and similar structures in different regions
(quiet Sun, active regions, coronal holes, etc., Figure~\ref{fig:wave_hist}).
Why? There are at least two potential explanations for this.
One being that the variations are due to the differing local
plasma conditions of each structure, while another is
that different driving mechanisms are responsible for the waves in differing
structures. A combination of the two is also possible. A detailed derivation of
power spectra in the different structures may help to shed some
light on this question.
\item Why do measurements of uni-directional motions (i.e., upwardly or
downwardly propagating) give
significantly higher values for the mean velocity amplitude than
the measurement of periodic variations? There seems no
clear explanation for this at present. One potential option
is that uni-directional motions are not waves and just the
result of the relocation of the magnetic field by some unknown process.
\item What is driving the diverse variety of MHD waves? Discovering what the main
driving mechanism is for incompressible waves, and how this interplays with
the omnipresent photospheric $p$-mode compressible waves, will play a
vital role in assessing the validity of various wave heating
models. Ultimately, understanding this will also allow the total amount of
energy flux available for atmospheric heating to be determined reliably.
\item What happens to the wave energy? Again, this is
another important question for assessing current wave heating models.
However, this question may be much harder to answer. The
length-scales for the dissipation of wave energy may
be much smaller than current resolution limits and even those
planned for the near future, so it is unlikely that direct evidence
for dissipation will be found. Indirect signatures of wave
dissipation may potentially be sought but, to the best of our
knowledge, non have been suggested. One fate suggested for
incompressible kink waves is that they are mode-converted into torsional
Alfv{\'{e}}n motions due to resonant coupling
(e.g, \citealp{TERetal2010c}). However, no direct evidence for this
transfer of energy has yet been
documented, requiring both imaging and spectroscopic data.
While some studies of this nature are under way, it is
unclear how straightforward it will be to
interpret the Doppler-shift signatures (see, e.g., \S{\,}\ref{difficulties_section}).
\end{itemize}

\vspace{3mm}
It must be stressed that the above questions are not listed in order of importance.
However, we feel that those listed have an overarching central importance when
attempting to understand the long-standing problems of atmospheric heating
and energy transfer through the chromosphere. In addition, we
also highlight that there is huge potential for insights into the
chromosphere via solar magnetoseismology (SMS) approaches. Almost
every structure in the chromosphere shows detectable wave motion, and therefore
there is a wealth of data that is currently awaiting to be exploited using SMS
techniques. Thus, the answers to all of the key science questions
detailed above will only arise through the novel use of high-resolution
chromospheric datasets alongside the rapidly developing field of
MHD seismology.

%\begin{table}
%\caption{Observations of Alfv\'enic oscillations}
%\begin{tabular}{llll}
%\hline\noalign{\smallskip}
%Instrument & Wavelength &  Period (seconds) & Wave Mode  \\
%Hill Top/CoMP & 10474\AA (Fe XIII) & 300  &  Transverse  \\
%Hinode/SOT & 3968\AA (Ca II H) & 100 - 500  & Transverse \\
%Hinode/SOT & 3968\AA (Ca II H) & 50  & Transverse \\
%Hinode/SOT & 3968\AA (Ca II H) & 170 - 250  & Transverse \\
%DST/ROSA & 3933\AA (Ca II K) \& 6563\AA\ (H$\alpha$) & 65 - 220 & Transverse \\
%SST/SOUP & 6563\AA (H$\alpha$) & 120 - 700 & Torsional \\
%\end{tabular}
%\end{table}

\begin{acknowledgements}
D.B.J. wishes to thank the UK Science and Technology
Facilities Council (STFC) for the award of an Ernest
Rutherford Fellowship alongside a dedicated Research Grant.
R.J.M is grateful to Northumbria University for their 
support via the award of an Anniversary Fellowship and 
the Leverhulme Trust (UK) for the award of an Early Career Fellowship.
G.V. acknowledges the support of the Leverhulme Trust (UK).
V.F. and D.B.J. are grateful to The Royal Society 
(Scientific Seminar Scheme) for support in organisation of the India--UK 
scientific seminar where the ideas contained within this paper were 
first discussed.
S.D.T.G. thanks the Northern Ireland Department for
Employment and Learning for a PhD studentship. 
I.G. would like to thank the
University of Sheffield for a SHINE studentship.
\end{acknowledgements}

%\bibliographystyle{aps-nameyear}      % basic style, author-year citations
%\bibliography{example}   % name your BibTeX data base
%\nocite{*}


\end{document}